\documentclass{article}

\usepackage{epsfig} 
\usepackage{amsmath}
\usepackage{amsfonts}
\usepackage{graphicx}

 \usepackage{amssymb}

\date{\today} 
 
\newcommand{\bnabla}{\mbox{\boldmath $\nabla$}}
\newcommand{\bOmega}{\mbox{\boldmath $\Omega$}}

\newcommand{\insertplot}[5]{\begin{figure}
 \hfill\hbox to 0.05in{\vbox to #5in{\vfill
 \inputplot{#1}{#4}{#5}}\hfill}
 \hfill\vspace{-.1in}
 \caption{#2}\label{#3}
 \end{figure}}
 \newcommand{\inputplot}[3]{
 \special{ps: plotfile #1}
}
\newcounter{fig}

\begin{document}

\title{ 
Einstein--Yang-Mills--Chern-Simons solutions 
\\
in $D=2n+1$ dimensions
} 
  
\author{{\large Yves Brihaye,}$^{\dagger}$
{\large Eugen Radu}$^{ \ddagger \star }$
and {\large D. H. Tchrakian}$^{\ddagger \star  }$ \\  
$^{\dagger}${\small Physique-Math\'ematique, Universit\'e de Mons, Mons, Belgium} 
\\
$^{\ddagger}${\small School of Theoretical Physics -- DIAS, 10 Burlington
Road, Dublin 4, Ireland }
\\
$^{\star}${\small  Department of Computer Science,
National University of Ireland Maynooth,
Maynooth,
Ireland}}

\newcommand{\ta}{\theta}
\newcommand{\Si}{\Sigma}
\newcommand{\vf}{\varphi}
\newcommand{\dd}{\mbox{d}}
\newcommand{\tr}{\mbox{tr}}
\newcommand{\la}{\lambda}
\newcommand{\ka}{\kappa}
\newcommand{\f}{\phi}
\newcommand{\al}{\alpha}
\newcommand{\ga}{\gamma}
\newcommand{\de}{\delta}
\newcommand{\si}{\sigma}
\newcommand{\bomega}{\mbox{\boldmath $\omega$}}
\newcommand{\bsi}{\mbox{\boldmath $\sigma$}}
\newcommand{\bchi}{\mbox{\boldmath $\chi$}}
\newcommand{\bal}{\mbox{\boldmath $\alpha$}}
\newcommand{\bpsi}{\mbox{\boldmath $\psi$}}
\newcommand{\brho}{\mbox{\boldmath $\varrho$}}
\newcommand{\beps}{\mbox{\boldmath $\varepsilon$}}
\newcommand{\bxi}{\mbox{\boldmath $\xi$}}
\newcommand{\bbeta}{\mbox{\boldmath $\beta$}}
\newcommand{\ee}{\end{equation}}
\newcommand{\eea}{\end{eqnarray}}
\newcommand{\be}{\begin{equation}}
\newcommand{\bea}{\begin{eqnarray}}

\newcommand{\ii}{\mbox{i}}
\newcommand{\e}{\mbox{e}}
\newcommand{\pa}{\partial}
\newcommand{\Om}{\Omega}
\newcommand{\vep}{\varepsilon}
\newcommand{\bfph}{{\bf \phi}}
\newcommand{\lm}{\lambda}
\def\theequation{\arabic{equation}}
\renewcommand{\thefootnote}{\fnsymbol{footnote}}
\newcommand{\re}[1]{(\ref{#1})}
\newcommand{\R}{{\rm I \hspace{-0.52ex} R}}
\newcommand{\N}{{\sf N\hspace*{-1.0ex}\rule{0.15ex}%
{1.3ex}\hspace*{1.0ex}}}
\newcommand{\Q}{{\sf Q\hspace*{-1.1ex}\rule{0.15ex}%
{1.5ex}\hspace*{1.1ex}}}
\newcommand{\C}{{\sf C\hspace*{-0.9ex}\rule{0.15ex}%
{1.3ex}\hspace*{0.9ex}}}
\newcommand{\eins}{1\hspace{-0.56ex}{\rm I}}
\renewcommand{\thefootnote}{\arabic{footnote}}

\def\theequation{\thesection.\arabic{equation}}

\maketitle

\begin{abstract}
We investigate finite energy solutions 
of the Einstein--Yang-Mills--Chern-Simons system in odd spacetime dimensions, $D=2n+1$, with $n>1$.
Our configurations are static and spherically symmetric, approaching at infinity a Minkowski spacetime background.
In contrast with the Abelian case, the  contribution of the  Chern-Simons term is nontrivial already
in the static,   spherically symmetric  limit.  
Both globally regular, particle-like solutions and black holes are constructed numerically 
for several values of $D$.
These solutions carry a nonzero electric charge and have finite mass.
For globally regular solutions, the value of the electric charge is fixed by the Chern-Simons
coupling constant.
The black holes can be thought as non-linear superpositions of Reissner-Nordstr\"om
and non-Abelian configurations.
A systematic discussion of the solutions is given for $D=5$,
in which case  the  Reissner-Nordstr\"om black hole
becomes unstable and develops
non-Abelian hair. 
We show that some of these non-Abelian configurations are  stable 
under  linear, spherically
symmetric perturbations.
A detailed discussion of an 
exact $D=5$ solution  describing extremal 
black holes  and solitons is also provided.

 \end{abstract}
\medskip
\section{Introduction}
\setcounter{equation}{0}

In recent years the interest in the properties of gravity in more than $D=4$ 
spacetime dimensions has increased significantly.
This interest was enhanced by the development of string theory, 
which requires a ten-dimensional spacetime,
to be consistent from a quantum theoretical viewpoint.
Even in the absence of matter, solutions to the Einstein equations in dimensions higher than $3+1$ exhibit properties
which are strikingly different. 
For example, in a  $4 + 1$ dimensional asymptotically
flat vacuum spacetime with a given ADM mass and angular momentum, the geometry need not
necessarily be that of the Myers-Perry~\cite{Myers:1986un} black hole.
Notably, in this case there is the black ring~\cite{Emparan:2001wn} solution whose horizon topology is $S^2\times S^1$,
in contrast to $S^3$ of the former~\cite{Myers:1986un}.

The rapid progress in the last decade has
provided a rather extensive picture of the 
landscape of  solutions for the five dimensional case \cite{Emparan:2006mm},
including configurations with Abelian matter fields
\cite{Elvang:2003yy}-\cite{Chng:2008sr}.
Although the situation for $D> 5$ is more patchy, 
analytical \cite{Emparan:2009at} and numerical \cite{Kleihaus:2010pr} results suggest that the non-standard 
solutions found in $D= 5$ have higher dimensional generalisations; moreover, even 
more complex configurations are likely to exist as the spacetime dimension  increases.
 
In the  case of higher dimensional gravitating systems of 
nonlinear matter fields, in
particular with non-Abelian (nA) gauge fields, 
black hole and regular solutions are still relatively
scarcely explored. This is an important direction since
the theory of gravitating nA gauge fields can be regarded as the most
natural generalisation of Einstein-Maxwell theory.
Moreover, for the better known case of a $D=3+1$
dimensional spacetime, the results in the literature show that
various well-known, and rather intuitive, features of self-gravitating
solutions with Maxwell fields are not shared by their counterparts with nA gauge fields. 
For example, the Einstein-Yang-Mills (EYM)
equations admit black hole solutions that are not uniquely
characterised by their mass, angular momentum  and YM charges,
thus violating the no-hair conjecture
\cite{Volkov:sv}. Therefore the uniqueness theorem for electrovacuum
black hole spacetimes  ceases to apply for EYM systems.
Also, in contrast with the Abelian situation,
self-gravitating Yang-Mills (YM) fields can form particle-like configurations
\cite{Bartnik:1988am}. 
Another surprising result is the existence
of nA solutions which are static but not spherically symmetric \cite{Kleihaus:1996vi}.
However, since it turns out that all these asymptotically flat solutions are unstable, 
their physical relevance is obscure\footnote{For the
sake of completeness, one should mention that
the picture is very different once one gives up the assumption of asymptotic flatness.
For example, in anti-de Sitter  (AdS) $3+1$ dimensional spacetime, stable nA solutions have been shown to exist
\cite{Winstanley:1998sn}; there are also monopole and dyon solutions even in the absence
of a Higgs field.
 As found in  \cite{Gubser:2008zu},
some of the AdS nA solutions may provide a model of holographic superconductors.  
Also, the non-asymptotically flat nA solutions in \cite{Chamseddine:1997nm} 
(with a dilaton field possesing a Liouville potential), have found interesting applications
in providing  gravity
duals of ${\cal N}=1$ super-Yang-Mills theory.
}.
(A detailed review of $D=4$  gravitating particle-like and black hole solutions with nA gauge fields 
can be found in  Ref. \cite{Volkov:1998cc}.)

The study of  $D>4$  black hole 
solutions with non-Abelian matter fields is 
only in its beginnings.
Based on the experience with Einstein-Maxwell 
solutions,
it is natural to expect that higher dimensions $D > 4$ allow for a rich landscape
of  solutions that do not have four dimensional counterparts.
At the same time, considering such configurations is a legitimate task, 
since the gauged supersymmetric models 
generically contain non-Abelian fields.

Most of the solutions displayed so far in the literature
are spherically symmetric (an exception, being the results in \cite{Radu:2007jb}). 
As a new feature and in contrast with the situation in the $D=4$ case,
 a generic property of the  asymptotically flat
higher dimensional EYM solutions is that their mass  and action,
as defined in the usual way, diverge \cite{Volkov:2001tb}, 
\cite{Okuyama:2002mh}, \cite{Radu:2005mj}, \cite{Brihaye:2005tx}.
This can be understood heuristically by noting that the  
Derrick scaling requirement
\cite{Derrick:1964ww}
 is not fulfilled in
spacetimes for dimension five and higher. Finite energy solutions exist only when the
usual YM system is augmented with higher derivative corrections
in the nA action  \cite{Brihaye:2002hr}. Such terms can occur in the low energy effective action of
string theory and represent the gauge field counterparts of the
Lovelock gravitational hierarchy \cite{Lovelock:1971yv} (for a review of these aspects, see \cite{Radu:2009rs}).

In a recent work~\cite{Brihaye:2010wp}, a different way of regularising the mass
of a $D=4+1$ dimensional asymptotically flat, gravitating nA solutions was proposed. This was done 
by introducing a Chern-Simons (CS) term  in the action. The CS density is a higher order term in the YM
curvature and connection, and as such can be viewed as an $alternative$ to the 
higher order curvature terms of the YM hierarchy employed previously in \cite{Radu:2005mj,Brihaye:2002hr}.
It turns out that this prescription $does$ result in finite mass globally regular and black hole
solutions and leads to a variety of new features as compared to the well known case of $D=4$  EYM  solutions.
For example, these configurations cary an electric charge, emerging as perturbations of the Reissner-Nordstr\"om
(RN) black holes.
Moreover, in contrast to all other known asymptotically flat nA black holes without scalars,
some of these solutions in \cite{Brihaye:2010wp} were found to be stable
under linear, spherically symmetric perturbations.
Also, for a particular value of the CS coupling constant, it was possible to construct both solitons
and extremal black hole solutions, by exploiting the model in \cite{Gibbons:1993xt}. 

In this work we propose a general framework for the study of Einstein--Yang-Mills--Chern-Simons (EYMCS) solutions
for an arbitrary  $D=2n+1$ spacetime dimension.
Our configurations are static and spherically symmetric, approaching at infinity a Minkowski spacetime background.
Based on numerical results for $D=5$ and $D=7,9$, we conjecture that the presence of a CS term in the action
allows for finite energy solutions for any $D=2n+1$, with $n>1$. (The case $D=3$ is special, 
since it requires the presence of a negative cosmological constant.)
Most of the numerical results in this paper are for the $D=5$, in which case, we provide a systematic discussion of the
solutions in \cite{Brihaye:2010wp}. We have also discussed some results for $D=7,9$, which reveal some 
new features of the solutions.

The paper is structured as follows: in Section {\bf 2} we present the general framework and analyse the field equations
for an $SO(D+1)$ gauge group. 
In Section {\bf 3}, the general features a consistent truncation  of the general model for an $SO(D-1)\times SO(2)$
gauge group are discussed.
Numerical results for $D=5$ and $D>5$, respectively, are presented in Sections {\bf 4} and {\bf 5}. 
We conclude with Section {\bf 6} where the significance of, and further consequences arising from, the solutions
we have constructed are discussed.

\section{The general model}
\setcounter{equation}{0}
\subsection{The action and field equations}

In odd spacetime dimensions, the usual gauge field action can be augmented  
by a (dynamical) Chern-Simons (CS) term. Restriction to odd dimensions follows from the fact that Chern-Pontryagin (CP)
densities are defined only in even dimensions, and the CS density is defined formally in one dimension lower than the CP
density. The resulting odd dimensional space is then interpreted as the spacetime on which the dynamical CS term appears
in the Lagrangian.

Such  terms appear in various supersymmetric theories,
the ${\cal N}=8,~D=5$  gauged supergravity model 
\cite{Gunaydin:1985cu}, \cite{Cvetic:2000nc} being perhaps the best known case,
due to its role in the conjectured AdS/CFT correspondence.  
However, in this work we shall restrict ourselves to a simple
EYMCS model, which does not seem to correspond to
a consistent truncation of any gauged supergravity model.
Also, our solutions approach asymptotically
the Minkowski spacetime background.
In the case of an Abelian gauge group in $D=5$, a CS term 
leads to some new features\footnote{Note that the situation can be different for
charged magnetic branes, see $e.g.$
the asymptotically AdS$_5$ Abelian solutions with a CS term in  \cite{D'Hoker:2009bc}.} 
only for rotating black holes ~\cite{Gauntlett:1998fz}.
However, we shall see that for a nA gauge group,
the CS term  can  affect the properties of solutions even in the static, spherically symmetric case.  

We consider the following action for the EYMCS model in $D=2n+1$ dimensions
\bea
\label{action}
S= \int_{ \mathcal{M}}  d^Dx \sqrt{-g} \left( \frac{R}{16\pi G} 
-{\cal L}_{\rm{YM}}  \right)
-\kappa \int_{ \mathcal{M}}  d^D x~ {\cal L}_{\rm{CS}}^{(D)} ,
\eea
where 
\begin{eqnarray}
&&{\cal L}_{\rm{YM}}=
\frac{1}{2}\,\mbox{Tr}\bigg\{ F_{\mu\nu}F^{\mu\nu} \bigg \},
\end{eqnarray}
is the usual Yang-Mills lagrangian, 
with 
\begin{eqnarray}
F_{\mu\nu}=\partial_\mu A_\nu-\partial_\nu A_\mu+e [A_\mu,A_\nu]
\end{eqnarray}
 the gauge field
strength tensor, 
while $e$ and $\kappa$ are the gauge and the CS coupling constant, respectively. 
(The value of 
$\kappa$ is fixed in supersymmetric theories.
However, in this work we shall treat $\kappa$  as a free input parameter.
This   has been also motivated by the study in \cite{Kunz:2006yp} of the Einstein-Maxwell-CS system, 
which revealed a nontrivial dependence of the properties of the solutions on the value
of $\kappa$.)

The definition of the Chern-Simons density on $D-$dimensional spacetime follows from that of the corresponding
Chern-Pontryagin
density on $D+1$ (even) dimensions. The latter is, by definition, a total divergence
\[
\bnabla\cdot\bOmega=\mbox{Tr} 
\bigg\{ 
F\wedge{F}\dots\wedge{F}
\bigg\}
,\qquad n\ {\rm times}
\]
in $2n-$dimensions, and the CS density on a $D=2n-1$ dimensional spacetime is formally defined
as one of the $2n$ components of the density $\bOmega$.

The CS densities ${\cal L}_{\rm{CS}}^{(D)}$ thus defined are $gauge\ variant$. The explicit expressions of
first three, in $D=3,5$ and $7$ dimensional spacetimes, are
\bea
{\cal L}_{\rm CS}^{(3)}&=&
\vep^{\la\mu\nu}\mbox{Tr}\,
\bigg\{A_{\la}\left[F_{\mu\nu}-\frac23 e_{~} A_{\mu}A_{\nu}\right]
\bigg\},
\label{CS3}
\\
{\cal L}_{\rm CS}^{(5)}&=&
\vep^{\la\mu\nu\rho\si}\mbox{Tr}\,
\bigg\{A_{\la}\left[F_{\mu\nu}F_{\rho\si}-e_{~} F_{\mu\nu}A_{\rho}A_{\si}+
\frac25 e^2_{~} A_{\mu}A_{\nu}A_{\rho}A_{\si}\right]
\bigg\},
\label{CS5}\\
{\cal L}_{\rm CS}^{(7)}&=&
\vep^{\la\mu\nu\rho\si\tau\ka}
\mbox{Tr}\,
\bigg\{
A_{\la}\bigg[F_{\mu\nu}F_{\rho\si}F_{\tau\ka}
-\frac45 e_{~} F_{\mu\nu}F_{\rho\si}A_{\tau}A_{\ka}
-\frac25 e_{~}^2 F_{\mu\nu}A_{\rho}F_{\si\tau}A_{\ka}
\nonumber
\\
\label{CS7}
&&\qquad\qquad\qquad\qquad\qquad\qquad
+\frac45 e_{~}^3 F_{\mu\nu}A_{\rho}A_{\si}A_{\tau}A_{\ka}
-\frac{8}{35} e_{~}^4 A_{\mu}A_{\nu}A_{\rho}A_{\si}A_{\tau}A_{\ka}\bigg]
\bigg\}
\,{},
\eea
which are all manifestly $gauge\ variant$. Remarkably however, the Euler--Lagrange variations of these densities are actually
$gauge\ covariant$. Indeed, these variational terms are expressed in $gauge\ covariant$ form for arbitrary $D=2n+1$ as
\[
(n+1)\
\vep^{\mu_{1}\mu_{2}\mu_{3}\mu_{4}\dots\mu_{2n-1}\mu_{2n}}F_{\mu_{1}\mu_{2}}F_{\mu_{3}\mu_{4}}\dots F_{\mu_{2n-1}\mu_{2n}}\,.
\]
Perhaps what is still more relevant in our case, where we restrict attention to $static$ solutions only, is the fact that the
CS densities \re{CS3}-\re{CS7}, $etc.$, in that case reduce to a very useful form which can be expressed for the arbitrary
$D=2n+1$ case. 
Working in a gauge such that $\partial_t A_{\mu}=0$,
one can show that,
up to a total divergence term (which we ignore here since we are only interested in the Euler-lagrange equations),
the effective arbitarary $n$ CS Lagrangian is
\[
{\cal L}_{\rm CS}^{(2n+1)}=(n+1)\,\vep^{i_1i_2i_3i_4\dots i_{2n-1}i_{2n}}\mbox{Tr}
\bigg\{ 
\,A_0\,F_{i_1i_2}\,F_{i_3i_4}\dots
F_{i_{2n-1}i_{2n}}
\bigg\}
\,.
\]

The field equations are obtained by varying the action (\ref{action}) with
respect to the field variables $g_{\mu \nu},A_{\mu}$ 
\begin{eqnarray}
\label{Einstein-eqs}
&&R_{\mu \nu}-\frac{1}{2}g_{\mu \nu}R =
 8 \pi G ~T_{\mu \nu},
\\
\nonumber
&&D_{\mu}\left(\sqrt{-g}\,F^{\mu\tau}\right)=
 {\kappa}  \frac{(D+1)}{2 \sqrt{-g} } \vep^{\tau \la\mu\dots \nu\rho}   F_{\la\mu}\dots F_{ \nu\rho} ,
\end{eqnarray}
where 
\begin{eqnarray}
\label{Tij}
T_{\mu\nu} =
    2
    \mbox{Tr}\bigg \{F_{\mu\alpha} F_{\nu\beta} g^{\alpha\beta}
   -\frac{1}{4} g_{\mu\nu}  ~ F_{\alpha\beta} F^{\alpha\beta}
  \bigg \},
\end{eqnarray}
is the energy momentum tensor.
 One can show that this tensor is covariantly conserved ($i.e.$ $\nabla_\mu T^{\mu\nu}=0$)
 for solutions of the YMCS equations.
  
In what follows, we shall be seeking to construct finite mass/energy solutions of the
above equations. This is made possible by the fact that the
energy density functional arising from the Lagrangian \re{action} actually satisfies Derrick scaling~\footnote{Strictly speaking
Derrick scaling \cite{Derrick:1964ww} applies only in flat space background. However,
 in practice it works also for gravitating configurations in
asymptotically flat spaces  \cite{Heusler:1996ft} (at least in the spherically symmetric
case).} by virtue of the presence
of the CS term in it. This is because the CS term, which in $D=2n+1$ dimensional spacetime scales as $L^{-(2n+1)}$,
balances the Yang-Mills term which scales as $L^{-4}$. Specifically, in the $2n=D-1$ space dimensions, $2n+1\ge 2n\ge 4$,
for the cases of interest here, namely for $n\ge 2$. 

Inasfar as the CS term here plays the role of regularising the energy by providing the required Derrick balance, this makes it
an alternative to employing YM higher order curvature terms~\cite{Radu:2005mj,Brihaye:2002hr} for this purpose. The latter is of
course more versatile since its use is not restricted, as in the CS cases, to $2n+1$ dimensional spacetimes.
 
\subsection{A spherically symmetric Ansatz }

\subsubsection{The metric}

In $D$-dimensional spacetime, we restrict to static fields that
are spherically symmetric in the $D-1$ spacelike dimensions with a
general metric Ansatz
\begin{eqnarray}
\label{metric-gen} 
ds^{2}=f_1(r)dr^2 +f_2(r) d\Omega^2_{D-2}-  f_0(r) dt^{2},
\end{eqnarray} 
where
 $r$ and $t$ are the radial and time coordinates, while $d\Omega^2_{D-2}$ is the metric on the round $(D-2)$-sphere 
 (note that this Ansatz has still
 some freedom in the choice of the radial coordinate).
 
The numerical work has been done  for a metric gauge choice $f_2(r)=r^2$ and 
\begin{eqnarray}
\label{formN}
f_1(r)=N(r),~~~f_0(r)=N(r)\sigma^2(r)~~{\rm where}~~ N(r)=1-\frac {m(r)}{r^{D-3}} ,
\end{eqnarray}
the function $m(r)$ being related to the local mass-energy density (as defined in the standard way) up to some
 $D$-dependent factor.
 
 Another convinient metric gauge choice used in the literature
 is
 \begin{eqnarray}
\label{formf}
f_1(r)=\frac{f_2(r)}{r^2}=\frac{m(r)}{f(r)},~~~f_0(r)=f(r) ,
\end{eqnarray}
corresponding to an isotropic coordinate system (the $D=5$ exact solution
discussed in Section {\bf 4.4} is found  for this choice of coordinates).

\subsubsection{The YM fields}

The construction of a  static static, spherically symmetric  
YM Ansatz in $D=5$ spacetime dimensions leading to a nonzero 
CS term has been discussed in \cite{Brihaye:2009cc}.
 In what follows we present an extension of that result  for a generic $D=2n+1$ case.

There is some arbitrariness in the choice of the gauge group.
The only restriction is that it should be large enough 
to accomodate for a static spherically symmetric Ansatz,
with a nonvanishing  electric potential\footnote{For the most interesting
case $D=5$, this condition rules out the possibility of using the minimal 
non-Abelian gauge group $SO(3)$ \cite{Okuyama:2002mh}.}.
In $D=2n+1$ dimensions, the smallest simple gauge group supporting a nonvanishing
CS term is $SO(2n+2)$.
Here we shall  take the $SO(2n+2)$ YM fields in one or other
chiral representation of $SO_{\pm}(2n+2)$.
Our spherically
symmetric Ansatz is expressed in terms of the representation matrices
\be
\label{sigmap}
\Sigma_{\al\beta}^{(\pm)}=-\frac{1}{4}\left(\frac{1\pm\Gamma_{2n+3}}{2}\right)
[\Gamma_{\al} ,\Gamma_{\beta}]\quad,\quad \al,\beta=1,2,...,2n+2\ ,
\ee
$\Gamma_{\al}=(\Gamma_{i},\Gamma_{M})$, with the index $M=(2n+1,2n+2)$, being
the gamma matrices in $2n+2$ dimensions and $\Gamma_{2n+3}$,
the corresponding chiral matrix\footnote{Thus, the fact that we are using an antihermitian representation for the
$SO(D+1)$ algebra matrices leads to a factor of $i$ 
in front of ${\cal L}_{CS}^{(D)}$.}.
We shall adopt a normalization convention such that
\bea
2 \mbox{Tr}\,
\bigg\{ \Sigma_{\al\beta}^{(\pm)},\Sigma_{\al ' \beta '}^{(\pm)} \bigg\}
=\delta_{\alpha \alpha'}\delta_{\beta \beta'}~.
\eea
Our construction of a spherically symmetric gauge field Ansatz is based on the formalism of A. Schwarz
~\cite{Schwarz:1977ix}.  
An alternative formalism, \cite{Forgacs:1979zs}, is familiar in the literature, but the calculus of
\cite{Schwarz:1977ix} was found to be more convenient for the purposes of this work.

Our spherically symmetric Ansatz for the YM connection $A_{\mu}=(A_0,A_i)$ is
\bea
A_0&=&
\frac{1}{e}
\bigg \{
-(\vep\chi)^M\,\hat x_j\,\Sigma_{jM}^{(\pm)}-
\chi^{2n+3}\,\Sigma_{2n+1,2n+2}^{(\pm)}
\bigg \},
\label{a0p}
\\
A_i&=&\frac{1}{e}
\bigg \{
\left(\frac{\f^{2n+3}+1}{r}\right)\Sigma_{ij}^{(\pm)}\hat x_j+
\left[\left(\frac{\f^M}{r}\right)\left(\delta_{ij}-\hat x_i\hat x_j\right)+
(\vep A_r)^M\,\hat x_i\hat x_j\right]\Sigma_{jM}^{(\pm)}+
\nonumber
\\
&&\qquad\qquad\qquad\qquad\qquad\qquad\qquad\qquad +A_r^{2n+3}\,
\hat x_i\,\Sigma_{2n+1,2n+2}^{(\pm)}
\bigg \},
\label{aip}
\eea
(with $x^i$ the usual Cartesian coordinates on $R^{D-1}$ and $x^0=t$)
in which the summed over indices $M,N=2n+1,2n+2$ run over two values such that
we can label the functions $(\f^M,\f^{2n+3})\equiv\vec\f$,
$(\chi^M,\chi^{2n+3})\equiv\vec\chi$ and $(A_r^M,A_r^{2n+3})\equiv\vec A_r$
like three isotriplets $\vec\f$, $\vec\chi$ and $\vec A_r$, all depending on
the $2n$ dimensional spacelike radial variable $r$ and time $t$. $\vep$ is the two
dimensional Levi-Civita symbol, while $\hat x_i=x_i/r$  (with $x_ix_i=r^2$).

In what follows we are interested in configurations without a dependence on time.
Then
the parametrisation used in the Ansatz \re{a0p}-\re{aip} results in a gauge
covariant expression for the YM curvature $F_{\mu\nu}=(F_{ij},F_{i0})$
\bea
F_{ij}&=&\frac{1}{e}
\bigg \{
\frac{1}{r^2}\left(|\vec\f|^2-1\right)\Sigma_{ij}^{(\pm)}+
\frac1r\left[D_r\f^{2n+3}+\frac1r\left(|\vec\f|^2-1\right)\right]
\hat x_{[i}\Sigma_{j]k}^{(\pm)}\hat x_{k}+
\frac1rD_r\f^M\hat x_{[i}\Sigma_{j]M}^{(\pm)}
\bigg \}
,
\label{fijp}
\\
F_{i0}&=&
\frac{1}{e}
\bigg \{
-\frac1r\,\f^M(\vep\chi)^M\,\Sigma_{ij}^{(\pm)}\hat x_j+\frac1r\,
\left[\f^{2n+3}(\vep\chi)^M-\chi^{2n+3}(\vep\f)^M\right]\Sigma_{iM}^{(\pm)}
\nonumber
\\
&-&\left [ (\vep D_r\chi)^M+\frac1r\,
\left[\f^{2n+3}(\vep\chi)^M-\chi^{2n+3}(\vep\f)^M\right]\right ]
\hat x_i\hat x_j\Sigma_{jM}^{(\pm)}
-D_r\chi^{2n+3}\,\hat x_i\,\Sigma_{2n+1,2n+2}^{(\pm)}
\bigg \}
,
\label{fi0p}
\eea
in which we have used the notation
\be
\label{covp}
D_r\f^a=\pa_r\f^a+\vep^{abc}\,A_r^b\,\f^c\quad,\quad
D_r\chi^a=\pa_r\chi^a+\vep^{abc}\,A_r^b\,\chi^c
,
\ee
as the $SO(3)$ covariant derivatives of the two triplets
$\vec\f\equiv\f^a=(\vec\f^M,\f^{2n+3})$,
$\vec\chi\equiv\chi^a=(\vec\chi^M,\chi^{2n+3})$, with respect to the
$SO(3)$ gauge connection $\vec A_r\equiv A_r^a$.


After taking the traces over the spin matrices, it is convenient to relabel the triplets of radial function as
\begin{eqnarray}
\vec\f\equiv(\f^M,\f^{3}), ~~~~\vec\chi\equiv(\chi^M,\chi^{3})~~~ {\rm and}~~~~\vec A_r\equiv(A_r^M,A_r^{3}),
\end{eqnarray}
 with $M=1,2$ now.
 
 The triplet,
$\vec A_r(r)$, plays the role of a connection in the residual one dimensional system after the imposition of symmetry,
and encodes the $SO(D-1)$ arbitrariness of this Ansatz. In one dimension there is no curvature hence it can be
gauged away in practice~\cite{Brihaye:2009cc}.
  
However, 
finding solutions within the YM Ansatz \re{a0p}, \re{aip}  (which after setting $\vec A_r=0$ still
features six independent functions), is technically a difficult task. 
A further consistent trucation of the general Ansatz is $\f^2=\chi^2=0$, leading to an EYMCS system with six unknown
functions, four of them being gauge potentials parametrising the gauge field, and, two metric functions.  
Indeed, the
two gauge functions suppressed are redundent and would only be excited in an eventual stability analysis of our
solutions. 
To make connection with previous results on  EYM solutions, we shall note
\begin{eqnarray}
\label{red1}
\phi^1(r)=\tilde w(r),~~\phi^3(r)=w(r),~~
\chi^1(r)=\tilde V(r),~~\chi^3(r)=V(r), ~~~{\rm with}~~~\vec A_r=0.
\end{eqnarray}
The Ref. \cite{Brihaye:2009cc} gave numerical evidence for the existence of 
 solutions within the above Ansatz
for the case of $D=5$ ($i.e.$ an $SO(6)$ gauge group)
and AdS asymptotics.
Some of the features discussed there are generic.
For example, the resulting system has some residual symmetry under a rotation of the 'doublets'
$w(r),\tilde w(r)$ and $V(r),\tilde V(r)$ with the same constant angle
$u$ ($e.g.$ $w\to w \cos u+\tilde w \sin u$ etc.).
Note  that for configurations with $\tilde w(r) = \tilde V(r) =0$
the gauge potentials are unvariant under the "chiral" transformations
generated by $\Sigma_{2n+1,2n+2}^{(\pm)}$. The configurations with $w(r)=V(r)=0$ instead change
just by a sign under the same transformations.
Also, this Ansatz is invariant under the parity transformation $\phi^a \to -\phi^a,~\chi^a \to -\chi^a $.

A further simplification of the YM Ansatz consists in taking 
\begin{eqnarray}
\label{red2}
 \tilde w(r)=\tilde V(r)=0,
\end{eqnarray}
which is a consistent truncation, $SO(D-1)\times SO(2)$,  of
the general Ansatz.

\subsection{The equations}


The truncated YM Ansatz (\ref{red1}) together with the generic metric element (\ref{metric-gen})
leads to the following set of EYMCS equations:
\begin{eqnarray}
\label{eqf0}
f_0''-\frac{f_0'^2}{2f_0}-\frac{f_0'f_1'}{2f_1}
+\frac{D-3}{2f_2} (f_0' f_2'+2 f_0 f_1-\frac{f_0f_2'^2}{2f_2})
-\frac{\alpha^2}{4 f_2}
\bigg(
(D-2)(2D-7) f_1 (\tilde V w-V \tilde w)^2
\\
\nonumber
+2(2D-5)f_2 (V'^2+\tilde V'^2)
+\frac{3(D-2)(D-3)f_0f_1}{f_2}(w^2+\tilde w^2-1)^2
+2(D-2)f_0 (w'^2+\tilde w'^2) 
\bigg)=0,
\\
\label{eqf1}
\frac{f_0' f_2'}{f_0} 
+\frac{(D-3)f_2'^2}{2f_2}-(D-3)f_1
+ \alpha^2 
\bigg(
 \frac{f_2}{f_0}(V'^2+\tilde V'^2)
-(D-2)(w'^2+\tilde w'^2)
\\
\nonumber
+\frac{(D-2)(D-3)f_1}{2f_2}(w^2+\tilde w^2-1)^2
-\frac{(D-2) f_1}{2f_0}(\tilde V w-V \tilde w)^2
\bigg)
=0,
\\
\label{eqf2}
f_2''-\frac{f_1'f_2'}{2f_1}+\frac{(D-5)f_2'^2}{4 f_2}
-(D-3)f_1
+\frac{\alpha^2}{2}
\bigg(
(D-2)(w'^2+\tilde w'^2) +\frac{f_2}{f_0}(V'^2+\tilde V'^2)
\\
\nonumber
\frac{(D-2) f_1}{2f_0}(\tilde V w-V \tilde w)^2
+\frac{(D-2)(D-3)f_1}{2f_2}(w^2+\tilde w^2-1)^2
\bigg)
=0,
\end{eqnarray}
\begin{eqnarray}
\label{eqw}
w''+\frac{1}{2}(\frac{f_0'}{f_0}-\frac{f_1'}{f_1}+\frac{(D-4)f_2'}{f_2})w'
+\frac{f_1}{2f_0}\tilde V( \tilde V w-V \tilde w)
-\frac{(D-3)f_1}{f_2}w(w^2+\tilde w^2-1)
\\
\nonumber
-\kappa  \frac{(D^2-1)f_1}{(D-2)f_2^{(D-4)/2}\sqrt{f_1f_0}}(w^2+\tilde w^2-1)^{(D-5)/2} 
\bigg(
(w^2+\tilde w^2-1)   V'+(D-3)( V \tilde w-\tilde V  w)\tilde w'
\bigg)
=0,
\\
\label{eqtw}
\tilde w''+\frac{1}{2}(\frac{f_0'}{f_0}-\frac{f_1'}{f_1}+\frac{(D-4)f_2'}{f_2})\tilde w'
+\frac{f_1}{2f_0}  V(V \tilde w- \tilde V w)
-\frac{(D-3)f_1}{f_2} \tilde w(w^2+\tilde w^2-1)
\\
\nonumber
-\kappa\frac{(D^2-1) f_1}{(D-2)f_2^{(D-4)/2}\sqrt{f_1f_0}}(w^2+\tilde w^2-1)^{(D-5)/2} 
\bigg(
(w^2+\tilde w^2-1) \tilde V'+(D-3)(\tilde V w-V \tilde w)w'
\bigg)
=0,
\end{eqnarray}
\begin{eqnarray}
\label{eqV}
V''+
\bigg
(
\frac{(D-2)f_2'}{2f_2}-\frac{f_0'}{2f_0}
-\frac{f_1'}{2f_1}
\bigg)
V'
-\frac{(D-2)f_1 \tilde w}{2f_2}(V \tilde w-\tilde V w)
\\
\nonumber
-\kappa (D^2-1) \sqrt{f_0f_1}f_2^{1-D/2}(w^2+\tilde w^2-1)^{(D-3)/2}w'=0,
\\
\label{eqtV}
\tilde V''+
\bigg
(
\frac{(D-2)f_2'}{2f_2}-\frac{f_0'}{2f_0}
-\frac{f_1'}{2f_1}
\bigg)
\tilde V'
-\frac{(D-2)f_1 \tilde w}{2f_2}(\tilde V   w-  V \tilde w)
\\
\nonumber
-\kappa (D^2-1) \sqrt{f_0f_1}f_2^{1-D/2}(w^2+\tilde w^2-1)^{(D-3)/2}\tilde w'=0,
\end{eqnarray}
where a prime denotes a
derivative with respect to $r$.
Also, to simplify the expression of the above relations, we note
\begin{eqnarray}
\alpha^2=\frac{16\pi G}{(D-2)e^2}
\end{eqnarray}
 and absorb a factor of $1/e^{(D-3)/2}$ in the expression of $\kappa$.

After fixing a metric gauge,
the eq. (\ref{eqf1}) becomes an Hamiltonian constraint.
There is also a constraint  equation for the gauge fields,
\begin{eqnarray}
\label{constr1} 
\frac{f_2^{\frac{D-2}{2}}}{\sqrt{f_0f_1}}(\tilde V V'-V \tilde V')
+(D-2)f_2^{\frac{D-4}{2}} \sqrt{\frac{f_0}{f_1}}(w \tilde w'-\tilde w  w' )
-  \kappa  (D^2-1)(\tilde V w-V \tilde w)(w^2+\tilde w^2-1)^{\frac{D-3}{2}}=0\,,{~~}
\end{eqnarray}
which originates from the variational equation for $\vec A_r$ (one can show that (\ref{constr1}) is a first integral of the
eqs. (\ref{eqw})-(\ref{eqtV})).

Also, in what follows, we shall restrict to a dimension $D\geq 5$.
The case $D=3$ should be discussed separately, since the existence of physically interesting solutions
requires the presence of a cosmological constant.

\subsection{The asymptotics and a truncation }
Numerical evidence for the existence of
asymptotically AdS$_5$ solutions within the full Ansatz (\ref{red1}) was given in Ref. \cite{Brihaye:2009cc}. 
However, it seems that the presence in that case of a negative cosmological
constant was crucial in arriving at that result\footnote{This is not an entirely surprising
result. We recall that already in $D=4$ dimensions and a gauge group $SO(3)$, the presence
of a negative cosmological constant $\Lambda$ leads to some new qualitative features \cite{Winstanley:1998sn}.
In particular, $\Lambda<0$ allows for
EYM static solutions with a non-vanishing electric potential, which is not the case for
 asymptotically flat configurations \cite{Galtsov:1989ip}.
}. In the asymptotically flat case,
we could not find such solutions
(with four essential functions) but only configurations within 
the restricted $SO(D-1)\times SO(2)$ Ansatz (\ref{red2}).

Although we do not have a rigurous proof of that, some analytical
indications in this direction
come from the study of the asymptotics of the general solutions close to the horizon 
and at infinity.
Here it is convenient to use the metric Ansatz (\ref{formN}), with two functions $N(r)$
and $\sigma(r)$.
For black hole solutions, the horizon is located at $r=r_h>0$,
with $N(r_h)=0$ and $\sigma(r_h)>0$, while $N'(r_h)>0$ in the nonextremal case.

Then we suppose that all functions admit the following behaviour as $r \to r_h$: 
\begin{eqnarray}
\nonumber
&&w(r)= \cos U_1 \sum_{k=0}^\infty w_k(r-r_h)^k,~~\tilde w(r)= \sin U_1 \sum_{k=0}^\infty \tilde w_k(r-r_h)^k,
\\
\label{eh}
&&
V(r)= \cos U_1\sum_{k=0}^\infty V_k(r-r_h)^k,~~\tilde V(r)= \sin U_1\sum_{k=0}^\infty \tilde v_k(r-r_h)^k,
\\
\nonumber
&&N(r)=  
 \sum_{k=1}^\infty \bar N_k(r-r_h)^k,~~~\sigma(r)= \sum_{k=0}^\infty \bar \sigma_k(r-r_h)^k,
\end{eqnarray}
which satisfy the regularity condition $w  \tilde V - \tilde w V \to 0$ as $r \to r_h$. 
The cofficients $w_k$, $\tilde w_k$,  $\tilde V_k$,
$m_k$ and $\sigma_k$ are computed order by order by substituting 
this expansion in the field equations.
It turns out that the only free parameters are $\sigma(r_h)$, $w(r_h)$ and $v_1$.
Moreover, we have verified that, at least up to order four\footnote{Beyond this order, the involved relations were too
complicated to deal with.}
$w_k/\tilde w_k=V_k/\tilde V_k=1$.

Interestingly, a similar analysis for large values of $r$ leads to the same conclusions.
Here we suppose the solutions admit a  power series expansion with
\begin{eqnarray}
\nonumber
&&w(r)=\cos U \sum_{k=0}^\infty \frac{W_k}{r^k},~~\tilde w(r)=\sin U \sum_{k=0}^\infty \frac{\tilde W_k}{r^k},~~
 V(r)=\cos U \sum_{k=0}^\infty \frac{V_k}{r^k} ,~~\tilde V(r)=\cos U \sum_{k=0}^\infty \frac{\tilde V_k}{r^k},
\\
\label{infinity}
&&
N(r)=1+\sum_{k=D-3}^\infty \frac{M_k}{r^k},~~
\sigma(r)=1+\sum_{k=1}^\infty \frac{\sigma_k}{r^k}~.
\end{eqnarray}
After plugging this expansion in the field equations, we have found that, 
$W_k=\tilde W_k$ and $V_k=\tilde V_k$,  at least up to order $D+5$.
Moreover, the only free parameters in the above expressions are $W_2$ and $v_2$.

This result, together with the corresponding one 
for the near horizon expansion (\ref{eh})
strongly suggests that the functions $w(r),\tilde w(r)$ and $V(r),\tilde  V(r)$
have a constant ratio for any $r>r_h$ for any physical solution of (\ref{eqf0})-(\ref{eqtV}).
Although we do not have a rigurous proof,   
this conjecture has been confirmed by our numerics and 
all black hole solutions we have found have in fact 
only two essential gauge functions\footnote{Although we could construct
$D=5$ black hole solutions
within the general ansatz (\ref{red1}), it turns out that,
 within the numerical accuracy,
the ratios $w(r)/\tilde w(r)$ and $V(r)/\tilde  V(r)$
were in fact always constant.}.
This applies also for asymptotically flat solitons ($i.e.$
without an event horizon), 
in which case we have also failed to find 
asymptotically solutions within the general Ansatz  (\ref{red1}).
In asymptotically AdS$_5$ spacetime,
the asymmetry 
between $w,\tilde w$
and $V,\tilde V$ explictly
appears in the large-$r$ behaviour, 
being introduced by the cosmological term,
see the results in Section {\bf 2} of Ref. \cite{Brihaye:2009cc}.

Then, for the remaining of this work we shall
deal with the case of solutions within the restricted $SO(D-1)\times SO(2)$
 Ansatz with two esential functions: a magnetic potential,
 $w(r)$,
 and an electric one, $V(r)$.

\section{The $SO(D-1)\times SO(2)$ model. General properties}
\setcounter{equation}{0}
\subsection{The equations and scaling properties}
The equations of the model
simplify drastically for the truncation (\ref{red2}).
Their form within the metric parametrization  (\ref{formN}),
reads
\begin{eqnarray}
&&
\nonumber
w''
+(\frac{D-4}{r}+\frac{N'}{N}+\frac{\sigma'}{\sigma})w'
-\kappa\frac{(D^2-1)}{(D-2)}\frac{(w^2-1)^{\frac{D-3}{2}}}{\sigma N r^{D-4}}V'
+\frac{(D-3)w(1-w^2)}{r^2 N}=0,
\\
&&
\label{red-eqs}
V''+(\frac{D-2}{r}-\frac{\sigma'}{\sigma})V'
-\kappa\frac{(D^2-1)}{r^{D-2}}\sigma (w^2-1)^{\frac{1}{2}(D-3)}w'=0,
\\
&&
\nonumber
m'=\frac{\alpha^2}{2} r^{D-2}
\left(
\frac{V'^2}{\sigma^2}+\frac{(D-2)N w'^2}{r^2}
+(D-2)(D-3)\frac{(1-w^2)^2}{2r^4}
\right),
\\
\nonumber
&&
\sigma'=\alpha^2 (D-2)\frac{\sigma w'^2}{2r}
,
\end{eqnarray} 
the gauge constraint (\ref{constr1}) vanishing identically.
These equations can also be derived from the effective action
\begin{eqnarray}
\label{Leff}
S_{eff}=\int dt dr \bigg  \{
\sigma m' -\frac{1}{2}\alpha^2 
  \bigg [
r^{D-2} 
\bigg(
(D-2)\frac{N \sigma  w'^2}{r^2}
-\frac{V'^2}{\sigma}
 +\frac{(D-2)(D-3)}{2 r^4}\sigma (1-w^2)^2
 \bigg )
 \\
 \nonumber
{~~~~~~}-2\kappa(D^2-1)V (w^2-1)^{\frac{D-3}{2}}w'
 \bigg ]
\bigg \},
\end{eqnarray} 
(note that, as required, there is no coupling with the geometry for the term proportional with $\kappa$).

A generic feature of the YMCS model within the $SO(D-1)\times SO(2)$
truncation is the existence of a 
 first integral  for the electric potential $V(r)$,
\begin{eqnarray}
\label{fi}
V'=\frac{\sigma}{r^{D-2}}
\left(
\frac{P}{\alpha^2}+(D^2-1)\kappa F(w)
\right ),
\end{eqnarray}
where $P$ is an integration constant (we shall
see that,  for globally regular solutions, it's value is fixed by $\kappa$).
 The function $F(w)$ has the following general expression in terms of the hypergeometric function
 $_2F_1$:
\begin{eqnarray}
F(w)=(-1)^{\frac{1}{2}(D+1)}~_2F_1(\frac{1}{2},\frac{3-D}{2},\frac{3}{2};w^2)w,
\end{eqnarray} 
its explicit form for several dimensions being
\begin{eqnarray}
\nonumber
&&F(w)= w,~~~{\rm for}~~D=3;
~~~~~F(w)= -w+\frac{1}{3}w^2,~~~{\rm for}~~D=5,
\\
\nonumber
&&F(w)= w-\frac{2}{3}w^3+\frac{w^5}{5},~~~{\rm for}~~D=7,
\end{eqnarray} 
and
\begin{eqnarray}
\nonumber
&&F(w)= -w+w^3-\frac{3}{5}w^5+\frac{1}{7}w^7,~~~{\rm for}~~D=9.
\end{eqnarray} 
\\
One should also note that the eqs. (\ref{red-eqs}) together with the first integral (\ref{fi}) are invariant under the scaling
\begin{eqnarray}
\label{ss1}
 r\to \lambda r, ~~m\to \lambda^{D-3} m,~~\sigma \to \sigma,~w\to w,~~V\to V/\lambda,~~
 P\to \lambda^{D-2} P,~~{\rm and}~~\alpha\to \lambda \alpha,~~\kappa \to \lambda^{D-4} \kappa, 
\end{eqnarray}
with $\lambda$ an arbitrary positive parameter. 
There is also a second scaling symmetry of the  equations
(\ref{red-eqs})
\begin{eqnarray}
\label{ss2}
V\to \tilde \lambda V,~~\sigma \to \tilde \lambda \sigma,
\end{eqnarray}
together with $t\to  t/\tilde \lambda$,
all other variables remaining unchanged.
This symmetry is lost after setting $\sigma(\infty)=1$ as a boundary condition.

The last symmetry of the equations of the model consists in simultaneously
changing the sign of the CS coupling constant together with the electric or magnetic potential
\begin{eqnarray}
\label{ss3}
\kappa \to -\kappa,~~V\to - V,~~~~{\rm or}~~~\kappa \to -\kappa,~~w\to - w,
\end{eqnarray}
(the first integral (\ref{fi}) implies also $P\to -P$ in the first case).
In what follows, we shall use this symmetry to study solutions with a positive 
$\kappa$ only.

\subsection{The behaviour at infinity}
Unfortunately, it is not possible to find an exact solution of the equations 
(\ref{red-eqs}) with a nontrivial magnetic gauge potential $w(r)$,
 except for a special value of $\kappa$ in $D=5$ dimensions.
However, one can write   an approximate form of the solutions as
a power series with a finite number of undetermined constants, both
at infinity and at the horizon/origin of the coordinate system.
This analysis allows us to obtain some information 
on the possible global behaviour of solutions.

In deriving the asymptotic form of solutions as $r\to \infty$, 
we assume that  
 the spacetime approaches the Minkowski background at 
infinity, while the configurations have a finite ADM mass.
Then we arrive at the following  expansion of the solutions  at $r\to \infty$:
\begin{eqnarray}
\label{asympt-inf}
&&
m(r)=M_0-\frac{1}{2}\alpha^2(D-3)\frac{Q^2}{r^{D-3}}+\dots,
~~
\sigma(r)=1-\frac{\alpha^2 (D-3)^2J^2}{4 r^{2(D-2)}}+\dots,
\\
&& 
\nonumber
w(r)=\pm 1-\frac{J}{r^{D-3}}+\dots,
V(r)=V_0-\frac{Q}{r^{D-3}}+\dots~.
\end{eqnarray}
In the above relations,  $J,~M_0,~V_0$ are parameters given by numerics which fix all higher order terms, 
while $Q$ is a constant fixing the electric charge of the solutions,
\begin{eqnarray}
Q=\frac{1}{D-3}
\left(
\frac{P}{\alpha^2}+(D^2-1)\kappa F(\pm 1)
\right),
\end{eqnarray}
where
\begin{eqnarray}
\nonumber
F(-1)=-F(1)=(-1)^{\frac{D-1}{2}}\sqrt{\pi} \frac{\Gamma(\frac{D-1}{2})}{2\Gamma(\frac{D}{2})}.
\end{eqnarray}
The set  (\ref{asympt-inf}) of boundary conditions is shared by both globally regular and
black hole solutions.

\subsection{Soliton solutions: the expansion at $r=0$}

These are perhaps the simplest possible solutions of the system (\ref{red-eqs}) 
and can be viewed as higher dimensional generalizations of the
Bartnik-Mckinnon solutions \cite{Bartnik:1988am}, though dressed with 
an electric charge.
The most striking feature here is that the electric charge of the solitons is fixed by the 
value of the CS coupling constant.
Technically, this results from the fact that the term 
$ {P}/{\alpha^2}+(D^2-1)\kappa F(w)$ in the first integral (\ref{fi})
should vanish as $r\to 0$.
Since $w(0)=1$ for regular solutions, the parameter $P$ is fixed to be 
\begin{eqnarray}
P=2 \alpha^2(-1)^{\frac{(D-1)}{2}}\kappa \sqrt{\pi}\frac{\Gamma(\frac{D+3}{2})}{\Gamma(\frac{D}{2})}.
\end{eqnarray} 
One finds $e.g.$, $P=16 \alpha^2 \kappa$, $-128\alpha^2 \kappa/5$ and $256 \alpha^2 \kappa/7$
for $D=5,7$ and $9$, respectively.

Then, for globally regular soliton-type solutions, the electric charge parameter 
which enters  the far field expression (\ref{asympt-inf})
 is fixed by
\begin{eqnarray}
\label{Qc}
Q=Q^{(c)}=\kappa  \frac{4(-1)^{\frac{D-1}{2}}\sqrt{\pi}}{(D-3)}
\frac{\Gamma(\frac{D+3}{2})}{\Gamma(\frac{D}{2})}.
\end{eqnarray}
Its expression for the values of $D$ considered in numerics is 
$Q^{(c)}=16  \kappa$, $-64\kappa/5$ and $256\kappa/21$
for $D=5,7$ and $9$, respectively.

Also, one finds that the globally regular configurations  have the following expansions near the origin  $r=0$:
\begin{eqnarray}
\label{origin}
&&w(r)= 1-br^2+O(r^4), 
~~V(r)= (-1)^{\frac{D-1}{2}}2^{\frac{D-3}{2}} b^{\frac{D-1}{2}}(D+1) \kappa \sigma_0 r^2+O(r^4),~~ 
\\
\nonumber
&& m(r)= (D-2)\alpha^2 b^2 r^{D-1}+O(r^D),~~
\sigma(r)=\sigma_0+ (D-2)\alpha^2 b^2 \sigma_0 r^2+O(r^4).
\end{eqnarray}
The free parameters here are $b=-\frac{1}{2}w''(0)$ and $\sigma_0=\sigma(0)$.
The coefficients of all higher order terms in the $r\to 0$ expansion are fixed by these parameters.

\subsection{Black holes: the near horizon solution}
For the
solutions in this work, the event horizon is located at a constant value of the radial coordinate
$r = r_h$, with $N(r_h) = 0$.
In this case,
  one can write also an approximate form of the solutions
near the horizon, as a power series in $r-r_h$.
In the nonextremal case, the first terms in this expansion are
\begin{eqnarray}
\nonumber
&&m(r)=r_h^{D-3}+m_1(r-r_h)+\dots,
~
\sigma(r)=\sigma_h+\frac{3 \sigma_h w_1^2}{2 r_h}(r-r_h)+\dots,
\\
\label{exp-eh}
&&w(r)=w_h+w_1(r-r_h)+\dots,~
 V(r)=v_1(r-r_h)+\dots,
\end{eqnarray} 
where
\begin{eqnarray}
\nonumber
v_1=\frac{\sigma_h}{\alpha^2 r_h^{D-2}}
\left(
P+\alpha^2(D^2-1)\kappa F(w_h) 
\right),
~~
m_1=\frac{\alpha^2}{2}r_h^{D-2}
\left(
\frac{ v_1^2}{\sigma_h^2} +(D-2)(D-3)\frac{(1-w_h^2)^2}{2r_h^4} 
\right ),
\\
\label{rel-new1}
w_1= \frac{(D-3)w_h(w_h^2-1)}{r_h(D-3-\frac{m_1}{r_h^{D-4}})}
+\frac{(D^2-1)\kappa r_h v_1 (w_h^2-1)^{\frac{D-3}{2}}}{(D-2)\sigma_h(-m_1+(D-3)r_h^{D-4})},
~~
\sigma_1=\frac{\alpha^2 (D-2)}{2 r_h}\sigma_h w_1^2,
\end{eqnarray} 
while $N(r)=N'(r_h)(r-r_h)+\dots,$ with $N'(r_h)=(D-3-m_1/r_h^{D-4})/r_h$.
The obvious condition $N'(r_h)>0$
implies the existence of a lower bound on the event horizon radius $r_{\min}$,
for given values of $Q$ and $\kappa$.
The only free parameters in the expansion above are $\sigma_h$ and $w_h$.
In a numerical approach, their values are found by matching the near horizon  
form of the solutions (\ref{exp-eh})  with the asymptotic
expansion (\ref{asympt-inf}).

As $r\to r_{min}$, the function $N(r)$
develops a double zero at the horizon, $i.e.$ $N(r)=N_2(r-r_h)^2+\dots$,
and the black holes become extremal.
We shall see that such solutions exist indeed, emerging as limiting configurations of a branch of
nonextremal black holes.
The near horizon expansion is  more constrained in this case. 
Supposing $w_h^2\neq 1$,
one finds that the event horizon radius $r_h$ is the largest positive solution of the equation
 \begin{eqnarray}
\label{wh-rh-extr}
1-\frac{\alpha^2(D-2}{4}(1-w_h^2)^2
\bigg (
\frac{1}{r_h^2}+
\frac{2(D-2)(D-3)r_h^{2(D-5)}w_h^2 }{(D^2-1)^2 \kappa^2(w_h^2-1)^{D-3}}
\bigg )=0,
\end{eqnarray} 
the value of the electric charge parameter being also fixed (with $w(\infty)=\pm 1$):
\begin{eqnarray}
\label{Q-extr}
Q=\frac{(D-2)r_h^{2(D-4)}w_h}{(D^2-1)\kappa (w_h^2-1)^{\frac{D-5}{2}}}
\bigg(
\frac{(D^2-1)^2\kappa^2 (w_h^2-1)^{\frac{(D-5}{2}}}{(D-2)(D-3)w_hr_h^{2(D-4)}}
(F(\pm 1)-F(w_h))-1
\bigg)~.
\end{eqnarray} 
Also, the coefficient $N_2$ in the leading order expansion of the metric function $N(r)$ is
\begin{eqnarray}
N_2=\frac{(D-3)(D-4)}{2r_h^2}
\bigg [
1+\frac{\alpha^2(D-2)(D-6)(w_h^2-1)^2}{4(D-4)r_h^2}
\left(
\frac{2(D-2)^2(D-3)r_h^{2(D-4)}w_h^2}{(D-6)(D^2-1)\kappa^2 (w_h^2-1)^{D-3}}-1
\right)
\bigg ].
\end{eqnarray} 
The parameters $w_1,v_1$ and $\sigma_1$ have a similar expression as those found in the nonextremal
case (there one should replace the expressions (\ref{wh-rh-extr}) and (\ref{Q-extr}) for $r_h$ and $Q$, respectively).

For $D=5$ and $Q>Q^{(c)}$, we have found numerical evidence for the existence 
of a different type of extremal black holes, with $w(r_h)=1$.
As $r\to r_h$, these solutions have basically the same leading order expression as an extremal RN black hole.
 However, the next to leading order terms in the
near horizon expansion exhibits
 non-integer powers of $r-r_h$.
 These special $D=5$ configurations will be discussed in the Section {\bf 4.3} below.

\subsection{ An $AdS_2 \times S^{D-2}$ solution}
As expected, the near horizon structure of the  extremal solutions with $w_h^2\neq 1$
can be extended to a full AdS$_2\times S^{D-2}$
solution of the field equations. 
This is a new exact, essentially nA solution to the EYMCS field equations, with
\begin{eqnarray}
\label{ex1}
ds^2=v_1(\frac{dr^2}{r^2} -r^2 dt^2)
+v_2 d\Omega_{D-2}^2,~~~w(r)=w_0,~~V(r)= q r,
\end{eqnarray}
where
\begin{eqnarray}
q=-\frac{(D-2)(D-3)v_1 v_2^{(D-6)/2}}{(D^2-1)\kappa}\frac{w_0}{ (w_0^2-1)^{(D-5)/2}}.
\end{eqnarray}
The AdS radius satisfies the relation
\begin{eqnarray}
v_1= 8 v_2^2 
\bigg(
(D-3)(4(D-4)v_2+\alpha^2(D-2)(w_0^2-1)^2
(
6-D+\frac{2(D-2)^2(D-3)w_0^2v_2^{D-4}}{(D^2-1)^2 \kappa^2 (w_0^2-1)^{D-3}}
)
\bigg )^{-1},
\end{eqnarray}
where
the size  of the $S^{D-2}$ part of the metric results as a solution of the equation
\begin{eqnarray}
v_2=\frac{1}{4}\alpha^2(D-2)
(w_0^2-1)^2
\left(
1+\frac{2(D-2)(D-3)}{(D^2-1)\kappa^2}\frac{w_0^2}{(w_0^2-1)^{D-3}}v_2^{D-4}
\right ),
\end{eqnarray}
being a function of $w_0$ only.
Unfortunately, one can write a simple solution of the above equation, except for $D=5$.
In this case, the general solution reads
\begin{eqnarray}
v_1=\frac{1536 \alpha^2 \kappa^4 (w_0^2-1)^2}{(64 \kappa^2-\alpha^2 w_0^2)(64 \kappa^2+\alpha^2 w_0^2)},
~~
v_2=\frac{48 \alpha^2 \kappa^2 (w_0^2-1)^2}{64 \kappa^2-\alpha^2 w_0^2},
~~
q=\frac{32 \sqrt{3} \alpha \kappa^2 w_0(w_0^2-1)}{\sqrt{64 \kappa^2-\alpha^2 w_0^2}(64 \kappa^2+\alpha^2 w_0^2)}.
\end{eqnarray}
 
One can see that, for any $D$, the properties of the $AdS_2 \times S^{D-2}$ solution
are uniquelly specified by the constant $w_0$.
Note that for $w_0^2\neq 1$, the nA magnetic gauge field is nonvanishing;
the case $w_0=\pm 1$ is special and describe the near horizon geometry of the extremal 
RN solutions.

Although  finding local solutions in the vicinity of the horizon does not guarantee the
existence of the global solutions, the above result provides an argument that the EYMCS
system 
is likely to present asymptotically flat, extremal
black hole solutions. In $D=5$ and $D=7,9$,  this is confirmed by our numerical and analytic results.
If there are extremal black holes in the bulk, the parameter $q$ in (\ref{ex1}) is related 
to the bulk charge parameter $Q$ via
\begin{eqnarray}
q=\frac{v_1}{v_2^{(D-2)/2}}
\left(
(D-3)Q-(D^2-1)\kappa (F(w_0)+F(\pm 1)
\right),
\end{eqnarray}
with $w(\infty)=\pm 1$, the allowed values at infinity of the bulk magnetic gauge potential.

It would be interesting to consider these solutions in the context of the attractor mechanism 
and to compute their entropy functions 
(for a discussion 
of Sen's entropy function for $D=5$ 
supergravity models containing Abelian Chern-Simons terms, 
see e.g. \cite{Arsiwalla:2008gc}).

\subsection{Relevant parameters and global charges}
The only global charges associated with the 
solutions are the mass ${\cal M}$ and the electric charge
${\cal Q}$
\begin{eqnarray}
\label{mass}
{\cal M}=\frac{(d-2)V_{D-2} M_0}{16 \pi G}, ~~{\cal Q}=\frac{(D-3)V_{D-2} Q}{g},
\end{eqnarray} 
where $V_{D-2}=2 \pi^{(D-1)/2}/\Gamma((D-1)/2)$ is the area of the unit $D - 2$ sphere.
The mass ${\cal M}$ is the charge associated with the Killing vector $\partial/\partial t$;
the electric charge ${\cal Q}$ is associated  with the $U(1)$ gauge symmetry generated
by $\Sigma_{2n+1,2n+2}^{(\pm)}$.
For the boundary conditions in this work, the  electrostatic potential difference $\Phi$ between the horizon and infinity
is fixed by the value at infinity
of the electric potential $V(r)$, as read from (\ref{asympt-inf}). 
In a thermodynamical description of the system, $\Phi$ corresponds to the
chemical potential,
\begin{eqnarray}
\label{ep}
\Phi=\frac{V_0}{e}.
\end{eqnarray} 
The Hawking temperature and the entropy of the black holes are given by
\begin{eqnarray}
\label{THS}
T_H= \frac{1}{4 \pi} \sigma(r_h) N'(r_h) ,~~~~S=\frac{A_H}{4G},~~~~~{\rm with}~~A_H=V_{D-2} r_h^{D-2}.
\end{eqnarray} 
The solutions should also satisfy the First Law of thermodynamics,
\begin{eqnarray}
\label{first-law}
 d{\cal M}=T_H dS+\Phi d {\cal Q}.
\end{eqnarray}
In the discussion  of the black hole thermodynamics,
we shall restrict to configurations
in a canonical ensemble, the relevant potential being
the  Helmholz free energy
\begin{eqnarray}
\label{F}
F[T_H,{\cal Q}]=  {\cal M}-T_H S.
\end{eqnarray}
As usual, in practice, it is convenient to work with
quantities which are  invariant under the rescaling (\ref{ss1}) (note that
in the numerics
we set $\alpha=1$).
For solutions in a canonical ensemble, we normalize the global quantities with respect the
charge parameter $Q$ and define $e.g.$,  the dimensionless quantities 
\begin{eqnarray}
\label{adimF}
f=\frac{ F}{Q/g^2},~~t_H=T_H Q^{1/(D-3)},~~a_H=\frac{A_H}{Q^{\frac{D-2}{D-3}}}
~~{\rm and}~~~j=\frac{J}{Q}.
\end{eqnarray}
%
One should note that there is no conserved quantity associated with the parameter 
$J$ which appears in the large-$r$ asymptotics (\ref{asympt-inf}). Also, $J$ does not enter the First Law
(\ref{first-law}).

In addition to the mass and the electric charge, there is another global   charge
which has a topological origin. This is the volume integral ${\cal P}$
of the topological density calculated from the ``magnetic'' components, $F_{ij}$, namely the Chern--Pontryagin (CP) density
defined on the $(D-1)-$dimensional space dimensions with Euclidean signature. 
The correct expression for this CP density
must take account of the fact that the gauge group for  $F_{ij}$ is $SO(D-1)$, and not~\footnote{Note that the $SO(D-1)$ curvature
consists of two $SO_{\pm}(D-1)$ curvatures, each contributing $\pm 1$ CP charge in the spherically symmetric case.
If the appropriate factor of $\Si_{D,D+1}=\ga_D$ in \re{CPSO4} is not accounted for in the CP density, these two charges
will cancel each other out. This is discussed in detail in the context of the $SO(D)$ monopole on $\R^D$ in
\cite{Tchrakian:2010ar}.} one or other of the two chiral algebras $SO_{\pm}(D-1)$. Using the  spherically
symmetric components of the curvature $F_{ij}$
given by \re{fijp},   this charge density is calculated to be
\be
\label{CPSO4}
\vep_{i_{1}i_{2}i_{3}i_{4}\dots i_{D-2}i_{D-1}}\,\mbox{Tr}\,
\big \{\Si_{D,D+1}\,
F^{i_{1}i_{2}}\,F^{i_{3}i_{4}}\dots F^{i_{D-2}i_{D-1}}
\big \} 
\simeq -\frac{1}{r^{D-2}}\,(w^2-1)^{\frac{D-3}{2}} w'\,
.
\ee
Usually, 
for spherically symmetric soliton solutions, the integral of the above quantity is suitably normalised such that 
it yields $unit$ ``magnetic'' CP charge, ${\cal P}=1$.
However, noticing that $(w^2-1)^{\frac{D-3}{2}} w'$ is just the derivative of
the function $F(w)$ which enters the first integral (\ref{fi}), 
 it is more interesting to use a nonstandard normalization
and to define the ``magnetic'' CP charge  ${\cal P}$ directly as the integral of (\ref{CPSO4}).
Then, taking into account the boundary conditions\footnote{Note that we take $w(\infty)=-1$, 
which was the case for all solutions we could find numerically.}  (\ref{asympt-inf}), (\ref{origin}), it follows that
the soliton solutions  exhibit an interesting 
connection between the 
electric charge and the 
``magnetic'' CP charge
\begin{eqnarray}
\label{QeP}
{\cal Q}=\kappa {\cal P}.
\end{eqnarray}

\subsection{The Reissner-Nordstr\"om  solution}
The Reissner-Nordstr\"om (RN) solution plays  
an important role in what follows.
Thus, for completness, we briefly discuss here 
its basic thermodynamical properties.
This solution is recovered for 
 $w(r) \equiv \pm 1$, in which case
the magnetic components of the field strength  vanish identically. 
Then the configuration becomes essentially Abelian 
and one finds the following  exact solution
\begin{eqnarray}
\label{RN}
m(r) =M_0 - \frac{\alpha^2(D-3) Q^2}{2 r^{D-3}}~,
~~
\sigma(r)=1, ~~w(r) =\pm 1,~~V(r)=V_0-\frac{Q}{r^{D-3}},
\end{eqnarray} 
with $V_0$ a constant which is usually chosen such that the electric potential
vanishes on the horizon, $i.e.$,$V_0=Q/r_h^{D-3}$.
The mass ${\cal M}$, electric charge ${\cal Q}$ and chemical potential $\Phi$
follow authomatically from (\ref{mass}), (\ref{ep}). 
 The  
Schwarzschild-Tangerlini vacuum black hole corresponds to the case $Q=0$.

The RN solution has an outer event horizon at $r=r_h$,
with
\begin{eqnarray}
\label{rh-RN}
r_h=
\left(
\frac{1}{2}
(M_0+\sqrt{M_0^2-2\alpha^2(D-3)Q^2}) 
\right )^{1/(D-3)},
\end{eqnarray} 
whose existence imposes an upper bound for the electric charge for a given mass.
The Hawking temperature and the entropy of this solution can be written in terms of $r_h,Q$ 
(which are the parameters used in numerics)
as
\begin{eqnarray}
\label{TH-RN}
 T_H=\frac{D-3}{4\pi r_h}
 \left(
 1-\frac{(D-3)}{2}\frac{\alpha^2 Q^2}{r_h^{2(D-3}}
 \right),~~~S=\frac{V_{D-2}r_h^{D-2}}{4G}.
\end{eqnarray} 
With these definitions, 
one can easily verify that the First Law (\ref{first-law}) is indeed fullfilled. 
In addition, the RN black holes satisfy the 
Smarr law
\begin{eqnarray}
\label{smarr}
{\cal M}=\frac{D-2}{D-3}T_H S+\Phi {\cal Q}.
\end{eqnarray} 

The relation (\ref{TH-RN}) shows that, 
for a given $Q$, there is a minimal value of the event horizon radius
at which an extremal black hole is approached,
\begin{eqnarray}
\label{rmin-RN}
r_h\geq r_h^{(min)}=
\left(\frac{1}{2}(D-3)\alpha^2 Q^2 \right)^{\frac{1}{2(D-3)}}.
\end{eqnarray} 
Thus the Hawking temperature vanishes in this limit,
while 
the event horizon area approaches a minimal value, with
\begin{eqnarray}
\label{aH-min-RN}
a_H^{(min)}=\frac{1}{2}(D-3)\alpha^2 V_{D-2}.
\end{eqnarray} 
After  expressing $r_h$ as a function of $A_H$ according to \re{THS},
one gets the following relation between the reduced variables $t_H$ and $a_H$:
\begin{eqnarray}
\label{tH-RN}
t_H=\frac{(D-3)}{4 \pi} 
\left (\frac{V_{D-2}}{a_H} \right )^{\frac{1}{D-2}}
\bigg (
1-\frac{\alpha^2(D-3)}{2}\left (\frac{V_{D-2}}{a_H} \right )^{\frac{2(D-3)}{D-2}}
\bigg).
\end{eqnarray}
Unfortunately, one cannot invert this relation to get $S(T_H,Q)$.
However, one can see that, for a given $Q$ the solutions exist only for $0\leq T_H\leq T_H^{(max)}$,
 the maximal value of the Hawking temperature being
 \begin{eqnarray}
\label{TH-max-RN}
T_H^{(max)}= \frac{1}{2^{\frac{(2D-7)}{(2D-6)}}\pi }
(D-3)^{\frac{4D-13}{2(D-3)}}(2D-5)^{-\frac{(2D-5)}{2(D-3)}}\
\frac{1}{(\alpha Q)^{\frac{1}{D-3}}},
\end{eqnarray}
the entropy at this turning point
being
 \begin{eqnarray}
\label{Sc-RN}
S(T_H^{(max)})=\frac{V_{D-2}}{G}
\left(
\frac{2^{\frac{5D-14}{2(D-2)}}}{\alpha\sqrt{(D-3)(2D-5)}Q}
\right)^{\frac{D-2}{D-3}}.
\end{eqnarray}
 
It is also of interest to express the free energy of the RN solution
as a function of $T_H,Q$.
The only relation we could find in this case reads
\begin{eqnarray}
\label{f-RN}
f= \frac{1}{t_H^{D-3}}\frac{(D-3)^{D-3}} {(D-2)2^{2D-7}\pi^{D-3} }\frac{ V_{D-2}} {\alpha^2}
\bigg (1+\sqrt{1-\frac{2(D-3)(2D-5)V_{D-2}^2}{\alpha^2(D-2)^2}\frac{1}{f^2}} \bigg )^{-1}
\\
\nonumber
\times 
\bigg [
1-f^2 \frac{\alpha^2 (D-2)^2 }{2(2D-5)^2(D-3)V_{D-2}^2 } 
   \left(1-\sqrt{1-\frac{2(D-3)(2D-5)V_{D-2}^2}{\alpha^2(D-2)^2}\frac{1}{f^2}} 
    \right)^2
\bigg ]^{D-3}.
\end{eqnarray} 
A study of the above relations shows the existence of two branches of Abelian solutions, each with different 
thermal properties. (The generic picture here is dimension independent; then the well-known $D=4$
result in \cite{Davies:1978mf} applies also in higher  dimensions.)
For a fixed $Q$, there is first a branch of large black holes 
whose entropy decreases with $T_H$, which therefore are unstable.
This branch stops in a critical configuration with a maximal value of $T_H$
given by (\ref{TH-max-RN}), where a secondary branch of small black hole emerges.
This branch has a positive specific heat and ends in an extremal configuration
with an event horizon area given by (\ref{aH-min-RN}).
These features are exhibited in Figures 7, 8 (see the RN curves there). The picture for $D>5$ is qualitatively
the same.

\subsection{The issue of perturbative static solutions around Reissner-Nordstr\"om black holes}
Since the RN black hole is a solution of the model for any $D$,
one might expect the existence of a branch of nA solutions connected with it.
Such solutions, if they exist, would emerge as $static$ perturbations 
around the RN background. However, as we shall argue, this is the case for $D=5$ only. 

Suppose that there is a perturbative solution of the equations (\ref{red-eqs}) around the RN black hole,
\begin{eqnarray}
\label{pert-RN}
&&m(r)=m_0(r)+\epsilon m_1(r)+\dots,~~\sigma(r)=1+\epsilon \sigma_1(r)+\dots,
\\
\nonumber
&&w(r)=\pm 1+\epsilon W_1(r)+\dots,~~V(r)= V_0(r)+\epsilon V_1(r)+\dots,
\end{eqnarray}
with $\epsilon$ a small parameter. In the above relations, 
$m_0(r)$, $V_0(r)$ are the functions which enter the RN solution (\ref{RN}).

After substituting (\ref{pert-RN}) in the  equations  (\ref{red-eqs}), one finds that to lowest order, 
the equation for $W_1(r)$
decouples. This equation is the only relevant one, it's general-$D$ expression being
\begin{eqnarray}
\label{stab10}
  (r^{D-4} N W_1')'=2 \left( (D-3)r^{d-6}  \pm 8 \kappa V_0' \delta_{D,5} \right)W_1 ,
\end{eqnarray}
(with $N=1-m_0(r)/r^{D-3}$).
Then it turns out that the case $D=5$ is special, since the CS term  gives a nonzero contribution to the $W_1$-equation
only for this dimension.
The perturbation $W_1(r)$ starts from some (arbitrary)
nonzero value at the horizon and vanishes at infinity, in order to be consistent with the asymptotic
behaviour (\ref{exp-eh}), (\ref{asympt-inf}). However, one can show that for $D>5$,
there is no solution of (\ref{stab1}) that satisfies this asymptotic behaviour.
To prove that, we rewrite the eq. (\ref{stab10}) in the equivalent form
\begin{eqnarray}
\label{stab1-2}
 \frac{1}{2}(r^{D-4} N (W_1^2)')'= \frac{1}{2}r^{D-4}N W_1'^2 +2(D-3)r^{D-6} W_1^2\,,
\end{eqnarray}
recalling that $D>5$ now. Then, after integrating from $r_h$ to infinity, one finds that $W_1$
necessarily vanishes identically\footnote{Basically, 
the r.h.s. of (\ref{stab1-2}) is greater or equal to zero, while
$N(r_h)=0$ and $W_1(\infty)=0$.}. The same argument applies when considering higher order terms 
in the expansion (\ref{pert-RN}).
Therefore we conclude that, for $D>5$, the RN solution is stable 
with respect to nA perturbations within the considered EYMCS model~\footnote{At the cost of replacing the usual
Yang-Mills term $F^2$ by $F^{2p}$ \cite{Tchrakian:1984gq}, one can expect that
 a brach of the static nA solution emerges as a perturbation around
the corresponding RN-type background, in the $D=4p+1$ case,
also.}.

For $D=5$, a similar reasoning implies that
the "mass" term  $1\pm \frac{8\kappa Q}{r^2}$ in the Eqn. (\ref{stab10}) 
should necessarily be negative in the vicinity of the horizon.
Then, one finds $\kappa Q>0$  for $w(\infty)=-1$ and $\kappa Q<0$ for $w(\infty)=1$.

\section{The results in $D=5$}
\setcounter{equation}{0}
\subsection{Numerical methods}

The   scaling   transformation (\ref{ss1})
  can be used to fix an arbitrary value\footnote{In principle, one can use (\ref{ss1})
  to fix instead the value of the CS coupling constant $\kappa$.
  However, this choice is less interesting.} for $\alpha$.
 The usual choice is $\alpha=1$, which
 is what we employ for all solutions in this work.
  This fixes  the EYM length scale
$L=\sqrt{16 \pi G/((D-2) e^2) }$, while the mass scale is fixed by $\mu =L^{(D-3)/2}/G$.
All other quantities get multiplied with suitable factors of $L$. 

To control the quality of the numerical results,
we have performed some of the calculations with two different methods, finding excellent agreement.
First, the  equations (\ref{red-eqs})
were solved with suitable boundary conditions which result from  (\ref{eh}),
(\ref{infinity}) using a standard solver \cite{colsys}.
This solver involves a Newton-Raphson method for 
boundary-value ordinary differential equations, 
equipped with an adaptive mesh selection procedure.
Typical mesh sizes include $10^3-10^4$ points.
The solutions in this work have a typical relative accuracy of $10^{-7}$. 
In this approach, the value of the electric potential at infinity $V_0$
is fixed, the electric charge resulting from numerics,
 $i.e.$, the configurations are in a grand canonical ensemble.
 (The first integral (\ref{fi}) has been used to verify the accuracy of the solutions.)

In addition to employing this algorithm, families of solutions with a fixed electric charge
 were constructed by using a standard Runge-Kutta
ordinary  differential equation solver. In this approach we 
evaluate the initial conditions at $r=r_h+10^{-5}$, for global tolerance $10^{-12}$,
adjusting  for shooting parameters and integrating  towards  $r\to\infty$.
In this case the electric charge $Q$ was fixed via the equation (\ref{fi}), the electrostatic potential
 $V_0$ resulting from numerics.
We have confirmed that there is good agreement between the results obtained with these two different methods.

Also, for both approaches, we have restricted our integration 
to the (physically more relevant) region outside 
of the horizon, $r \geq r_h$.

\subsection{Perturbative solutions: an instability of the RN$_5$ black hole}
The branch of $D=5$ nA solutions emerges as a perturbation of the RN black hole.
(In what follows, we shall suppose without any loss of generality that the RN black hole has a positive
electric charge, $Q>0$.)

The perturbative solutions are found by solving the Eqn. (\ref{stab10}),
which for $D=5$ reads
\begin{eqnarray}
\label{stab1-v2}
 r(r N W'_1)'-4(1 \pm \frac{8\kappa Q}{r^2})W_1=0~,
\end{eqnarray}
where $N=1-\frac{Q^2+r_h^4}{r_h^2r^2}+\frac{Q^2}{r^4}$. 
Although this linear equation does not appear to be solvable in terms
of known functions, one can construct an approximate solution
near the horizon and at infinity. For a vacuum choice $w \equiv -1$, 
one finds that, as $r\to r_h$
\begin{eqnarray}
\label{stab2}
W_1(r)=W_h+\frac{2W_h r_h(-8\kappa Q+r_h^2)}{r_h^4-Q^2}(r-r_h)+\dots,
\end{eqnarray}
with all higher order coefficients fixed by $W_h$.
Because (\ref{stab2}) is linear, one can take $W_h=1$,
  without any loss of generality.

\setlength{\unitlength}{1cm}
\begin{picture}(8,6)
\put(3,0.0){\epsfig{file=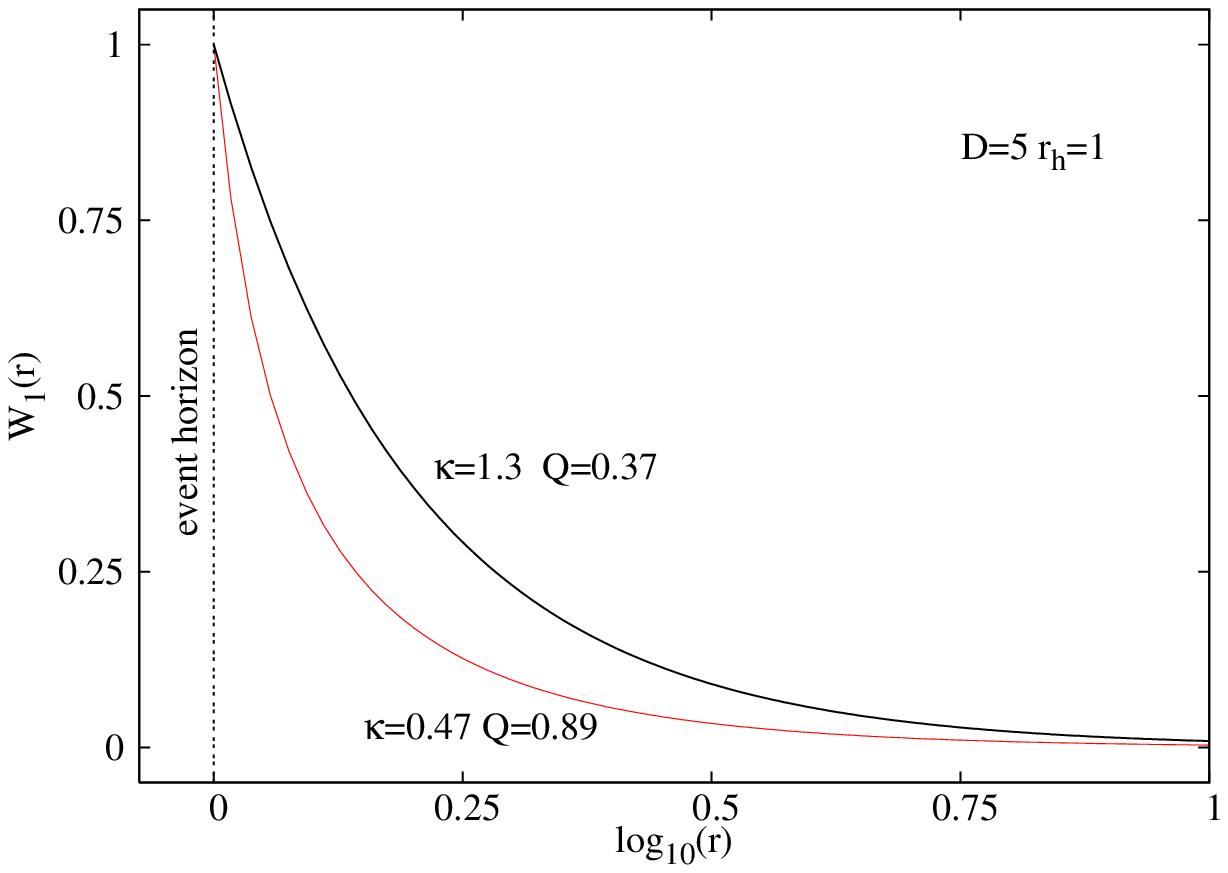,width=8cm}} 
\end{picture}
\\
\\
{\small {\bf Figure 1.}
The profiles of typical solutions of the $D=5$ perturbation equation (\ref{stab1-v2})
are presented as a function of the radial coordinate $r$ for two different values of $\kappa$. 
The values of  the charge parameter $Q$ for the corresponding critical Reissner-Nordstr\"om  solutions with $r_h=1$
are also shown.

}
\vspace{0.5cm}

 \setlength{\unitlength}{1cm}
\begin{picture}(8,6)
\put(3,0.0){\epsfig{file=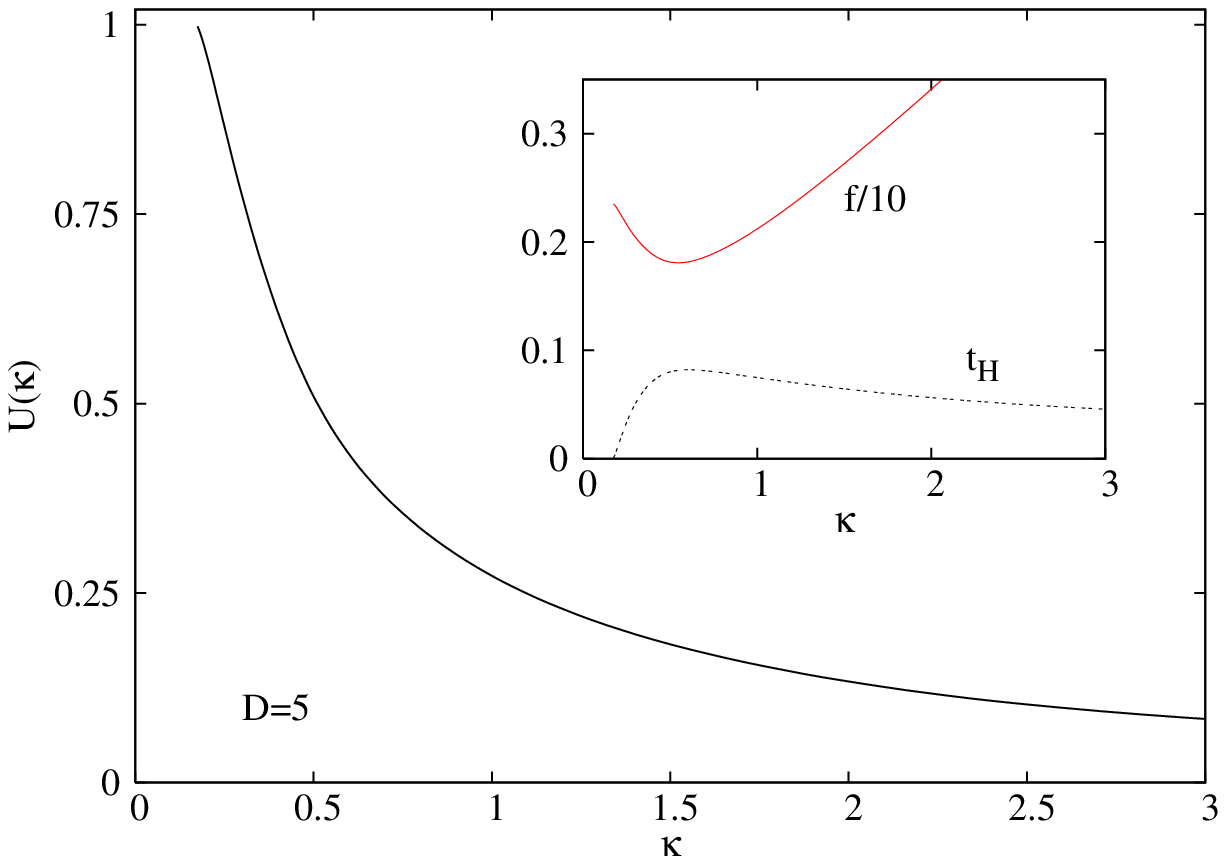,width=8cm}} 
\end{picture}
\\
\\
{\small {\bf Figure 2.}
The shape function $U(\kappa)=Q/r_h^2$ which 
gives the unstable Reissner-Nordstr\"om solution where a branch of non-Abelian
configurations emerges is shown as a function of 
the CS coupling constant $\kappa$.
The inlet shows the scaled free energy and Hawking temperature of
critical Abelian solutions.}
\vspace{0.5cm}
 
  At infinity, the only reasonable asymptotics reads
\begin{eqnarray}
\label{stab3}
W_1(r)= \frac{J}{r^2}+ \frac{2J(Q^2-4\kappa Qr_h^2+r_h^4)}{3r_h^2 r^4}+\dots,
\end{eqnarray}
in terms of a free parameter $J$.
The solutions interpolating between (\ref{stab2}) and (\ref{stab3})
are constructed numerically, typical results being shown in Figure 1.

 It turns out that such perturbative solutions do not exist for arbitray
values of $(r_h,\kappa)$. Restricting to solutions of (\ref{stab1-v2})
with monotonic behaviour\footnote{Note that there are also solutions of
(\ref{stab1-v2}) where $W_1(r)$ has nodes.} of $W_1(r)$, we find that for
given $(r_h,\kappa)$ for which such solutions exist, these pertain to a
fixed value of $Q$. This value results from the numerics.

 The existence of these configurations can be understood as follows. For $w(\infty)=-1$,
the second term in (\ref{stab10}) shows the existence of an effective mass term $\mu^2$ for $W_1$
near the horizon, with $\mu^2\sim 1-8\kappa Q/r_h^2$; 
all  solutions we could find have $\mu^2<0$.
Then it is also convenient to introduce the dimensionless function
\begin{eqnarray}
\label{funct-U}
U(\kappa)=\frac{Q}{r_h^2},
\end{eqnarray}
which uniquelly fixes the parameters of the 
critical RN solution. One finds $e.g.$
\begin{eqnarray}
\label{funct-U1} 
t_H=\frac{\sqrt{U}(1-U^2)}{2\pi},~~f=\frac{\pi}{8 U}(1+5 U^2), ~~{\rm while}~~~
\frac{\cal{M}}{Q}=\frac{3\pi}{8 U}(1+U^2).
\end{eqnarray}
 
  The shape of the function $U(\kappa)$ for the fundamental set of solutions of (\ref{stab1-v2}),
for which $W_1(r)$
are nodeless,   is shown in Figure 2.
The inlet there shows the scaled free energy and temperature of the 
critical RN solutions where static linear nA
perturbations arise,  as a function of $\kappa$.

  The function 
$U(\kappa)$ depends monotonically on the CS coupling constant $\kappa$. 
As $\kappa\to 1/8$, one finds $U=1$ $i.e.$,
the corresponding RN solution becomes extremal, while 
the $D=5$ RN black hole with the maximal value of the temperature (\ref{TH-max-RN})
is unstable for $\kappa \simeq 0.55$.
Also, one finds that the function $U$
decreases along the branch of large black holes, with
 $U\simeq 1/4\kappa$ for large $\kappa$. 
 
No physically reasonable
solutions of eq. (\ref{stab10})  
are found for $\kappa<1/8$, or for perturbations of the form $w(r)=+1+ \epsilon W(r)$,
 in which cases the effective mass for $W$ is always real (we recall we take $Q>0,~\kappa >0$). 
 
\subsection{Nonperturbative black hole solutions }
\subsubsection{General properties}

 The instability discussed above signals the presence of a symmetry breaking 
 branch of nA solutions bifurcating from
the RN black hole. 
 
 Indeed, 
our numerical results provide evidence  
for the existence of finite mass
black hole solutions of the EYMCS system
with nontrivial magnetic 
gauge fields outside the horizon.
These solutions smoothly interpolate 
between the asymptotics (\ref{exp-eh}), (\ref{asympt-inf})
(all $D=5$ configurations have $w(\infty)=-1$).
 
  On the basis of analytical and numerical results we have a pretty clear picture of
 the behaviour of the $D=5$ EYMCS black holes,
 this case being studied in a systematic way.
The properties of the solutions depend  on the value of the CS coupling constant $\kappa$.
 For a fixed $r_h$, the value of  the electric charge parameter $Q$ is also important,
 some basic features of the solutions depending on whether $Q$ is less or greater than $Q^{(c)}=16\kappa$.

As a general feature, we could not find configurations with multinodes of the function $w(r)$.
Heuristically, this can be understood as follows: the existence of 
such configurations would imply $w(r)=0$ as the  limit of multinodes.
However, such solutions would have infinite energy, with 
$m(r)=3/2\log r+M_0,~\sigma(r)=1$ and $V(r)=0$, which is not compatible
with the boundary conditions in the present work.

More importantly, since our solutions emerge as perturbations of RN black holes,
we notice the existence of configurations without nodes of the magnetic potential $w(r)$.
We shall see that some of these solutions are stable.
Also, different from the case of other asymptotically flat hairy black holes with nA fields \cite{Volkov:sv},
$w(r)$ may take values outside the  interval $[-1,+1]$.

Moreover,  it is possible to find more than one solution for the same value of $(\kappa,Q,r_h)$. 
In this case, apart from configurations with a monotonic behaviour of $w(r)$,
there are solutions solutions with local extrema of the magnetic gauge potential,
 see Figure 3  for such an example. However,
 it is likely that the  solutions with local extremal are always thermodynamically disfavoured 
because spatial oscillations in $w$ increase the total mass.
Thus in what follows we shall restrict to the study of the fundamental
branch of solutions  with a monotonic behaviour of the nA gauge function,
$i.e.$, $w'(r)<0$ everywhere.

Some of the general features of the solutions are shown in Figure 4 where we plot 
the values $w_h$ of the magnetic gauge potential
at the horizon and $V_0$ of the electric potential at infinity  as a functions of $r_H$ 
for a fixed $\kappa$ and several values of the charge parameter $Q$.
For example, one can see that the solutions
exist for $w_h^{(max)}\leq w_h\leq -1$, where the value $w_h^{(max)}$ increases with $Q$.
For  $Q<Q^{(c)}$ one finds $w_h^{(max)}<1$, while $w_h^{(max)}=1$ for $Q \geq Q^{(c)}$.

\setlength{\unitlength}{1cm}
\begin{picture}(8,6)
\put(-0.5,0.0){\epsfig{file=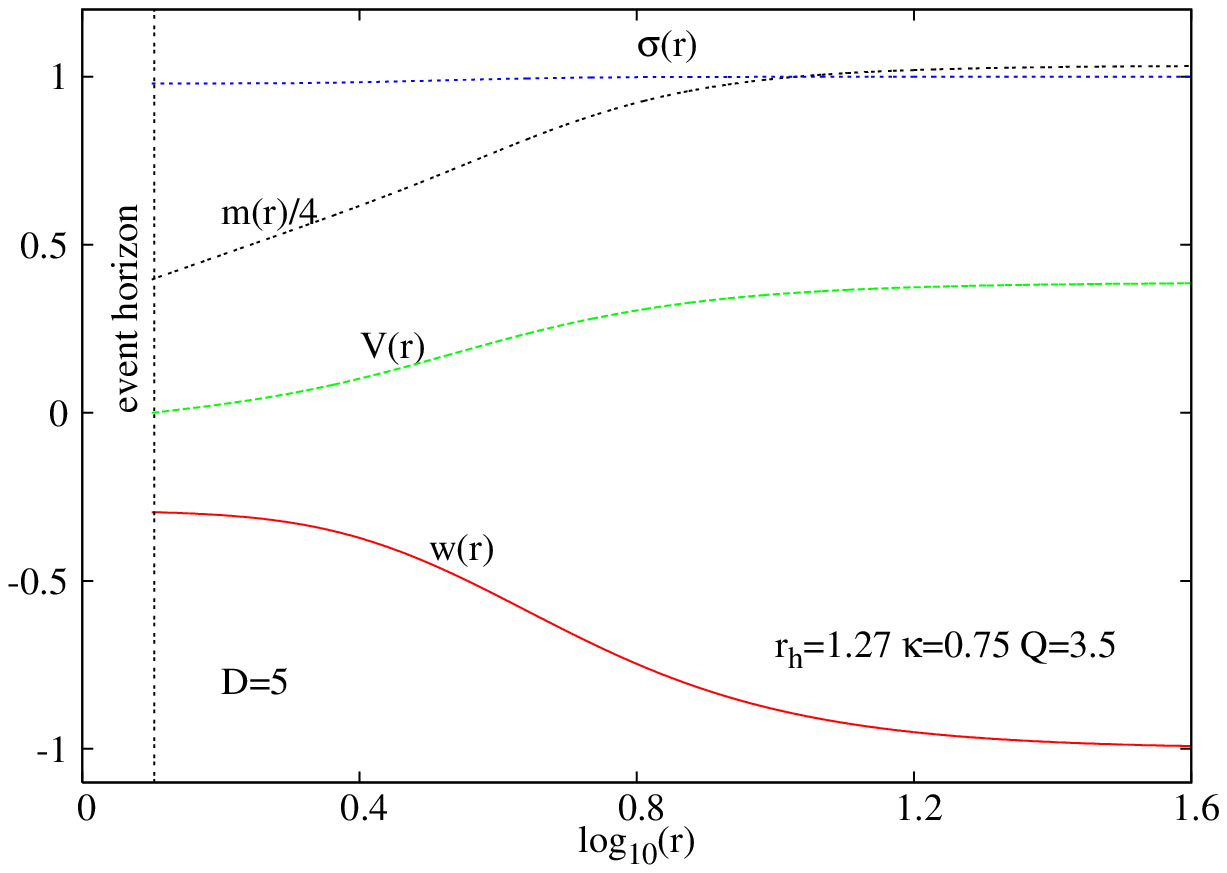,width=8cm}}
\put(8,0.0){\epsfig{file=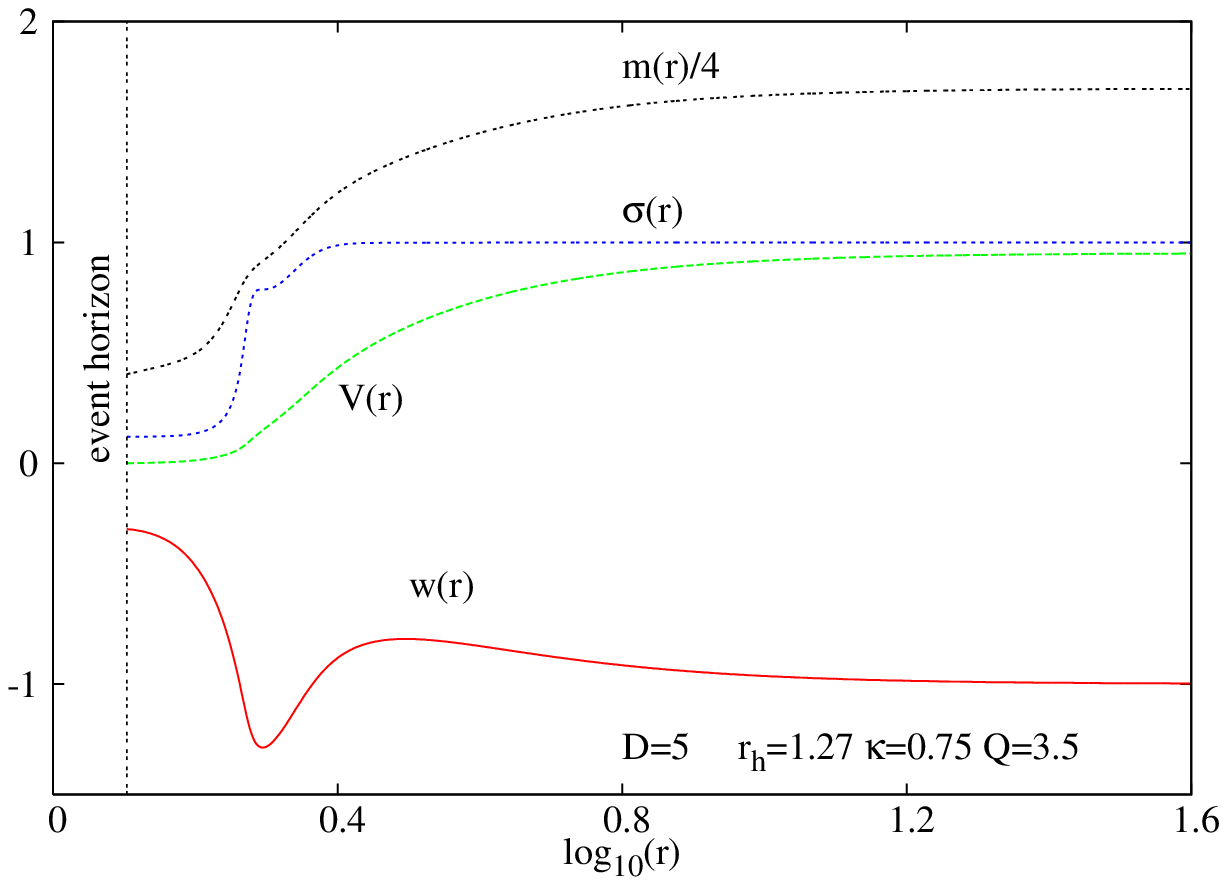,width=8cm}}
\end{picture}
\\
\\
{\small {\bf Figure 3.}
The profiles of two  EYMCS black hole solutions are presented as a function of the radial coordinate $r$. 
$m(r)$ and $\sigma(r)$ are metric functions, while $V(r)$
and $w(r)$ are electric and magnetic gauge potentials, respectively. }
 \vspace{0.5cm}

\setlength{\unitlength}{1cm}
\begin{picture}(8,6)
\put(-0.5,0.0){\epsfig{file=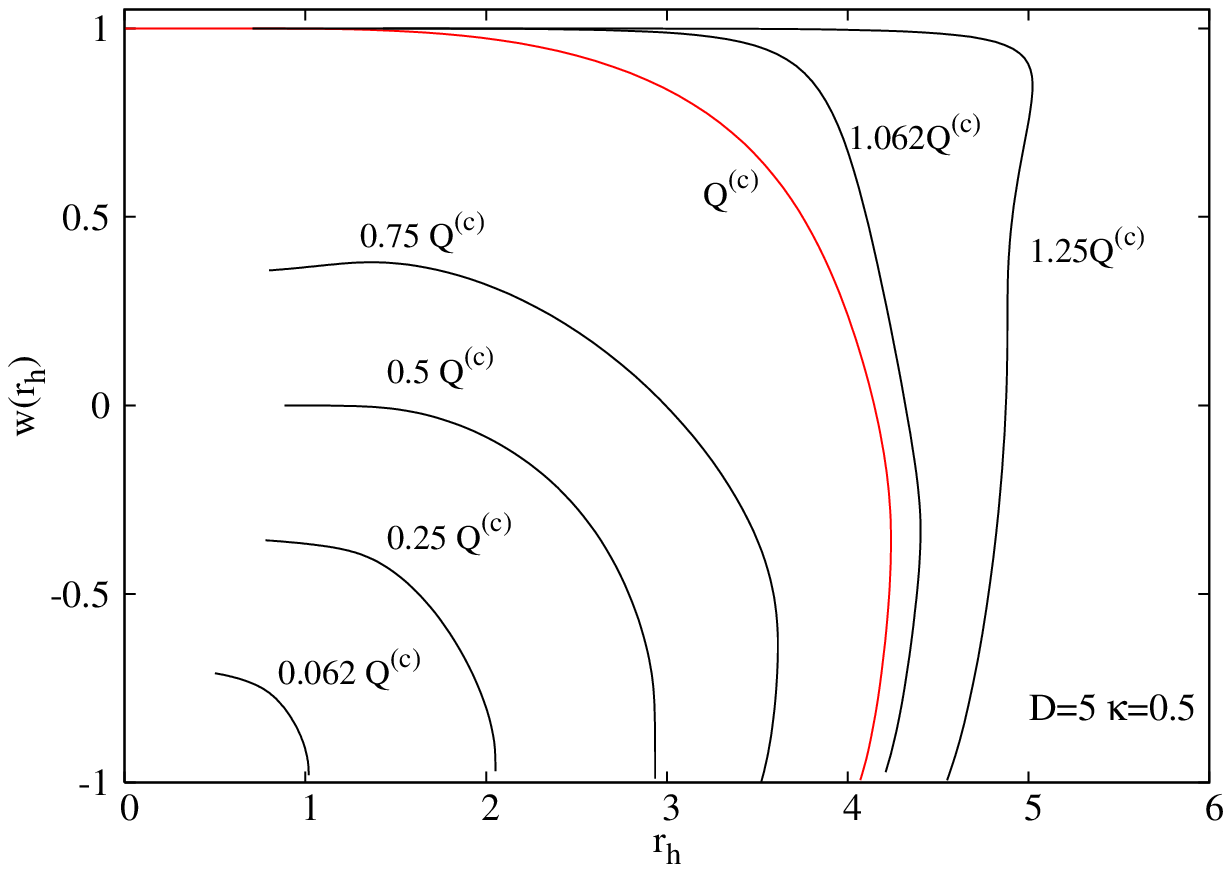,width=8cm}}
\put(8,0.0){\epsfig{file=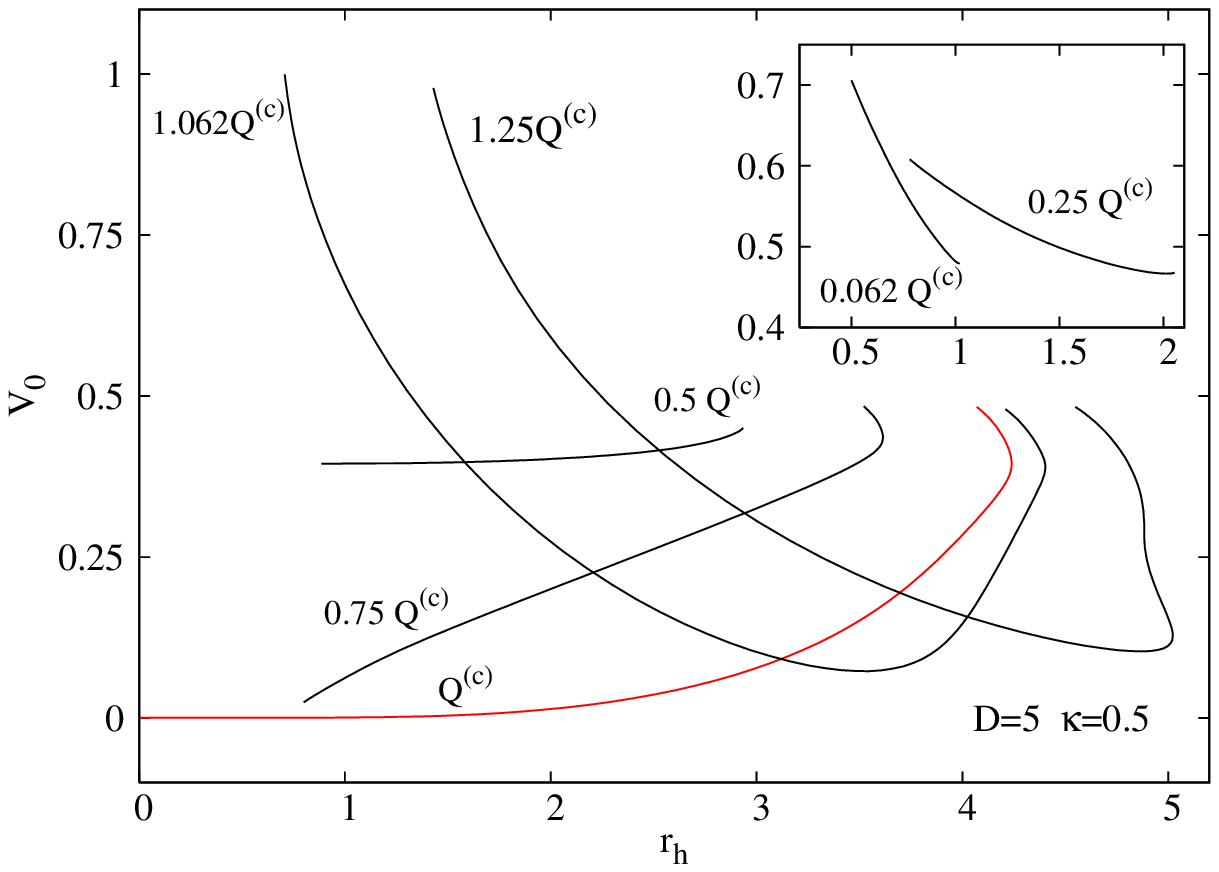,width=8cm}}
\end{picture}
\\
\\
{\small {\bf Figure 4.}
The value at the event horizon  of the magnetic gauge potential $w(r_h)$ (left)
and of the electric potential at infinity $V_0$ (right)
are shown as a functions of the event horizon radius $r_h$
for a fixed value of the Chern-Simons coupling constant
$\kappa$ and several values of the electric charge. 
 Here and in Figures 5-8, $Q^{(c)}=16\kappa$
is the critical value of the electric charge parameter.
}
\vspace{0.5cm}
 
 The behaviour of solutions as a function of the electric charge for a fixed 
event horizon radius ($i.e.$, at fixed entropy) is displayed in Figure 5. The value of the CS
coupling constant is also fixed there.
One can notice there the existence of both a maximal and a minimal value for $Q$.
The nA solution emerges as a perturbation of a RN black hole
for a minimal value of $Q$ given by $r_h^2 U(\kappa)$.
They exist up to a maximal value of $Q=r_h^2+16 \kappa$ where an extremal black hole is approached with $w(r_h)=1$.
Also, above some value of $Q$ close to $Q^{(c)}$, $w(r_h)$ becomes very close (but not equal) to $1$.

In Figure 6 we plot the order parameter $J$ which enters the first relevant term in the large$-r$ expansion
of the magnetic gauge potential as a function of the Hawking temperature ($i.e.$
a varying $r_h$) and several values of the electric charge parameter. Again, the behaviour of  $J$
depends crucially on the value of $Q$. For $Q<Q^{(c)}$, $J$
approaches a constant value as $T_H\to 0$. The behaviour is different for  $Q \geq Q^{(c)}$,
$J$ in that case increasing strongly with $T_H$ and taking always large values,
which makes difficult its accurate computation.

 \setlength{\unitlength}{1cm}
\begin{picture}(8,6)
\put(3,0.0){\epsfig{file=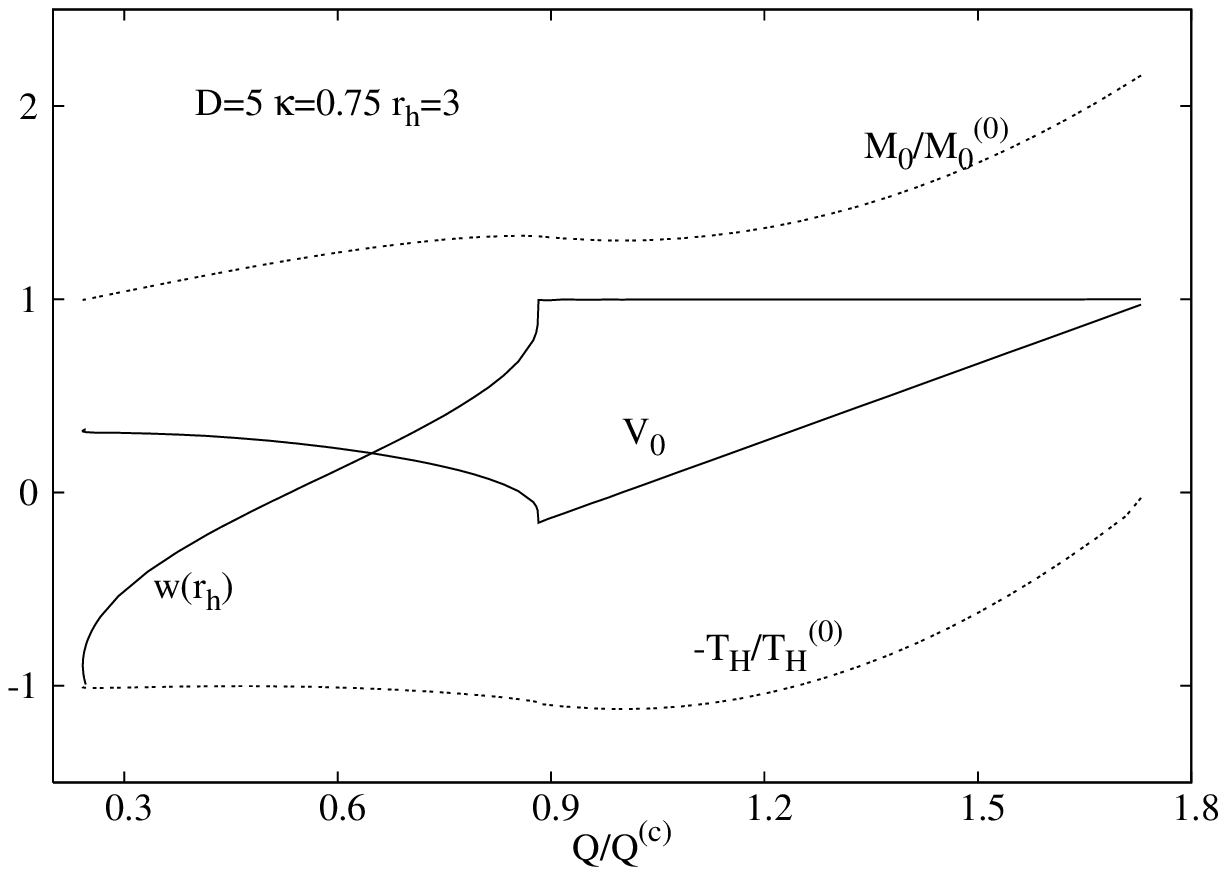,width=8cm}} 
\end{picture}
\\
\\
{\small {\bf Figure 5.}
Several relevant parameters are plotted as a function of the electric charge
parameter 
$Q$ for black hole solutions with a fixed value of the Chern-Simons coupling constant $\kappa$
and a given event horizon radius.
The value of the mass parameter $M_0$ and of the Hawking temperature $T_H$
are normalized with respect the critical Reissner-Nordstr\"om solution where a branch of non-Abelian
solutions emerges as a perturbation. }
\vspace{0.5cm}
   
 \setlength{\unitlength}{1cm}
\begin{picture}(8,6)
\put(3,0.0){\epsfig{file=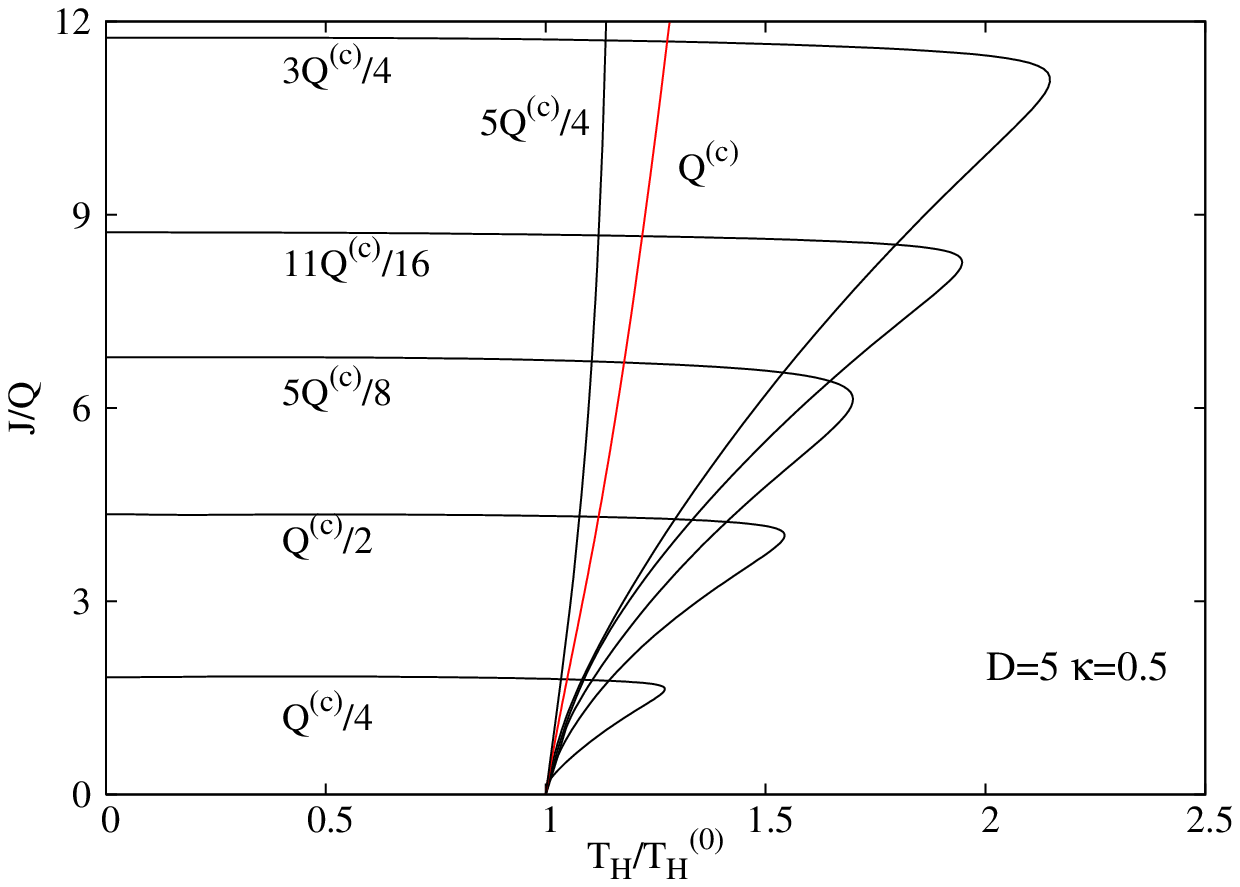,width=8cm}} 
\end{picture}
\\
\\
{\small {\bf Figure 6.}
The  order parameter $J$ which enters the asymptotics
of the magnetic gauge potential at infinity 
is shown as a function of the Hawing temperature $T_H$
for a fixed value of the Chern-Simons coupling constant
$\kappa$ and several values of the electric charge.
Note that $T_H$ is normalized with respect the temperature of the critical
Abelian solution where a branch of non-Abelian
solutions emerges as a perturbation. }
\vspace{0.5cm}

\subsubsection{The thermodynamics of solutions}
The nA solutions appear  to exist for values of the  CS coupling constant $\kappa \geq 1/8$.
Similar to the RN case,   the  nA black holes with  given $\kappa,Q$
are found only for a finite interval of  $r_h$ ($i.e.$, of the entropy) only. 
The detailed picture depends however on the ratio $Q/Q^{(c)}$.
(We recall that for $D=5$, $Q^{(c)}=16\kappa$.) 

For fixed $Q \neq Q^{(c)}$, the behaviour of the solutions is rather similar to the Abelian case and
the temperature reaches its maximum at some intermediate value of the event horizon radius,
an extremal black hole being approached for a minimal value 
of $r_h$.
A plot of the horizon area as a function of the temperature reveals  
the existence of several branches of nA solutions.
The typical picture can be summarised as follows.
For a given $\kappa>1/8$ and any value of $Q$,
a branch of non-Abelian solutions emerges as a perturbation of a
\setlength{\unitlength}{1cm}
\begin{picture}(8,6)
\put(-0.5,0.0){\epsfig{file=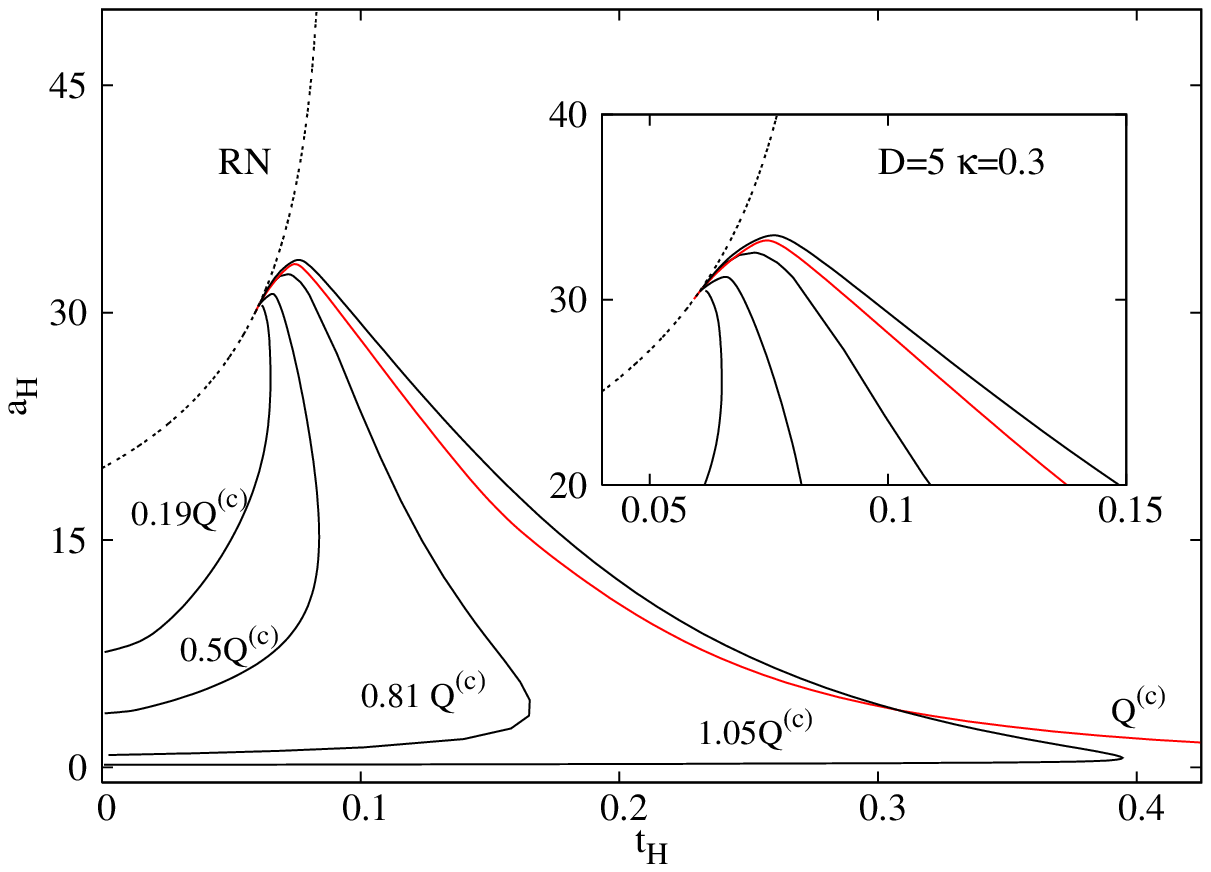,width=8cm}}
\put(8,0.0){\epsfig{file=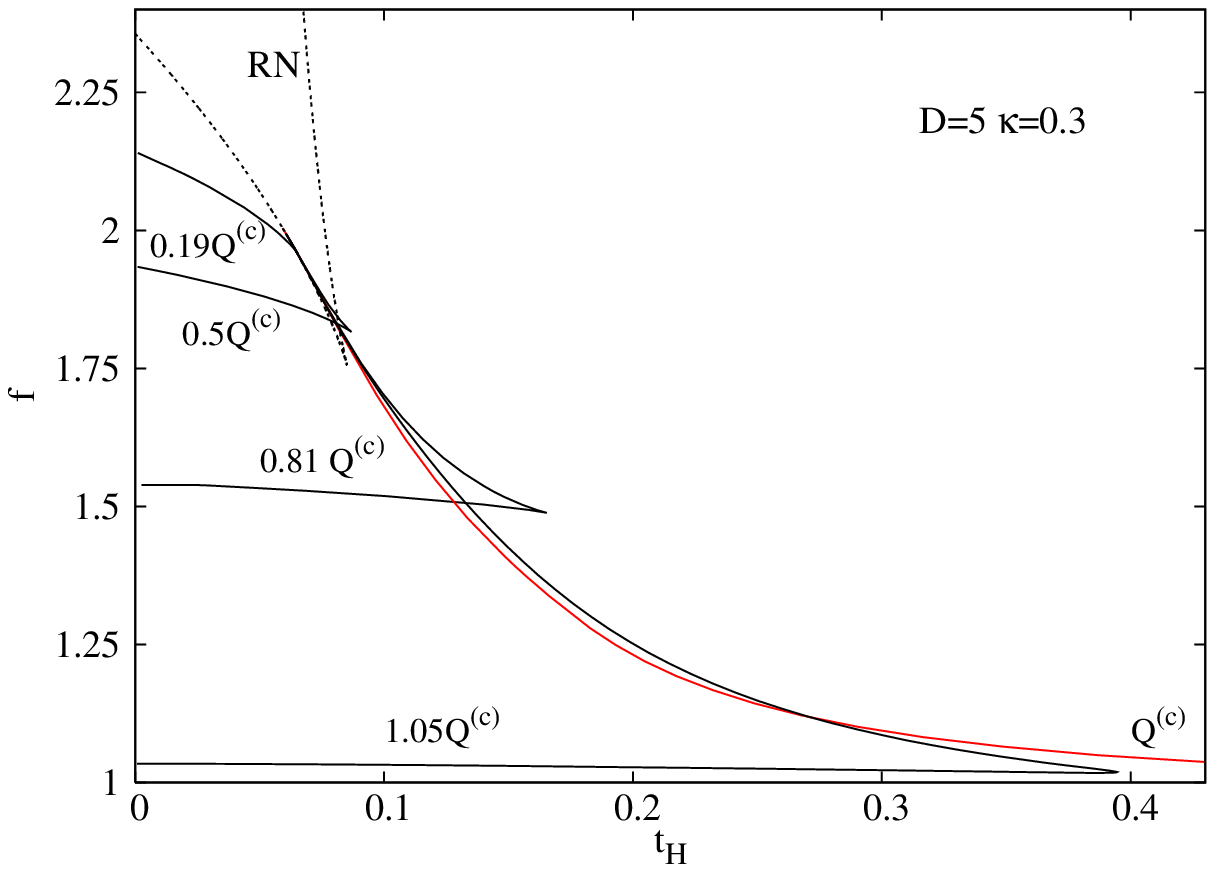,width=8cm}}
\end{picture}
\\
\\
{\small {\bf Figure 7.}
The scaled event horizon area $a_H$ (left) and free energy $f$ (right) are plotted $vs.$ the scaled temperature $t_H$ for
the non-Abelian  solutions with several values of the electric
charge and a given value of $\kappa$. 
The branch of Reissner-Nordstr\"om (RN) solutions is also shown.
 }
\vspace{0.5cm}
\\
\setlength{\unitlength}{1cm}
\begin{picture}(8,6)
\put(-0.5,0.0){\epsfig{file=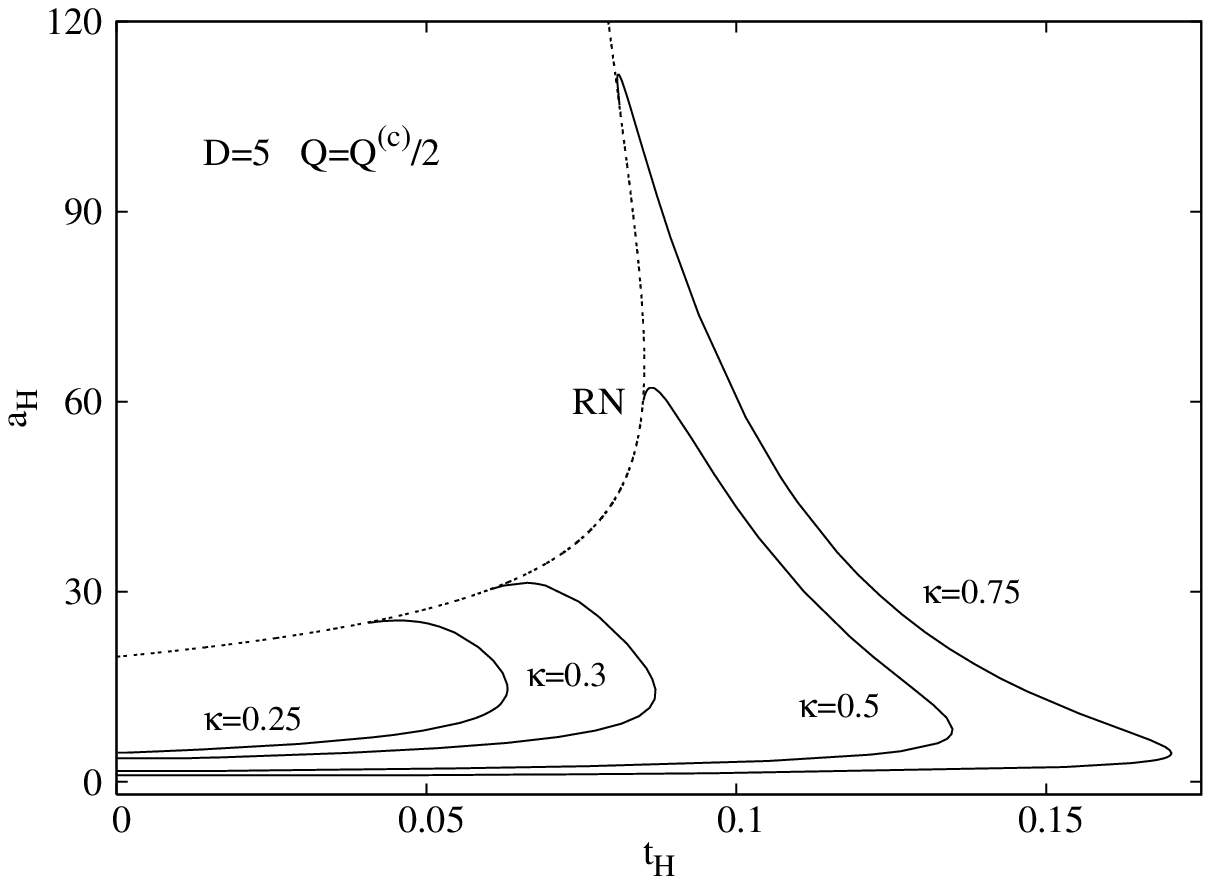,width=8cm}}
\put(8,0.0){\epsfig{file=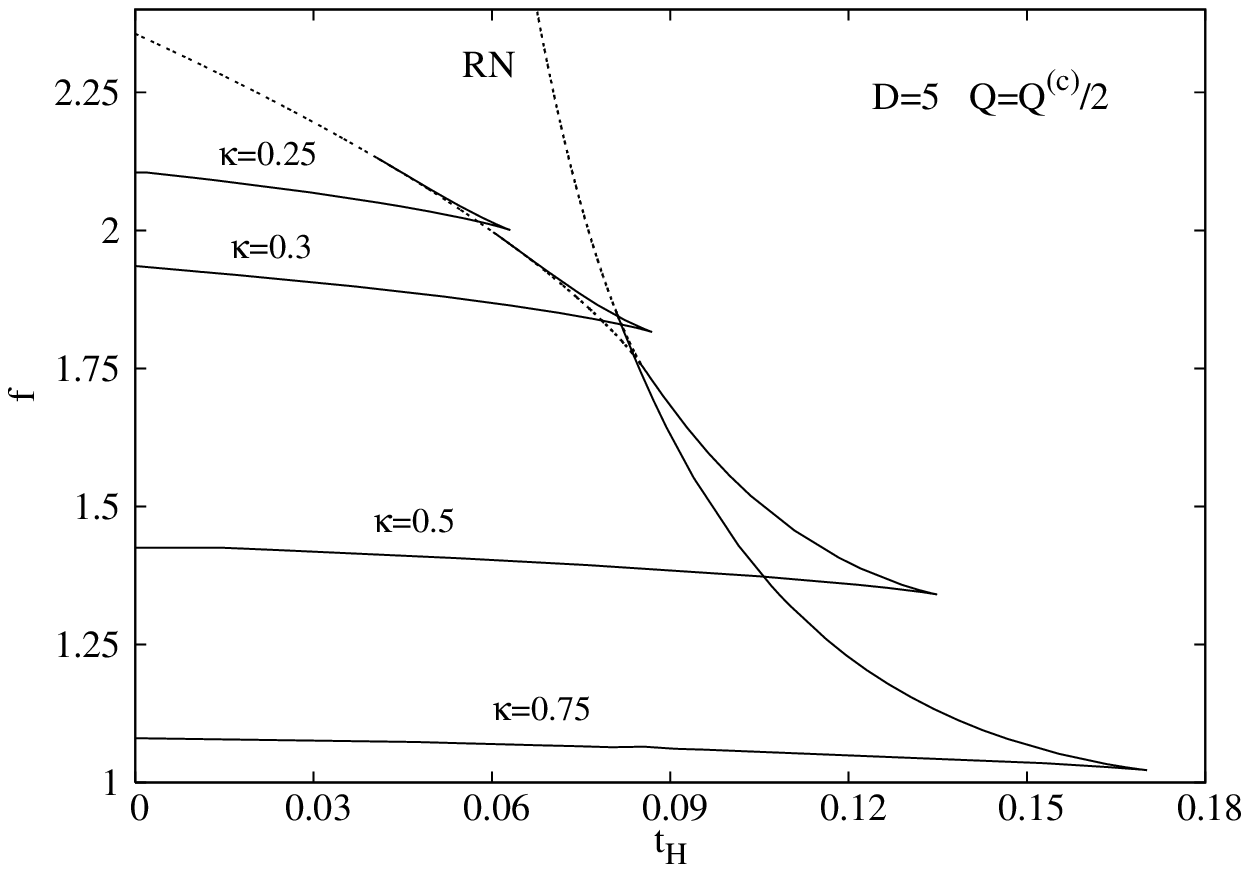,width=8cm}}
\end{picture}
\\
\\
{\small {\bf Figure 8.}
Same as Figure 7 for a fixed value of the electric charge parameter $Q=Q^{(c)}/2$
and several different values of the Chern-Simons coupling constant $\kappa$.
  }
\vspace{0.5cm}
\\
 critical RN configuration with $r_h=\sqrt{Q/U(\kappa)}$.
The entropy increases with temperature along this branch, which has, however,
a small  extension in both $a_H$ and $t_H$.
This branch continues in a secondary one, where the temperature 
still increases, while the event horizon area decreses. Thus, these solutions 
have negative specific heat.
For $Q \neq Q^{(c)}$, this branch ends for a maximal value of $t_H$ (whose value depends on $Q$)
where a third branch of  solutions emerge.
This branch extends backward in $(t_H,a_H)$ and has a positive specific heat. 
The Hawking temperature vanishes there 
for a minimal value  $r_h^{(min)}$  of the event horizon radius.

As $r\to r_h^{(min)}$, an extremal nA black hole solution  with an $AdS_2\times S^3$
near horizon geometry is approached. For $Q <Q^{(c)}$, the parameters of this extremal black hole can be read from 
from (\ref{wh-rh-extr}), (\ref{Q-extr}).
For example, the near horizon expansion of the solutions implies
\begin{eqnarray}
r_h^{(min)}=  4\sqrt{3}  \kappa \frac{(1-w_h^2)}{\sqrt{64 \kappa^2- w_h^2}},
\end{eqnarray}
where $w_h$ satisfies the cubic equation 
\begin{eqnarray}
(64 \kappa^2-w_h^2)Q^2+2\kappa(1+w_h)^2(128 \kappa^2(w_h-2)+w_h(w_h^2-2w_h+3))=0.
\end{eqnarray}
For $Q >Q^{(c)}$, the limiting extremal solution has 
\begin{eqnarray}
r_h=r_h^{(min)}=\sqrt{Q-Q^{(c)}},
\end{eqnarray}
and 
\begin{eqnarray}
\label{ne1}
&&H(r)=\frac{4}{r_h^2}(r-r_h)^2+\dots,~~\sigma(r)=\sigma_h\left(1+\frac{3}{2r_h}\frac{w_1^2k^2}{2k-1}(r-r_h)^{2k-1}\right)+\dots,
\\
\nonumber
&&V(r)=\frac{2}{r_h \sigma_h}(r-r_h)+\dots,~~w(r)=1+w_1(r-r_h)^k+\dots,
\end{eqnarray}
where $w_1$ and $\sigma_h$ are parameters fixed by numerics and 
\begin{eqnarray}
k=\frac{1}{2}\left(-1+\sqrt{5+32 \kappa} \right)>1.
\end{eqnarray}
Although close to the horizon,  due to the scaling relation (\ref{ss2}),  this extremal solution is essentially similar
to the RN one, their bulk form is different. 
The nA solution presents 
a magnetic hair (outside the horizon) and a  metric function $\sigma(r)\neq 1$.
Another interesting feature of the nA configurations with $Q >Q^{(c)}$ is that the magnetic flux lines 
are 'expelled' from the black holes as extremality is approached.
That is, one finds that the nA  magnetic field is vanishing 
on the horizon of the extremal black holes admitting the approximate expansion (\ref{ne1}), $F_{ij}=0$.
Thus these solutions seem to exhibit a sort of nA 'Meissner effect'
which is characteristic of superconductiong media.
This should be contrasted with the $Q<Q^{(c)}$ case, which possesses a
nontrivial nA magnetic field on the horizon.

  Some features of the nA black holes
   are shown in Figures 7 and 8 (left) where we plot the reduced area of the horizon $a_H=2\pi^2 r_h^3/Q^{3/2}$ 
as a function of the   dimensionless temperature 
$t_H$ for a fixed  CS coupling constant and several values of the charge parameter $Q$.
The branch of RN solutions as given by (\ref{tH-RN}), (\ref{f-RN}) is also shown there.

Furthermore, it turns out that the free energy $F={\cal M}-T_H S$  
of a RN solution  is larger than the free energy
of a lower branch nA solution with the same temperature and electric charge, except for 
configurations with $\kappa$ close to $1/8$ and small enough values of the charge, $Q \lesssim Q^{(c)}/3$.
Therefore the nA black holes are generically preferred.  
These aspects are exhibited in Figures 7, 8 (right) where the dimensionless free energy $f$  is 
plotted as a function of the  dimensionless temperature $t_H$. 
Moreover,   for the same values of the mass and electric charge, 
the RN solution typically  has a smaller event horizon radius (and thus a smaller entropy),
than the nA black hole \cite{Brihaye:2010wp}.
Note, however, that most of the nA configurations have no RN counterparts, see the Figures 7, 8.

Also, one can see that the interval of the scaled temperature $t_H$ covered by the set of nA solutions
shrinks to zero as $\kappa$ approaches the minimal value $1/8$.
As $\kappa \to 1/8$, all three branches described above collapse to a single point,
which is the extremal black hole solution.
This limiting solution admits a closed form expression and will be discussed separately. 

The overall picture is somehow different for  $Q=Q^{(c)}$, in which case,
despite the presence of an electric charge, the nA black holes behave in a 
similar way to the vacuum Schwarzschild-Tangherlini 
solution, with a single branch of thermally unstable configurations, see Figure 7. 
These solutions emerge again from a critical RN solution with  $r_h=\sqrt{Q^{(c)}/U(\kappa)}$
and can be continued for an arbitrarily small value of the event horizon radius.
As $r_h\to 0$, the  black holes with $Q=Q^{(c)}$ approach a set of globally regular particle-like solutions,
with $t_H$ diverging in that limit, as expected.

\subsection{On the existence of $D=5$ stable black hole solutions}

Typically, the existence of an unstable mode of a nA configuration is associated with the zeros of the magnetic
gauge potential $w(r)$. Thus the fact that we have found nodeless solutions suggests the existence
\setlength{\unitlength}{1cm}
\begin{picture}(8,6)
\put(-0.5,0.0){\epsfig{file=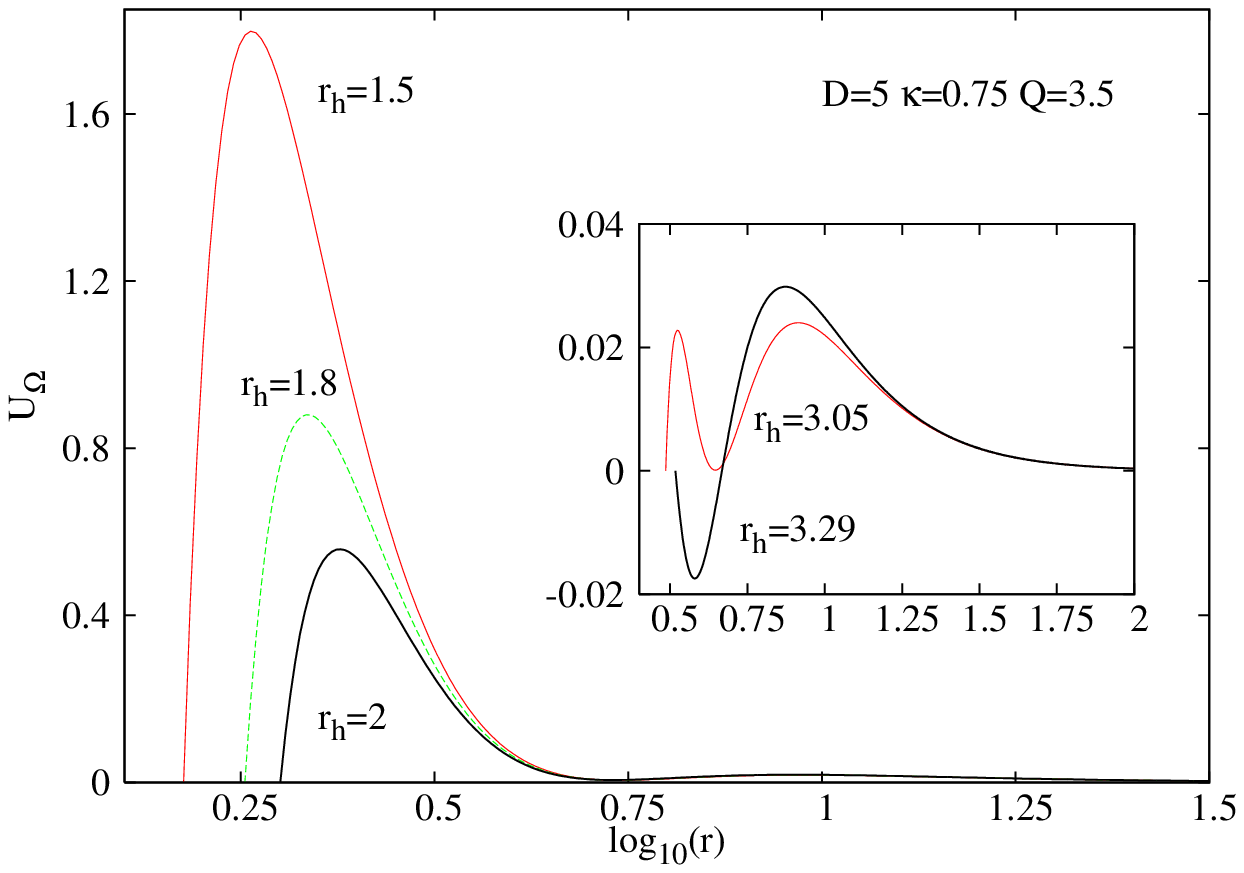,width=8cm}}
\put(8,0.0){\epsfig{file=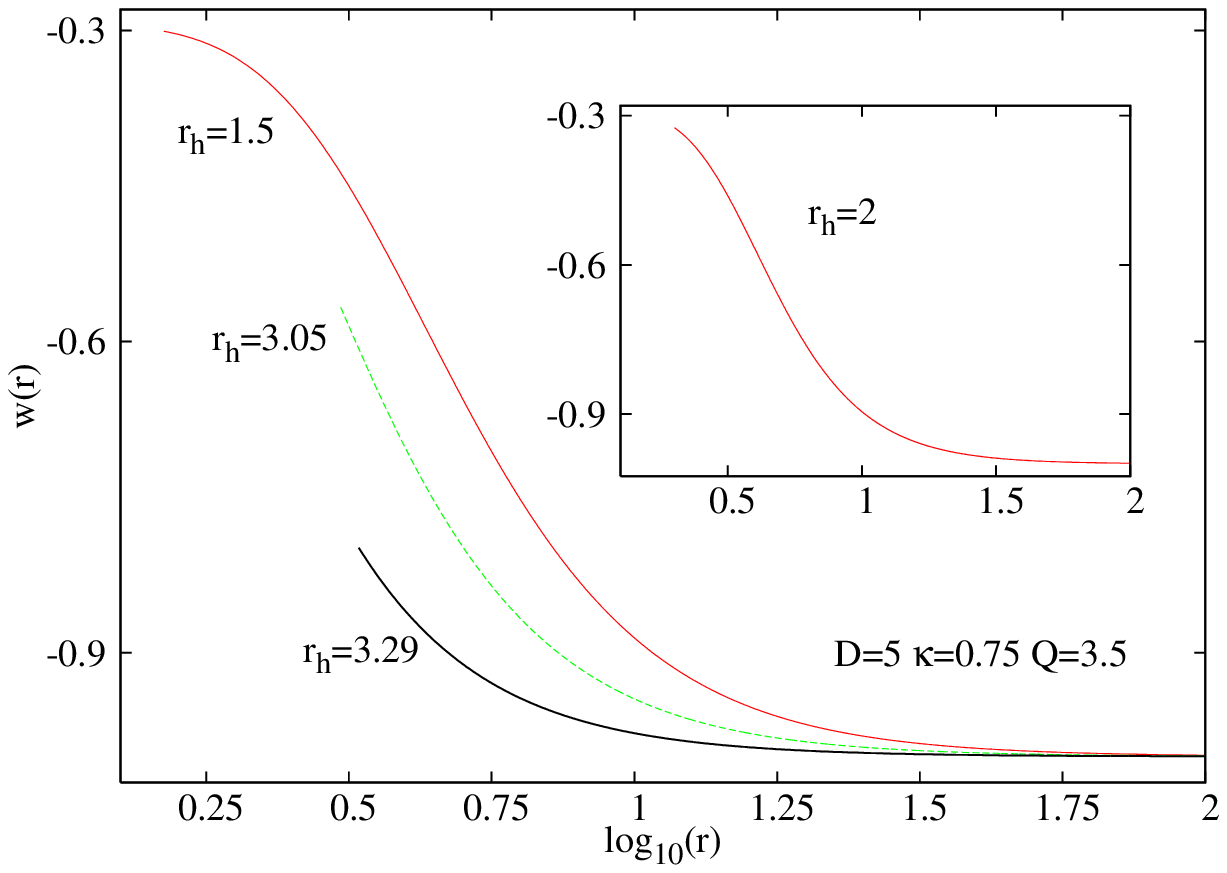,width=8cm}}
\end{picture}
\\
\\
{\small {\bf Figure 9.}
The potentials for the perturbation equation
(\ref{stab1}) are shown   together with the corresponding magnetic gauge potentials of the 
unperturbed solutions 
for several values of $r_h$ and  fixed values of  $\kappa$, $Q$.
For $r_h=3.29$, there is a negative region of the potential, $V(r)<0$.
 }
\vspace{0.5cm}
\\
 of  configurations
which are stable against spherically symmetric perturbations. Moreover, the nA solutions are arbitrarily close
to the Abelian RN configuration, which is known to be stable.

Thus in this subsection we address the issue of linear stability of the black holes discussed above.
For that purpose, we have to study the evolution 
of linear perturbations around the equilibrium configuration.
For the pure EYM case and with a gauge group $SU(2)$,
this has been studied by Okuyama and Maeda in \cite{Okuyama:2002mh},
who derived the corresponding pulsation equation.
Below we repeat their derivation with a slight modification due
to the presence of a CS term and hence also an electric potential.
Also, the purpose here is to show the existence of 
stable black hole solutions rather than study the unstable modes, which for our purposes here
is of secondary importance.
 
 In examining such time-dependent fluctuations, we consider the following metric Ansatz
 generalizing (\ref{formN}): 
\begin{eqnarray}
ds^2=\frac{dr^2}{N(r,t)}+r^2 d\Omega_3^2-N(r,t)\sigma^2(r,t)dt^2, ~~{\rm with}~~
N(r,t)=1-\frac{m(r,t)}{r^2}.
\end{eqnarray}
On the gauge field sector, we shall restrict our study to perturbations
within the considered $SO(4)\times SO(2)$ model.
For the $U(1)$ part, one can take without any loss of generality
an Ansatz with a single nonvanishing component $V(r,t)$.

The construction of a general time-dependent Ansatz for the $SO(3)$
gauge group in $D=5$ dimensions has been extensively discussed in \cite{Okuyama:2002mh}.
Interestingly, it turns out that the corresponding YM Ansatz is 
much more restricted in this case than in $D=4$ dimensions, containing only one 
potential $w(r,t)$.  
This agrees with the result found by taking an $SO(4)\times SO(2)$ truncation
of the general Ansatz (\ref{a0p}), \re{aip}.

 Then the perturbed variables can be written as
\begin{eqnarray}
&& m(r,t)=m(r)+\epsilon m_1(r)e^{i \Omega t}+\dots,~~
\sigma(r,t)=\sigma(r)(1+\epsilon \sigma_1(r)e^{i \Omega t})+\dots,
\\
\nonumber
&& w(r,t)=w(r)+\epsilon w_1(r)e^{i \Omega t}+\dots,~~
V(r,t)=V(r)+\epsilon V_1(r)e^{i \Omega t}+\dots,
\end{eqnarray}
with $m(r)$, $\sigma(r)$, $w(r)$ and $V(r)$
a static solution and $\epsilon$ a small parameter.

After replacing in the general EYMCS equations (\ref{Einstein-eqs}),
one finds the following relations valid to first 
%
\setlength{\unitlength}{1cm}
\begin{picture}(8,6)
\put(-0.5,0.0){\epsfig{file=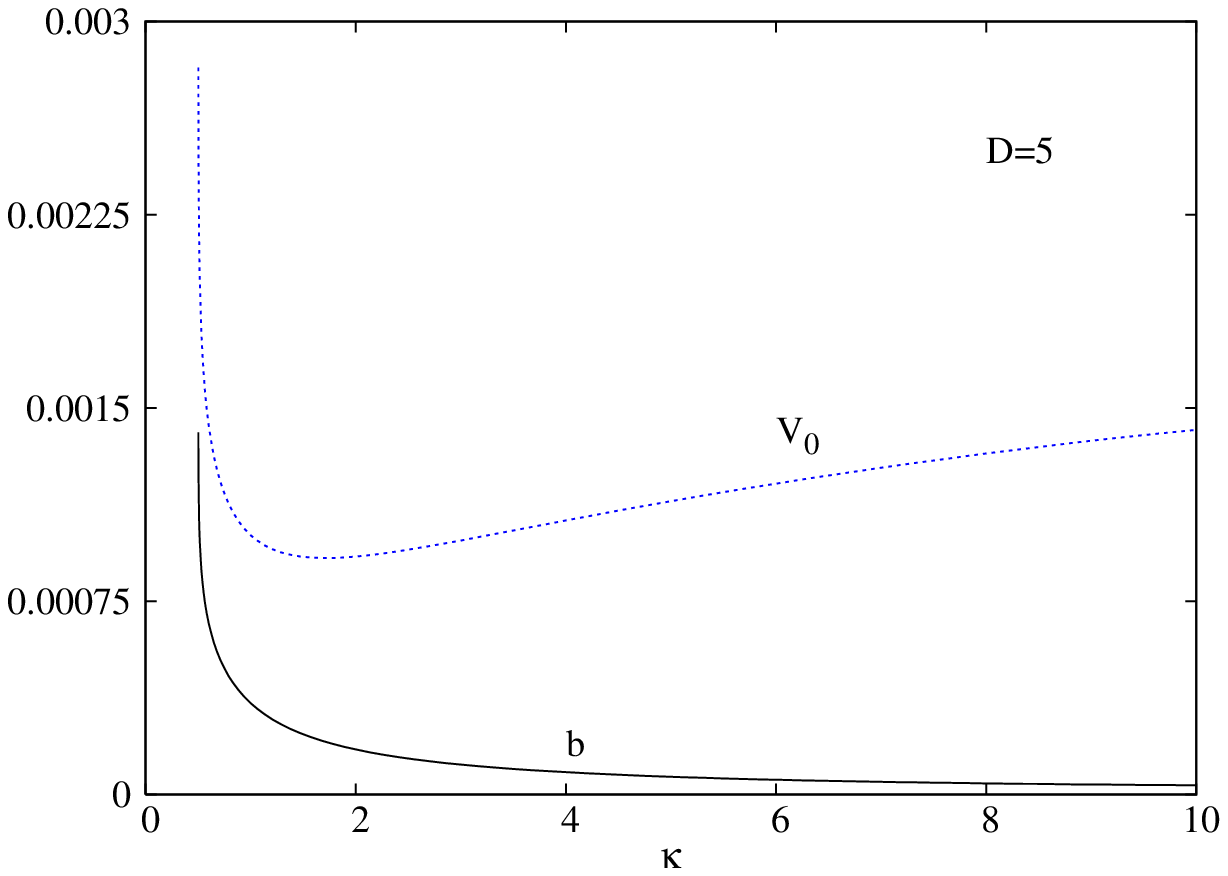,width=8cm}}
\put(8,0.0){\epsfig{file=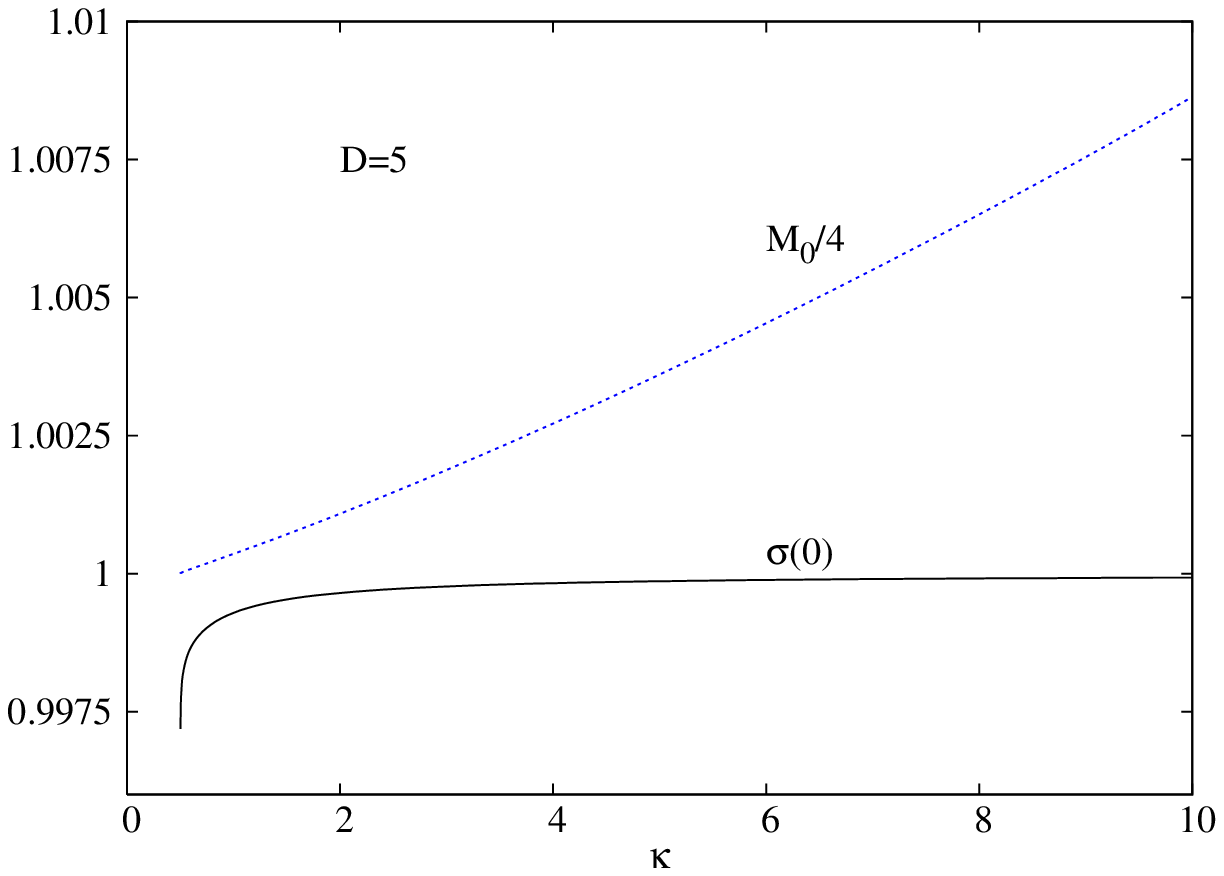,width=8cm}}
\end{picture}
\\
\\
{\small {\bf Figure 10.}
A number of parameters of the $D=5$
globally regular solutions are plotted as a function
 of the Chern-Simons coupling constant $\kappa$. 
 }
\vspace{0.5cm}
 \\order in $\epsilon$:
\begin{eqnarray}
\nonumber
&&
w_1''
+
\left(
\frac{1}{r}+\frac{N'}{N}+\frac{\sigma'}{\sigma}
\right)
w_1'
+\frac{8\kappa (1-w^2)}{r N\sigma}V_1'
-\frac{w'}{r^2 N}m_1'
+w'\sigma_1'
+\frac{1}{N}
\left(
\frac{\Omega^2}{N\sigma^2}+\frac{2}{r^2}(1-3w^2)
-\frac{16\kappa w V'}{r\sigma}
\right)
w_1
\\
&&{~~~~~~~~~~}
+
\left(
\frac{w'}{r^2 N}(1-\frac{r\sigma'}{\sigma})-\frac{w''}{r^2 N}
\right)
m_1
-\frac{8 \kappa(1-w^2)V'}{r N \sigma}\sigma_1=0,
\\
&&
\nonumber
m_1=3 r N w' w_1,~~~ \sigma_1'=\frac{3  \sigma w' }{ r }w_1',
~~V_1'=\frac{24 \kappa \sigma}{r^3}(w^2-1)w_1+\sigma V' \sigma_1.
\end{eqnarray}

Thus  the functions $m_1,V_1$ and $\sigma_1$
can be eliminated in favor of $w_1(r)$, leading to a single Schr\"odinger equation
for $w_1$: 
\begin{eqnarray}
\label{stab1}
-\frac{d^2 \chi}{d r_\star^2}+U_\Omega \chi =\Omega^2 \chi~,
\end{eqnarray}
\newline
where $\chi=w_1 \sqrt{r}$ and a new radial coordinate $r_\star$ 
is introduced via $\frac{d}{dr_\star}=N\sigma\frac{d}{dr}$.
The expression of the potential in (\ref{stab1}) is
\begin{eqnarray}
\nonumber
U_{\Omega} =\frac{N\sigma^2}{r^2}
\bigg[
6(w^2-w'^2-\frac{1}{6})
-\frac{5N}{4}
+12 (w^2-1)\frac{ww'}{r}
+\frac{(1-w^2)^2}{r^2}
(
\frac{9}{2}w'^2+
192 \kappa^2 -\frac{3}{4}
)
\\
\label{potential}
+\frac{16\kappa V'}{\sigma}(rw -3 (1-w^2)w')
-\frac{r^2 V'^2(1-6 w'^2)}{4\sigma^2}
\bigg].
\end{eqnarray} 
One can verify that for $\kappa=V=0$, the above relations reduce to those found in 
\cite{Okuyama:2002mh} for $D=5$ EYM system (note that, however, the unperturbed solutions there have infinite mass).

The potential above is regular in the entire range $-\infty<r_\star<\infty$.
Near the event horizon, one finds $U_{\Omega}\to 0$; for large values of $r_\star$ the potential
is positive and  bounded. Standard results from quantum mechanics \cite{Messiah} further imply that 
there are no negative eigenvalues for $\Omega^2$ (and then no unstable modes) 
if the potential $U_{\Omega}$ is everywhere positive.

Indeed, our numerical results show  the existence of 
black hole solutions with a positive potential $U(r_\star)$ for all value of $\kappa>1/8$ we have considered.
As expected, all stable solutions we could find in this way have no nodes of the magnetic
gauge potential $w(r)$ (see   the  Figure 9 for such configurations).
Therefore at least some of our solutions are linearly stable. 
The picture is however, quite complicated and depends on the values of
$\kappa$, $Q$ and of the event horizon radius.
For example, for solutions with $\kappa=0.3$ and $Q=0.187Q^{(c)}$,
all solutions between the critical extremal black hole configuration  (with $r_h\simeq 0.702$)
and $r_h \simeq 0.95$, have $U(r_\star)>0$.
However, for the same value of the $\kappa$, all configurations with $Q=1.05Q^{(c)}$ have $U(r_\star)<0$.

At the same time, we cannot predict anything if
$U(r_\star)$ is not positive definite.
In this case we have to solve  numerically the eq. (\ref{stab1}) as an eigenvalue problem.
Very likely, the full picture is complicated and a systematic study would represents a very complex task.
We note only that, by using a trial-function approach (see e.g. \cite{Boschung:1994tc}) we could prove that a number
of EYMCS solutions
with one node in $w(r)$ are indeed unstable.  
 
\subsection{The globally regular solutions}

All $D=5$ globally regular solutions emerge as zero event horizon radius limit of the black hole 
branch with $Q=Q^{(c)}$. In principle, a disconnected branch of solitons with $\kappa <1/8$
may exist; however, we could not find such configurations. Heuristically, this can be understood as follows.
Since, from (\ref{Qc}), the electric charge is proportional to the CS coupling constant, the existence of a
minimal value of $\kappa$ means that below that value the electrostatic repulsion is too small
compared with other interactions for a bound state to exist. 

The properties of these solutions are uniquely fixed by the CS parameter $\kappa$ and are somehow different from 
other nA solitons in the literature. For example, their mass is an almost linear function of $\kappa$, while the
value at the origin of the metric function $g_{tt}$ is very close to $-1$, see Figure 10.
The shooting parameter $b=-w''(0)/2$ and the value at infinity of the electric potential take also small
values. It would be interesting to find an analytical understanding of these numerical results.
A typical $D=5$ globally regular solution is shown in Figure 14 (left).

Concerning the stability of solitons, 
the formalism proposed above for black holes applies also in the zero event horizon case.
However, at this stage we cannot say something precise on their stability,
since all globally regular solutions we have investigated  have a negative potential.
Also, the fact that there are no nodeless solitons (since $w(0)=1$ and $w(\infty)=-1$) 
strongly suggests that all particle-like solutions are expected to be unstable.

\subsection{The case $\kappa=\alpha/8$: An exact solution}
 
In what follows, we find it interesting
to restore the $(G,e)$-factors in the general expressions.
For $\kappa=\alpha/8$, one finds the following exact solution of the EYMCS equations within the metric 
Ansatz (\ref{metric-gen})
employing isotropic coordinates\footnote{In principle,
this solution can also be written in the Schwarzschild-type
coordinate system (\ref{formN}).
For example, the relation between the Schwarzschild radial coordinate $\bar r$
and the radial coordinate $r$ in (\ref{ex-sol1}) is
$\bar r=r/\sqrt{f(r)}$.
However, this results in a much more complicated expression of the solution.}, and a gauge group $SO(4)\times SO(2)$:
\begin{eqnarray}
\label{ex-sol1}
&&
w(r)=\frac{J-2 r^2}{J+2 r^2},~~
V(r)=\frac{e}{4}\sqrt{\frac{3}{\pi G}}f(r) , ~~
f_0(r)=f^2(r) ,~~
f_1(r)=\frac{f_2(r)}{r^2}=\frac{1}{f(r)},
\\
\nonumber
&&{~~~~~~~~}{\rm with}~~ 
f(r)=\left [
1+\frac{ Q^{(c)2}}{2r^2}\left(\frac{Q}{Q^{(c)}}-\frac{J^2}{ (2r^2+J)^2} \right)
\right]^{-1},~
\end{eqnarray}
where $Q>Q^{(c)}$ and $J$ are arbitrary parameters, and $Q^{(c)}=16 \kappa/e$ is the critical
charge parameter, 
  $Q^{(c)}= \frac{8}{e}\sqrt{\frac{\pi G}{3} }$.
This describes an extremal black hole with nA hair, the regular event horizon being at $r=0$
 (in these isotropic coordinates).
The mass, electric charge, chemical potential and entropy of this solution
are 
\begin{eqnarray}
\label{ex-sol2}
&&
{\cal M}= \frac{\sqrt{3}\pi^{3/2}Q}{e\sqrt{G}} =\frac{8\pi^2}{e^2}\frac{Q}{Q^{(c)}},
~~{\cal Q}=\frac{4 \pi^2Q}{e} ,
~~\Phi=\frac{1}{4}\sqrt{\frac{3}{\pi G}}=\frac{2}{e Q^{(c)}},
\\
\nonumber
&& 
S=\frac{\pi^2}{2 G} 
\left(4\sqrt{\frac{\pi G}{3}}\frac{Q}{e}-\frac{32\pi G}{3 e^2} \right)^{3/2}=
\frac{\pi^2}{2 G}
\left (
\frac{1}{2} (Q-Q^{(c)} )Q^{(c)}  
\right )^{3/2}, 
\end{eqnarray}
such that ${\cal M}=\Phi {\cal Q}$.
This exact solution has a number of interesting properties. For example, one can show that the horizon is regular, with
the following behaviour as $r\to 0$
\begin{eqnarray}
\label{ex-sol3}
&&f_1(r)=\frac{f_2}{r^2}=\frac{(Q-Q^{(c)})Q^{(c)}}{2r^2}+1+\frac{2Q^{(c)2}}{J}+O(r^2),~~
~~f_0(r)= \frac{4 r^4}{(Q-Q^{(c)})^2Q^{(c)2}}+ O(r^6),
 \end{eqnarray}
for the metric functions, and
 \begin{eqnarray}
\label{ex-sol4}
 &&R=\frac{4}{Q^{(c)}(Q^{(c)}-Q)}+O(r^2),~~~
R_{\mu\nu\rho\sigma}R^{\mu\nu\rho\sigma}=\frac{304}{(Q^{(c)}-Q)^2Q^{(c)2} }+O(r^2),
 \end{eqnarray}
 for the curvature and Kretschmann scalars.
Also, the magnetic field is vanishing on the horizon  (this property is also shared by 
the extremal 
 black holes with $Q>Q^{(c)}$ and $\kappa>\alpha/8$), with $w(r)=1-4r^2/J+O(r^4)$
.

The RN solution is found by taking $J=0$ in (\ref{ex-sol1}).
The mass, electric charge and chemical potential of this solution are still given by 
(\ref{ex-sol2}), while the entropy of the RN solution is $S_{RN}=\frac{\pi^2}{2G}(\frac{Q Q^{(c)}}{2})^{3/2}$,
which is higher than the entropy of the corresponding nA configuration.
Thus the extremal Abelian solution is thermodynamically favoured, a feature
which seems to be shared by other extremal solutions with $\kappa>\alpha/8$
we have constructed numerically.

In the limit $Q=Q^{(c)}$, the solution (\ref{ex-sol1}) describes a particle-like soliton, in which case
 \begin{eqnarray}
\label{ex-sol5}
f(r)=\left [
1+\frac{2Q^{(c)2}(J+r^2)}{(J+2 r^2)^2 } 
\right]^{-1},~
\end{eqnarray}
while, from (\ref{ex-sol2}), 
${\cal M}=8 \pi^2/e^2$,
${\cal Q}=32\sqrt{G}\pi^{5/2}/(\sqrt{3}e^2)$,
and $\Phi=\sqrt{3/(\pi G)}/4$.
One can show that $r=0$ is a regular origin, with the following behaviour in 
that limit
\begin{eqnarray}
\label{ex-sol6}
&&f_1(r)=\frac{f_2}{r^2}= 1+\frac{2Q^{(c)2}}{J}+O(r^2),~~
~~f_0(r)= \frac{J^2}{(J+2Q^{(c)2})^2 }+ O(r^2),
 \end{eqnarray}
and
\begin{eqnarray}
\label{ex-sol7}
 &&R=\frac{48 Q^{(c)2}}{(J+2Q^{(c)2})^2 }+O(r^2),~~~
R_{\mu\nu\rho\sigma}R^{\mu\nu\rho\sigma}=\frac{5760 Q^{(c)4}}{(J+2Q^{(c)2})^4 }+O(r^2).
 \end{eqnarray}
 One should note that the parameter $J$ which appears $via$ the magnetic
 gauge potential $w(r)$
 does not enter any global quantity (although these nA solutions are supported by a nonzero $J$).
 However, $J$
 has a physical meaning, since it enters also the metric components.
 Then, although the mass and charge are the same, a local observer could distinguish between 
 the nA solution (\ref{ex-sol1}) and the corresponding RN solution $via$ the motion
 of a test particle.

The reason for the existence of the exact solution (\ref{ex-sol1}) can be understood
by noticing that the solutions reported in this work admit an 
interesting connection with a  model considered in \cite{Gibbons:1993xt}.
This connection follows from the observation that, for static configurations,
the action (\ref{action}) with an $SO(4)\times SO(2)$ gauge group reduces essentially to
a Einstein--Yang-Mills-Maxwell model, with a Chern-Simons-type coupling term between
the $U(1)$ and nA fields. Basically, with this group contraction down from $SO(6)$, the CS density \re{CS5} reduces
to the hybrid CS density in \cite{Gibbons:1993xt}. 
Moreover, as noticed in that paper, this model with $\kappa=\alpha/8$  corresponds to the coupling of the super-YM theory 
to the $D=5$ supergravity \cite{Gunaydin:1984ak}.

It is shown in \cite{Gibbons:1993xt} that for this value of the CS coupling constant, 
any flat space self-dual solution of the $D=4$ YM equation with a gauge group~\footnote{In \cite{Gibbons:1993xt} in fact,
the YM field is takes its values in one or other chiral represenation of $SO(4)$, namely for self- and antiself-dual
$SU(2)$ fields. Here, by contrast, our ``magnetic'' YM connection is fully $SO(4)$ valued, leading to the appearance
of $\Si_{5,6}=\ga_5$ in \re{sd-D4}.} $SO(4)$ can be uplifted to $D=5$ and promoted to soliton solutions of the full model. 

A slight generalisation of this construction can be summarised as follows.
Let us consider a $D=4$ seed configuration consisting in a geometry described by a Ricci-flat line element $d\sigma^2$
(parametrized by the $\vec x$-coordinates) with Euclidean signature, and a $SO(4)$ nA field
satisfying the self-duality equations\footnote{Thus all known $D=4$, $SU(2)$ YM instantons
can provide a solution of (\ref{sd-D4}).}
\begin{eqnarray}
\label{sd-D4}  
F_{ij}=\frac{1}{2}\sqrt{{\rm det} \sigma }\,\ga_5\,\vep_{ijkl}  F^{kl},
\end{eqnarray}  
in a metric background given by $d\sigma^2$.

Then this configuration can be uplifted to solutions of the 
$D=5$ EYMCS model in this work, for an $SO(4)\times O(2)$ gauge group. The five-dimensional line element is
\begin{eqnarray}
ds^2=\frac{1}{f(\vec x)}d\sigma^2-f^2  (\vec x)dt^2,
\end{eqnarray}
while the purely electric $SO(2)$ potential
reads
\begin{eqnarray}
\label{GT2}
V(\vec x)=\frac{e}{4}\sqrt{\frac{3}{\pi G}}f(\vec x). 
\end{eqnarray}
In the above relations, 
$f(\vec x)$ is a solution of the Poisson equation
\begin{eqnarray}
\label{GT1}
\nabla^2 (\frac{1}{f})=-\frac{4 \pi G}{3}
 \frac{1}{2}  \mbox{Tr}\,
\left \{ 
 F_{ij}\, F^{ij}
\right \}\,,
\end{eqnarray}
where the operator $\nabla^2$ is taken with respect to the four dimensional metric $d\sigma^2$.
The $D=5$ YM gauge field is $F=F_{ij} dx^{i}\wedge dx ^{j}+\left(dV\wedge dt\right)  \Sigma_{5,6}^{(\pm)}~$.
 
Then one can easily prove that the full set of equations (\ref{Einstein-eqs}) 
are satisfied, provided $\kappa$ takes the special value, 
\begin{eqnarray}
\label{kappa}
 \kappa=\frac{e}{2}\sqrt{\frac{\pi G}{3}}=\frac{\alpha }{8}.
\end{eqnarray}

In principle, this approach can be used to uplift to $D=5$
all four dimensional self-dual solutions of the YM equations, including configurations displaying no symmetries and
multi-center solutions. Note that only soliton solutions of the equation (\ref{GT1}) were considered in
Ref. \cite{Gibbons:1993xt}.
Here, this has been extended to the construction of black hole solutions 
by adding to $1/f$ an extra part which is a solution of the homogeneous equation $\nabla^2 (\frac{1}{f})=0$
with suitable boundary conditions.

The simplest case is found by taking
$d\sigma^2$ to be the four dimensional Euclidean space and 
$F_{ij}$ the BPST instanton  \cite{Belavin:1975fg}.
This leads to the spherically symmetric configuration (\ref{ex-sol1}) discussed above.
However, a variety of other physically interesting solutions can be obtained in a similar way
(this includes configurations with a $D=4$ non-flat base space, $e.g.$ the Euclideanised Schwarzschild metric and
the $D=4$ YM instantons in \cite{Charap:1977ww}, \cite{Brihaye:2006bk}).
Moreover, one can generalise this framework by including rotation in the $D=5$
metric Ansatz, in which case it is possible $e.g.$, to find a hairy generalisation of the
BMPV black hole \cite{Breckenridge:1996is}. Such solutions are found 
beyond the simple Ansatz in this work and will be reported elsewhere.

\section{On $D>5$ solutions}
\setcounter{equation}{0}

Our results confirm the existence of both black holes and soliton
solutions of eqs. (\ref{red-eqs}) for  $D=7,9$ as well.
Then we expect that the EYMCS model considered possesses
asymptotically flat, spherically symmetric solutions 
with finite mass for any $D=2n+1\geq 5$. 
The numerical methods employed in this case are similar 
to those discussed for $D=5$, though as expected
the numerical difficulties increase with $D$.

\setlength{\unitlength}{1cm}
\begin{picture}(8,6)
\put(-0.5,0.0){\epsfig{file=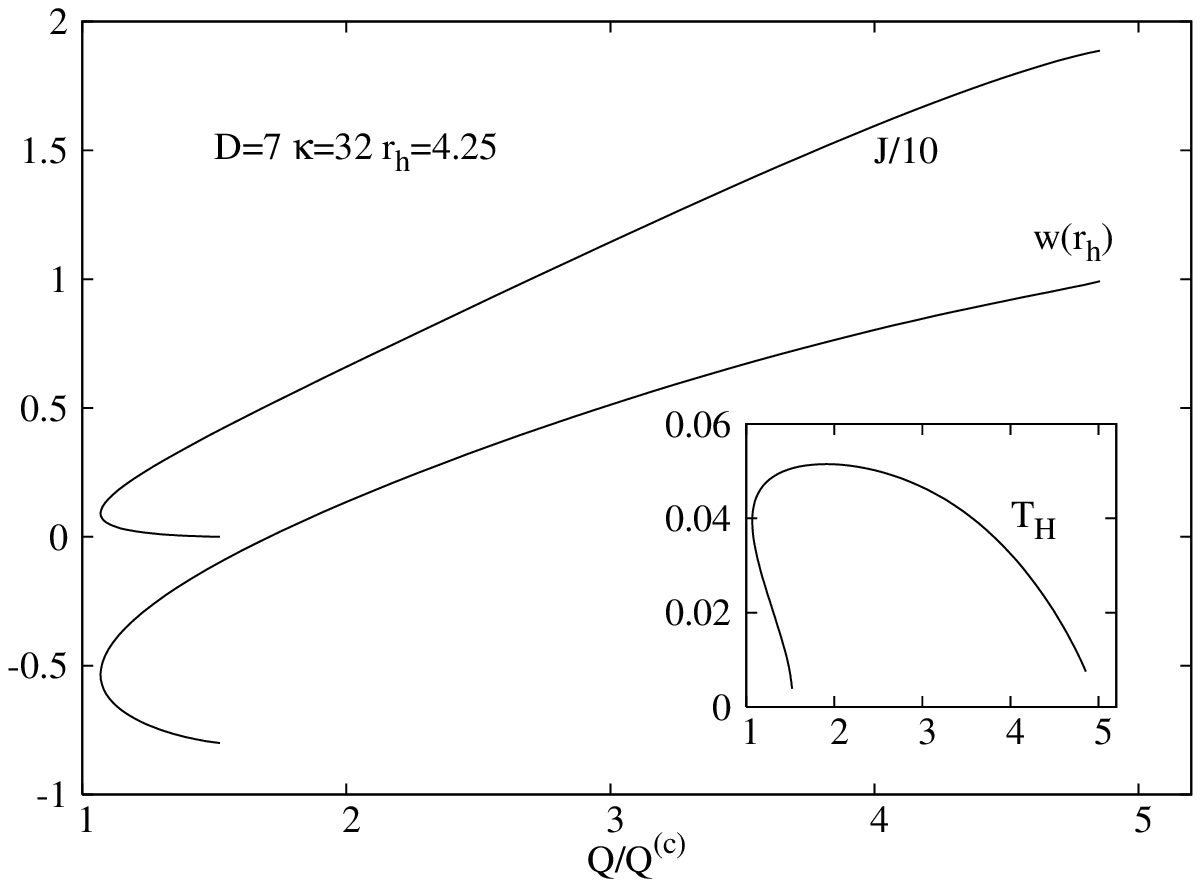,width=8cm}}
\put(8,0.0){\epsfig{file=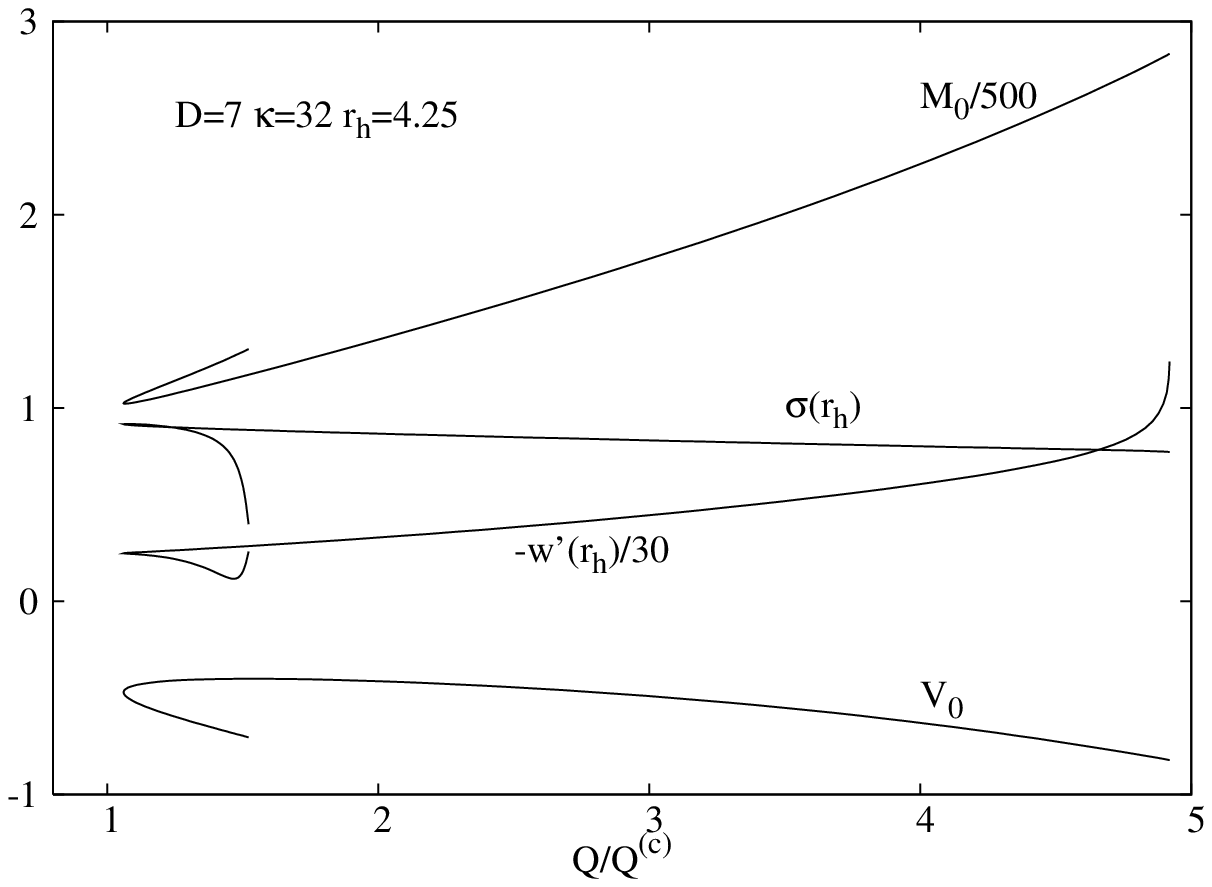,width=8cm}}
\end{picture}
\\
\\
{\small {\bf Figure 11.}
A number of parameters of  
$D=7$ black hole solutions are plotted as a function
 of the electric charge parameter $Q$ for fixed values of $(\kappa,~r_h)$. 
 Here and in Figure 12 $Q^{(c)}=-64\kappa/5$
is the critical value of the electric charge parameter. 
}
\vspace{0.5cm}
 
\setlength{\unitlength}{1cm}
\begin{picture}(8,6)
\put(-0.5,0.0){\epsfig{file=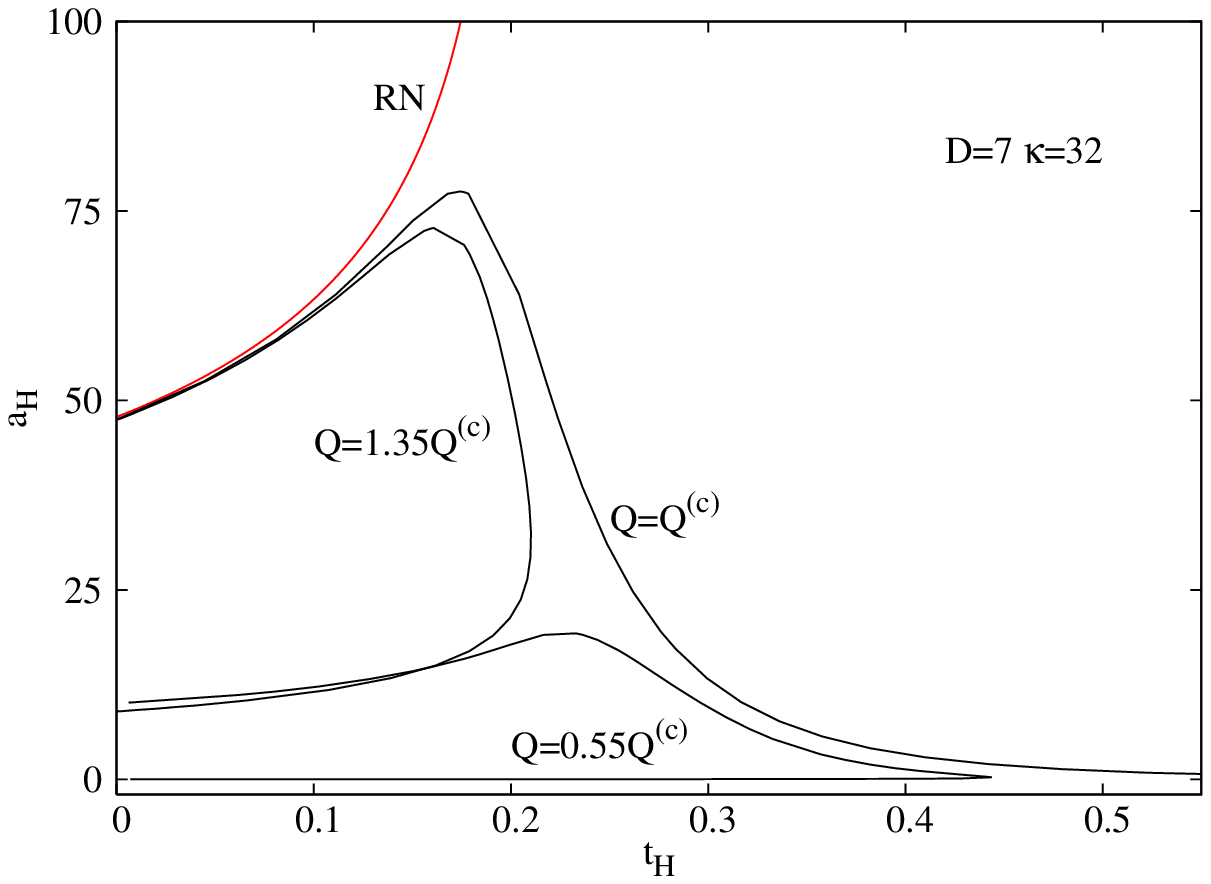,width=8cm}}
\put(8,0.0){\epsfig{file=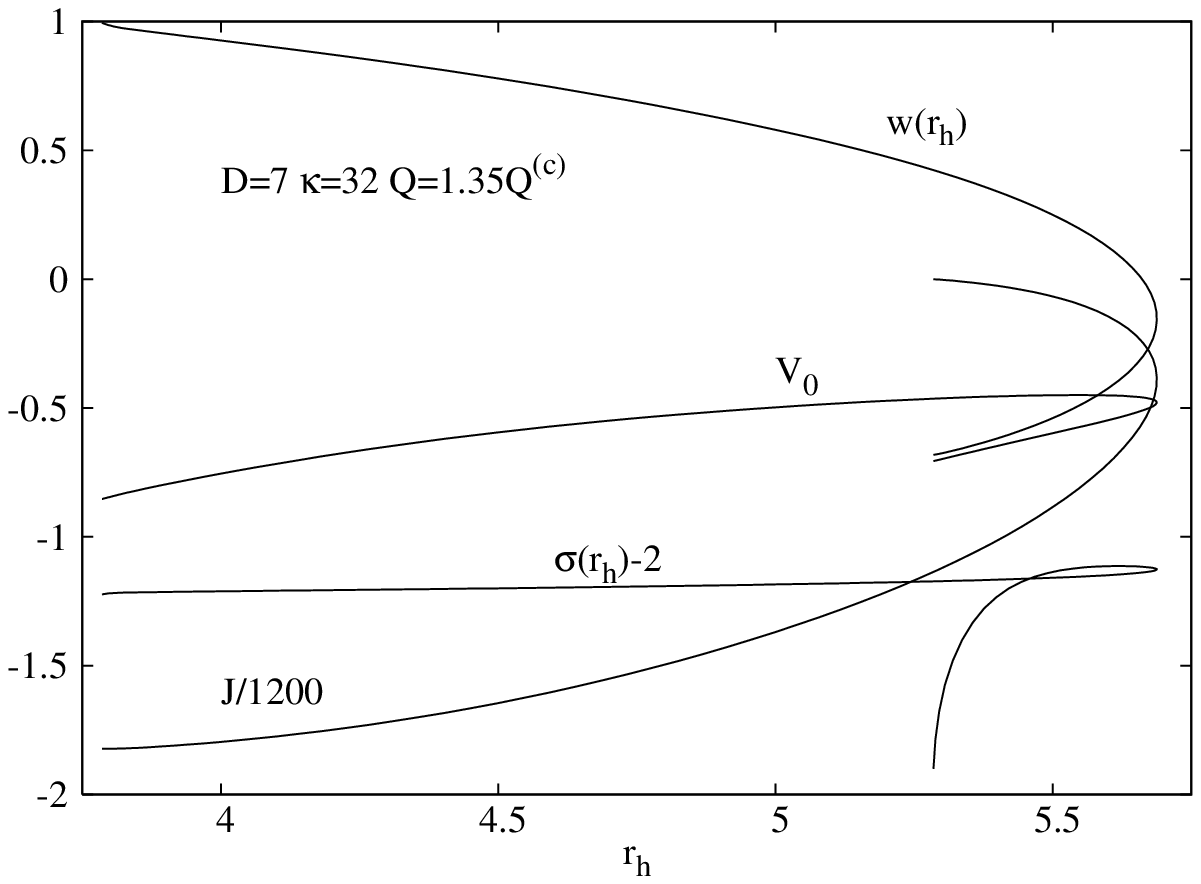,width=8cm}}
\end{picture}
\\
\\
{\small {\bf Figure 12.}
A number of parameters of  
$D=7$ black hole solutions are shown for fixed values of $\kappa,~Q$ and a varying $r_h$.}
\vspace{0.5cm}
\\
Determining the pattern of the $D>5$ solutions in the parameter space
represents a very complex task which is outside the scope of this paper.
Instead, we analysed in detail a few particular classes of $D=7,~9$ solutions which,
hopefully, reflect at least some of the relevant properties of the general pattern.

Not entirely surprisingly, it turns out that the case $D=5$ has some special properties.
For example we have seen already that for $D>5$, unlike in $D=5$, the branches of nA solutions
do not end in RN black holes.
Also, according to our description of the heuristic argument cancerning Derrick scaling at
the end of Section {\bf 2.2}, finite mass/energy solutions can be constructed in
spacetime dimensions $D\ge 7$ even in the absence of the gravitational term in the
Lagrangian, $i.e.$ for a fixed Minkowski background.

The reason is that only in $D=5$ is it necessary to have the Einstein--Hilbert term in the Lagrangian to satisfy the
(heuristic) Derrick scaling requirement. In that case, the Yang-Mills term scales as $L^{-4}$ so that in four spacelike
dimensions the scaling $L^{-5}$ of the CS term is not balanced in the absence of gravity, the latter scaling as $L^{-2}$.
Generally, the CS terms scales as $L^{-(2n+1)}$, in $2n$ spacelike dimensions.
\setlength{\unitlength}{1cm}
\begin{picture}(8,6)
\put(-0.5,0.0){\epsfig{file=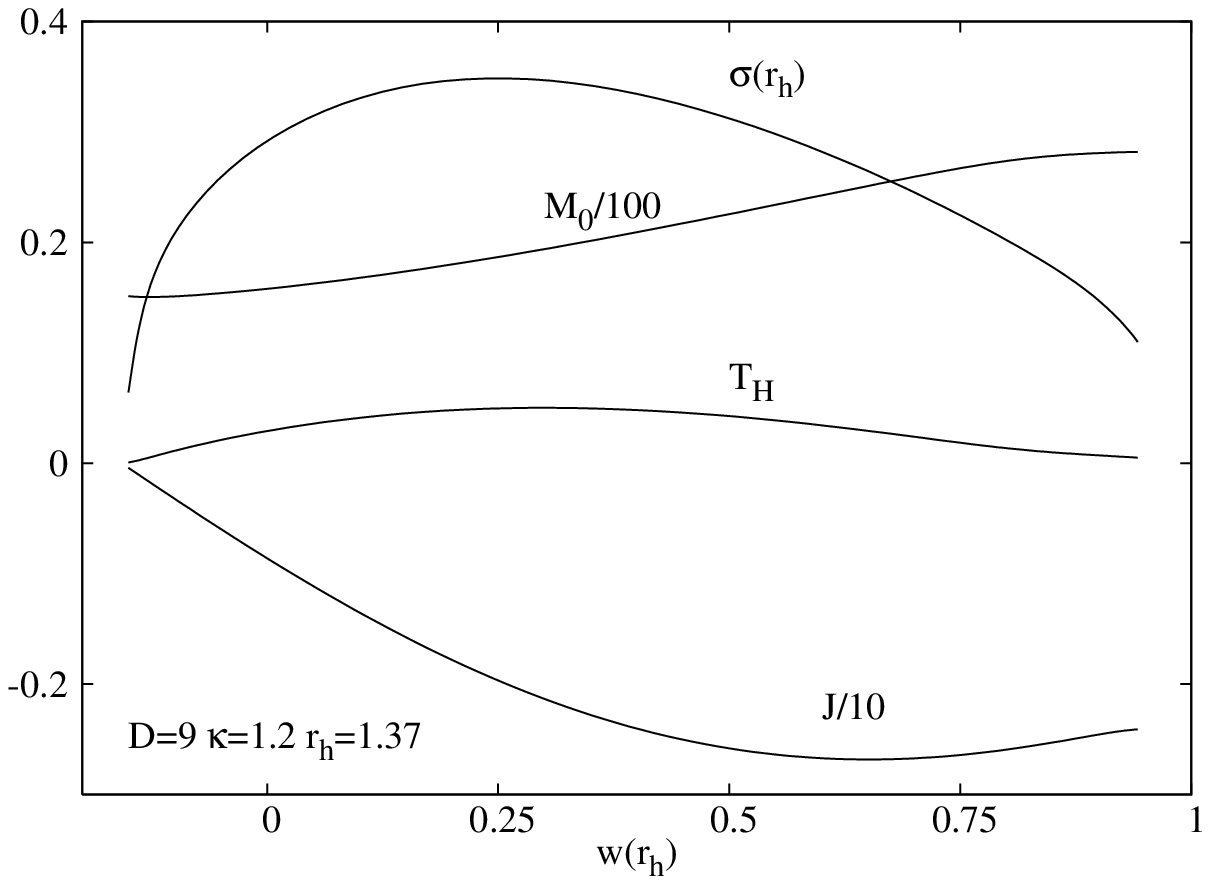,width=8cm}}
\put(8,0.0){\epsfig{file=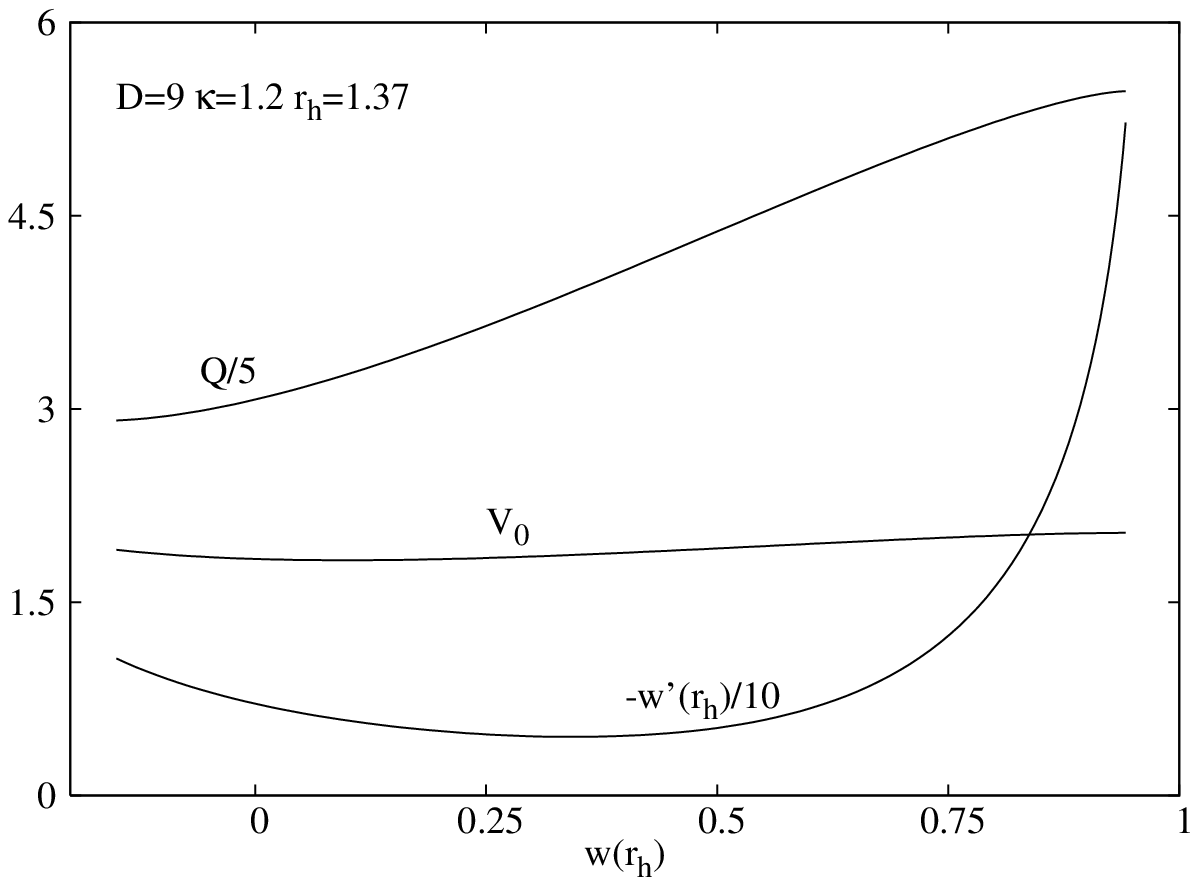,width=8cm}}
\end{picture}
\\
\\
{\small {\bf Figure 13.}
A number of parameters of  
$D=9$ black hole solutions are plotted as a function
 of the value $w(r_h)$ of the magnetic potential on the horizon for fixed values of $\kappa,~r_h$. }
\vspace{0.5cm}
 \\
\setlength{\unitlength}{1cm}
\begin{picture}(8,6)
\put(-0.5,0.0){\epsfig{file=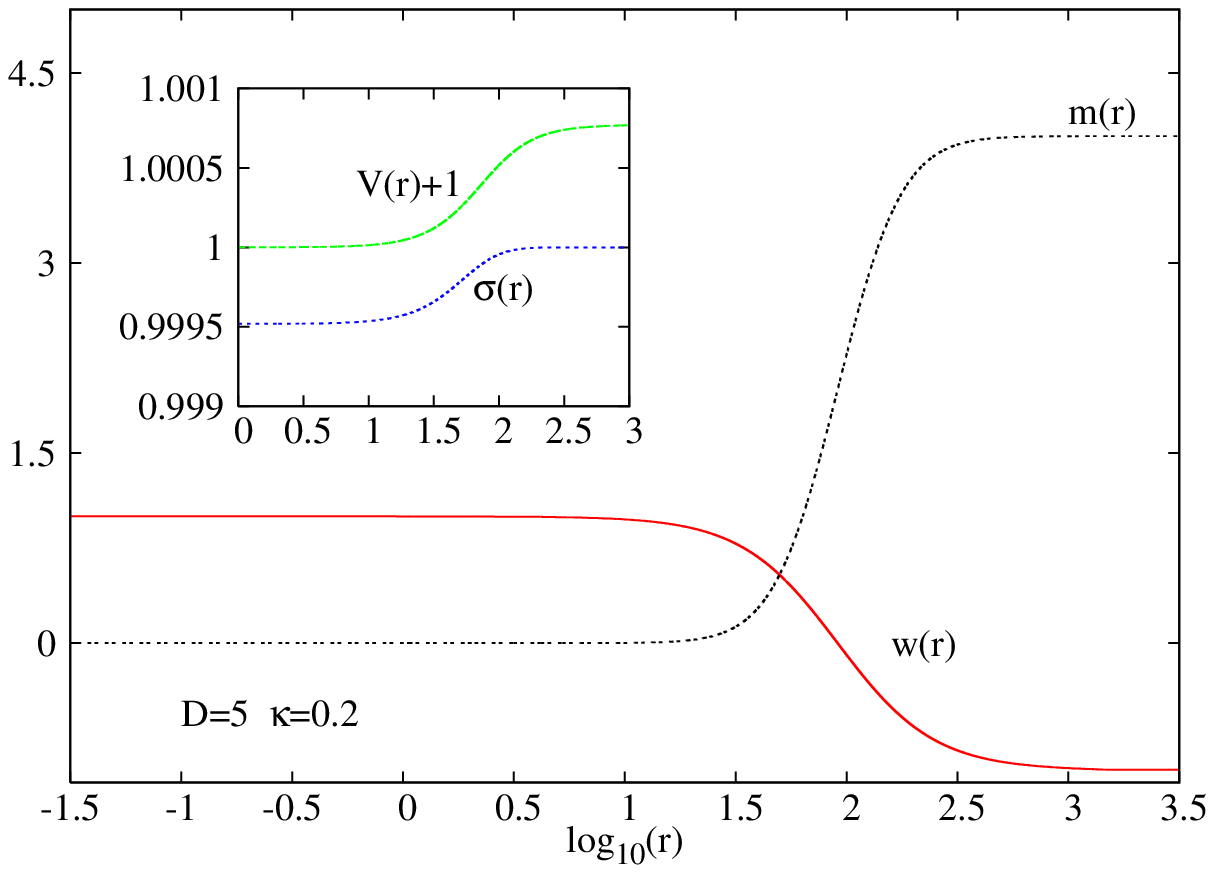,width=8cm}}
\put(8,0.0){\epsfig{file=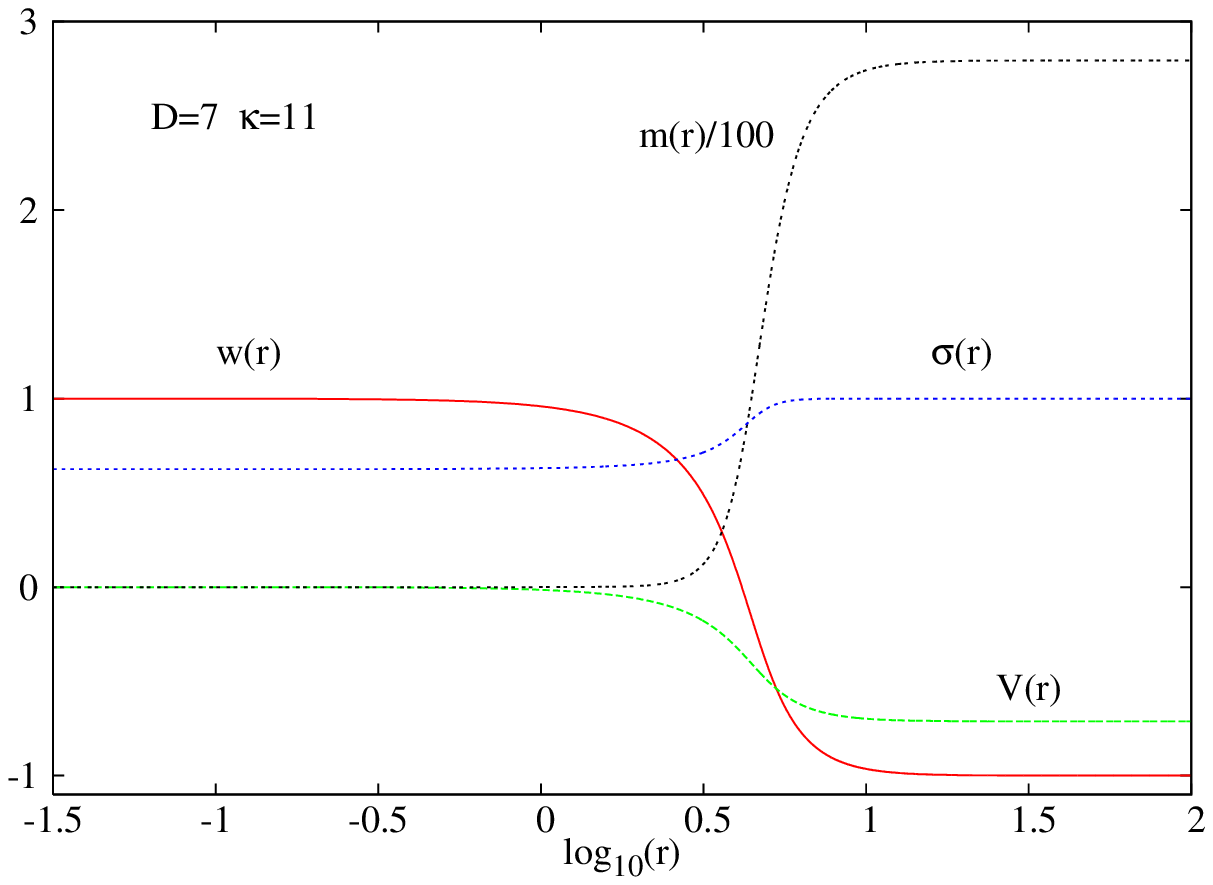,width=8cm}}
\end{picture}
\\
\\
{\small {\bf Figure 14.}
The profiles  of typical globally regular, particle-like EYMCS solutions  are presented 
as function of the radial coordinate $r$  
for $D=5,7$ spacetime dimensions. 
 }
\vspace{0.5cm}
\\
Then the Derrick scaling requirement can be
satisfied if the Yang-Mills term scales as $L^{-4p}$, provided that
\be
\label{Derrick}
2n+1>2n>4p\,,
\ee
$i.e.$, when $D>4p+1$. The Yang-Mills terms that scales as $L^{-4p}$ are the $p-$th members of the YM
hierarchy~\cite{Tchrakian:1984gq}
\be
\label{pYM}
{\cal L}_{\rm YM}^{(p)}=\frac{1}{2\cdot(2p)!}\mbox{Tr}\,\bigg\{ F(2p)^2\bigg\}\,.
\ee
Then the Lagrangian that satisfies the requirement (\ref{Derrick}) is
\be
\label{burz}
{\cal L}_{\rm matter}={\cal L}_{\rm YM}^{(p)}+\ka\,{\cal L}_{\rm CS}^{(2n+1)}\,,\quad n>2p\,.
\ee
This is possible to satisfy for $n\ge 3$, $i.e.$ $D\geq 7$. Thus for $D=7$ and $D=9$, the only choice of YM term is the
usual $p=1$ member
of the YM hierarchy ($i.e.$ the one in (\ref{action})), while for $D>9$ either one 
or both of $p=1$ and $p=2$ YM terms \re{pYM} are possible choices, $etc.$
All properties of these solitons are fixed by the value of the CS coupling constant, the relation (\ref{QeP})
being valid also in that case. A detailed discussion of the  
nongravitating YMCS solutions, including an existence
proof, will be presented elsewhere.

Returning to gravitating configurations, one notes that, however,
to some extent, the families of $D>5$ black holes resemble their five-dimensional counterparts.
In particular, the profiles of the solutions look similar to those exhibited in Section {\bf 4} and hence
will not be plotted here again. As in $D=5$, no multinode solutions of the magnetic gauge potential $w$
were observed for $D>5$.   There are, however, marked qualitative differences in some properties of the
gravitation solutions in $D>5$, $vs.$ those in $D=5$. These are discussed below.

 Starting with the dependence of black hole properties
 on the electric charge parameter $Q$ for a fixed event horizon radius $r_h$,
 we plot in Figure 11  the results
of the numerical integration for a given value of the CS coupling constant $\kappa$.
It seems that, for any $r_h$, the solutions exist for 
a finite range\footnote{Although
the solutions in Figure 11 have $Q>Q^{(c)}$, for other values of 
$r_h$ we could find solutions with $Q<Q^{(c)}$ as well.} 
of $Q$. 
 (Thus, as expected, one cannot find black holes
with an arbitrarily large electric charge.)
 
The limiting behaviour of these solutions at the limit of the
$Q$-interval is  qualitatively different from the case $D=5$. 
In terms of the value $w_h$ of the magnetic gauge potential on the horizon, the solutions exist for 
$w_h \in (w_{h,min},1)$ where the minimal value  $w_{h,min}$ depends on $r_h$ (although always with
$w_{h,min}\neq -1$, $i.e.$ the branch of nA solutions does not join the RN configuration).
The numerical results suggest that in this case the limiting configurations 
consist of extremal black holes with a regular horizon, see the Figure 11 (left).
Another feature of the solutions which is worth mentioning is that for intermediate values of the horizon 
  radius  there exists a large region of the parameter $w_h$ for which both 
$V'(r_h)$ and $w'(r_h)$ are very close to zero.
Moreover, for some intermediate values of $Q$, we notice the existence of 
two different solutions with the same charge parameter.

The behaviour of solutions as a function of the event horizon radius is shown in Figure 12
for several values of the charge parameter $Q$ and a given $\kappa$.
In terms of the scaled event horizon radius $a_H$ and scaled temperature $t_H$,
the sets of nA black holes with $Q\neq Q^{(c)}$ interpolate between two extremal configurations
and one can notice the existence of three branches of solutions. 
The first branch of nA solutions starts from an extremal black hole with $w(r_h)$ taking a minimal value.
This solution extends up to a maximal value of $r_h$, where a second
branch of nonextremal solutions emerges, extending backwards in $r_h$.
For the same $r_h$, the mass of one of these solutions is larger
than the corresponding mass on the first branch (when they both exist).  
Also, this secondary branch has a negative specific heat, while the entropy increases with the temperature 
for the first branch.
Finally, a third branch of solutions emerges for a critical $r_h$, which has again a positive specific heat.
This branch ends in an extremal configuration with a minimal $r_h$. 

The picture is different for $Q=Q^{(c)}$, in which case the third branch is absent
and the second branch of solutions extends to $r_h\to 0$, see Figure 12 (left). 
The limit of a vanishing event horizon radius corresponds to a globally regular, particle-like
configuration.
(This feature is similar to
the $D=5$ result.) 

The $D=9$ solutions we have studied possess a similar pattern. 
A number of results in this case are shown in Figure 13 as a function
of the parameter $w(r_h)$. 
One can see that the limits of the domain of existence corresponds to
extremal configurations with $T_H=0$.

We did not consider the issue of stability of $D>5$ solutions. In principle, this is a straightforward extension of the 
work in $D=5$, the problem reducing again to a single Schr\"odinger equation. 
The fact that for any $D$ we have found nodeless solutions suggests the existence  in all dimensions
of configurations
which are stable against spherically symmetric perturbations.

As mentioned already, 
similar to $D=5$, one finds also a different class of solutions describing globally regular solitons
with $Q=Q^{(c)}$. In Figure 14 (right) we show the profile of such a configuration in $D=7$ spacetime dimensions.
One can see some differences between this solution and the one in $D=5$, the distorsion of the spacetime geometry
being more pronounced in the higher dimensional case.

\section{Conclusions and further remarks}
\setcounter{equation}{0}
The main purpose of this work was to present a general study of 
static spherically symmetric, asymptotically Minkowskian, 
solutions in a simple EYMCS model in $D=2n+1$ dimensions. 
Our choice of gauge group is $SO(D-1)\times SO(2)$,
which is just sufficient to support a nonvanishing Chern--Simons density. While the smallest simplest gauge
group with this property is $SO(D+1)$, the asymptotic analysis suggested that there were no well behaved
solutions in that case. This is in contrast with the asymptotically AdS solutions encountered in \cite{Brihaye:2009cc},
which were found for the full $SO(D+1)$ gauge group (with $D=5$).

The CS term allows to avoid the usual Derrick-type scaling argument against the existence of $D>4$
static nA configurations with finite mass in the usual EYM system. This provides an $alternative$ to the 
higher order curvature terms of the YM hierarchy employed previously to reach the same result \cite{Brihaye:2002hr},
presenting at the same time a richer pattern.

While we could display a number of analytic results
valid for any (odd) values of $D>3$, our main numerical analysis was restricted to the case of
$D=5$ and  and to a lesser extent to $D=7$. 
In addition, we could confirm the existence of solutions for $D=9$
as well.
 One should emphasise that the properties of the EYMCS solutions in this work are 
strikingly different  from 
solutions to other nA models without a CS term, considered in the literature.
  
The main interesting results are for $D=5$, in which case they can be summarised as follows:
\begin{itemize}
\item 
The black hole solutions emerge as perturbations of the RN solution, which becomes
unstable when embedded in a larger gauge group.
\item
The nA solutions are generically thermodynamically favoured over the Abelian
configurations. 
\item
Some of the nA configurations were shown to be stable against small perturbations.
\item
A solution in closed form is found for the minimal value of the CS coupling constant, describing
an extremal black hole with nA hair. In the absence of a horizon, this becomes the soliton found in \cite{Gibbons:1993xt}.
\end{itemize}

Our results in $D=7$ and $9$ differ somehow from those in $D=5$. Most importantly, the RN solution
in these dimensions does not become unstable when embedded in the corresponding (larger) nA gauge group (in fact this property
holds for any $D>5$). 
As a result, the nA black holes do not emerge as perturbations of the the RN solutions. 
Instead, for a fixed value of the electric charge $Q\neq Q^{(c)}$, they appear to interpolate between two nA extremal black holes.
These qualitative differences are a consequence of the choice of EYMCS
model made here, which in any case does not seem to be a consistent truncation of a known  supergravity theory.
However, it is possible to search for different EYMCS models in higher (than five) dimensions, whose solutions are
likely to fulfill the properties of the $D=5$ solutions itemised above. For this we note that in $D=5$ the CS term scales
as $L^{-5}$, $versus$ the YM term, which scales as $L^{-4}$. In this respect, it would be useful if in the higher $2n+1$
dimensions where the CS term scales as $L^{-(2n+1)}$, the corresponding YM term would scale as $L^{-2n}$. This can be
achieved only in $D=4p+1$ dimensions, where the CS term scales as $L^{-(4p+1)}$, by replacing the usual YM term $F(2)^2$
(the $p=1$ member of the YM hierarchy) with the $p-$th member of the YM hierarchy which scales as $L^{-4p}$
\cite{Radu:2009rs}.  It is obvious that this can only be done when $n=2p$, thus the
properties itemised above cannot be duplicated in $D=4p+3$ dimensions.

This remark applies equally to the last of the properties itemised above, namely that closed form solutions can be
constructed in all $4p+1$ dimensions, for a special value of the CS coupling constant. 
While in the $p=1$ case discussed here the (usual) $p=1$ BPST instanton
is employed, in the case of a generic $p$ the instanton~\cite{Tchrakian:1984gq} of the $p-$th member of the YM
hierarchy is employed. This results in a general class of $p\geq 1$
exact solutions, whose properties   are similar to those of the $D=5$ configurations
discussed in Section {\bf 4.6}.

It is obvious that the black hole solutions in this work violate the no hair conjecture,
that is, two distinct solutions can exist for a given set of global charges.
Moreover, some of the nA solutions are really clasically stable,
because they have maximum entropy among the black holes with the same mass and charge.
This behaviour is somehow similar to that found in \cite{Lee:1991vy} for a family of monopole
black holes in $D=4$ Einstein--Yang-Mills--Higgs system.
There too, a branch of monopole nA black holes merges with the $magnetic$ RN solutions, which is unstable for some range of the 
parameters. Thus, similar to the case in \cite{Lee:1991vy}, the hairy black holes in this work may be relevant
for the issue of the final stage of an evaporating $D=2n+1$ RN black hole.

In principle, the study in this work can be extended in various directions.
For example, it will be interesting to consider more general asymptotics and solutions
describing black strings and $p-$branes. Another possible direction would be to include rotation.
However, even for the case of static solutions approaching the Minkowski background
at infinity, with a gauge group $SO(D-1)\times SO(2)$, we expect to find a variety
of interesting solutions. For example, we expect both EYMCS solitons and black holes to exist, 
which are static but not spherically symmetric. Indeed, such solutions were found in $D=4$
EYM system \cite{Kleihaus:1996vi}.

We close our discusion with some comments on another intriguing feature of the $D=5$ solutions.
Despite the different asymptotic structure of spacetime and the different 
horizon topology, these solutions have some similarities with the
colorful black holes with charge in AdS space \cite{Gubser:2008zu}, 
\cite{Manvelyan:2008sv}, \cite{Ammon:2009xh},
which provide a model of holographic superconductors. 
In both cases, an Abelian gauge symmetry is spontaneously
broken near a black hole horizon with the appearance of a condensate of nA gauge fields there,
leading to a  phase transition.  
Also, one can notice a striking similarity of the $J(T_H)$
curves shown in Figure 6 with some of those exhibited in the literature on AdS holographic superconductors.

It remains an interesting open problem to clarify if the asymptotically flat EYMCS black holes
may also provide useful analogies to phenomena observed in condensed matter physics.
The first step in this direction would be to compute the conductivity as
a function of frequency. This is obtained by perturbing the YM fields around the horizon.
However, given the presence of several branches, the general
picture is more complicated for asymptotically flat solutions,
already for the fundamental RN set of solutions. Also, in the absence of a cosmological
constant, the gauge/gravity duality (which seems to provide the deep reason behind the connection  
between general relativity solutions and condensed matter physics) is not yet understood.
At the same time, some of the features of the AdS
holographic duals of superconductors may occur for other asymptotics
as well, being generic properties of certain classes of hairy black holes.
We hope to return to a study of these aspects in a separate paper.

\section*{Acknowledgements}
 This work is carried out in the framework of Science Foundation Ireland (SFI) project
RFP07-330PHY. 
YB is grateful to the
Belgian FNRS for financial support.

\begin{small}

\end{small}

\end{document}